\documentclass[14pt,svgnames,psnames]{article}
\usepackage{epsf}
\usepackage{graphicx}
\usepackage{framed}
\usepackage{amsmath,amsfonts,amsbsy,amssymb}
\numberwithin{equation}{section}
\usepackage{indentfirst}
\usepackage{mathtools}
\usepackage{arydshln} 
\usepackage{framed}
\usepackage{xcolor}

\usepackage{hyperref}

\hypersetup{
    colorlinks=true,      
    linkcolor=DarkBlue,       
    citecolor=DarkRed,        
    urlcolor=DarkGreen         
}

\newcommand{\nn}{\nonumber}

\def\be{\begin{equation}}
\def\ee{\end{equation}}
\def\bse{\begin{subequations}}
\def\ese{\end{subequations}}

\def\bs{\boldsymbol}

\begin{document}

\begin{titlepage}

\def\slash#1{{\rlap{$#1$} \thinspace/}}

\begin{flushright} 

\end{flushright} 

\vspace{0.1cm}

\begin{Large}
\begin{center}

{\bf
Super Landau Model and Howe Duality: \\ From Supermonopole Harmonics to  Quantum Matrix Geometry}
\end{center}
\end{Large}

\vspace{1cm}

\begin{center}
{\bf Kazuki Hasebe}   \\ 
\vspace{0.5cm} 
\it{
National Institute of Technology, Sendai College,  
Ayashi, Sendai, 989-3128, Japan} \\

\vspace{0.6cm} 

{\sf
khasebe@sendai-nct.ac.jp}

\vspace{0.6cm} 

{\today} 

\end{center}

\vspace{1.4cm}

\begin{abstract}
\noindent

\baselineskip=18pt

Landau models serve as quantum mechanical systems for generating quantum matrix geometries. In this paper, we demonstrate that  Howe duality provides the underlying structure of the super Landau model,  reflecting a general feature of  coset-type  Landau models.  
 The (super) Howe duality relates different  Landau levels and accounts for the emergence of a dual fuzzy geometry. 
By employing super-spinor derivative operators,   
the supermonopole harmonics in both integer and half-integer Landau levels are explicitly constructed and the algebraic structure of the super-Hilbert space is revealed.  We  propose  a consistent  probabilistic interpretation for these  wavefunctions defined on a supermanifold. 
Through a level projection method, we derive the matrix coordinates of  fuzzy supersphere geometries for arbitrary Landau levels,  along with a precise determination of the non-commutative scale factor. 
It is  shown that the theta correspondence of  Howe duality  induces a geometric transformation between fuzzy objects.   
Finally, we  point out that  Howe duality realizes an internal-external space duality  and underlies quantum matrix geometries, suggesting that it may play a fundamental role in understanding  Matrix model geometries.

\end{abstract}

\end{titlepage}

\newpage 

\tableofcontents

\newpage 

\section{Introduction}

 Landau models provide the quantum mechanical foundation of the quantum Hall effect (QHE) and  a physical realization of non-commutative geometry (NCG). In the context of the QHE, NCG is not merely an underlying geometric framework but the central mechanism governing the dynamics of collective excitations \cite{Girvin-Jach-1984, Girvin-MacDonald-Platzman-1986, Ezawa-Tsitsishvili-Hasebe-2003}. Furthermore, NCG is essential for elucidating the geometric structures of D-branes and  Matrix theory \cite{Kabat-Taylor-1998, Taylor-2001}. Specifically, fuzzy geometries are  realized as classical solutions of Matrix models \cite{Myers:1999ps, Castelino:1997rv, Iso:2001mg,  Kimura:2001uk, Kimura:2002nq, Kimura:2003ab, Chaney:2015mfa,  Hasebe-2023-1, Gohara:2025zfh} and play a fundamental role in the recent  developments of emergent gravity  \cite{Steinacker:2016vgf, Sperling:2018xrm, Stern:2018wud, Steinacker:2023zrb}. 
 It is also known  that the  non-anticommutative structure of fermionic coordinates referred to as non-anti-commutative geometry (NACG) is induced in a graviphoton background  \cite{Ooguri-Vafa-2003-1, Ooguri-Vafa-2003-2, Boer-Grassi-vanNieuwenhuizen-2003, Seiberg-2003}. The frameworks of NCG and NACG are naturally unified within the supersymmetric (SUSY) NCG formalism \cite{Grosse-Klimcik-Presnajder-1997, Grosse-Reiter-1998, Ivanov-Mezincescu-Townsend-2003, hep-th/0511114}. An archetypical example of such a manifold is the fuzzy supersphere \cite{Grosse-Klimcik-Presnajder-1997, Grosse-Reiter-1998, Hasebe-2011}, which realizes covariant bosonic and fermionic matrix coordinates in a graded Lie algebraic fashion. The fuzzy supersphere has been shown to arise as a classical solution of a supermatrix model \cite{Iso-Umetsu-2004}.

  More than two decades ago, with a collaborator the author  introduced a supersymmetric generalization of the Landau model on a supersphere \cite{Hasebe-Kimura-2005}. He also  formulated the SUSY QHE \cite{Hasebe-2005-1} as  a natural SUSY counterpart of Haldane's spherical QHE \cite{Haldane-1983}. Just  as the fuzzy sphere geometry \cite{Hoppe-1982, Madore-1992} appears in the  lowest Landau level (LLL)  \cite{Hatsuda:2003ry, Hasebe-2010}, it was shown that  the fuzzy supersphere geometry emerges in the super LLL \cite{Hasebe-Kimura-2005}. The formulation of the SUSY Landau model was extended from a  supersphere to a super-plane \cite{Hasebe-2005-2, Ivanov-Mezincescu-Townsend-2006, Curtright-Ivanov-Mezincescu-Townsend-2007, Ivanov-2007, Melong:2025cis} and a  superhyperboloid \cite{Hasebe-2008-1}. A SUSY Chern-Simons field theory was also introduced as the effective field theory for the SUSY QHE \cite{Hasebe-2006}.  
  Interestingly, the SUSY brings  a unification of the Laughlin and Moore-Read wavefunctions in a single super Laughlin wavefunction \cite{Hasebe-2008-2}.  
Over the past two decades,  the studies of the SUSY QHE have evolved in various ways. The first line of the development concerns the SUSY generalization of the valence bond solid state: The super Laughlin-Haldane wavefunction is transformed to the SUSY valence bond solid state by translating the external superspace as an internal superspace  \cite{Arovas-Hasebe-Qi-Zhang-2009}.  The SUSY valence bond solid state exhibits a hidden  order \cite{Hasebe-Totsuka-2011} leading to the notion of  SUSY protected topological phases \cite{Hasebe-Totsuka-2013} and SUSY matrix product states \cite{Hasebe-Totsuka-2013-2}. Recent  developments along this line can be  found  in   \cite{Gao-Yang-Wang-2026, Totsuka-2023}. A more direct  application of the SUSY QHE is about the analysis of the neutral fermionic excitation in the  Moore-Read QH state. Moore-Read state possesses both QH and superconducting natures, and its bulk collective excitations are  bosonic magneto-roton excitation and neutral fermion excitation  regarded as  mutual superpartners. Based on the SUSY formalism, the wavefunction of the neutral fermion excitation was proposed \cite{Gromov-Martinec-Ryu-2020} and  its validity  was numerically confirmed  \cite{Pu-Balram-Fremling-et-al-2023}.    A non-relativistic supergeometry for the Moore-Read state has  been discussed in \cite{SalgadoRebolledo-Palumbo-2022}.  It is also reported that  the SUSY QHE is closely related to the M-theoretic origin of topological order \cite{Sati-Schreiber-2025}. Other significant results regarding the SUSY structure of the QHE can be found in the literature \cite{Dolan:2005mb, Tong-Turner-2015,  Sagi-Santos-2016, Bae-Lee-2021, Ma-Wang-Yang-2021, Nguyen-Prabhu-Balram-Gromov-2022}.   It is also worth mentioning the developments of superqubits \cite{Borsten-Dahanayake-Duff-Rubens-2010, Borsten-Bradler-Duff-2014, Borsten-Bradler-Duff-2016}.    Since both the superqubit and the SUSY QHE  are founded on the graded Hopf map, their SUSY generalizations  share many analogous properties.

In parallel, the NCG associated with Landau models has been extensively investigated over for more than   two decades \cite{Zhang-Hu-2001, Karabali-Nair-2002, Hasebe-Kimura-2003, Nair-Daemi-2004, Karabali-Nair-2006,  Daoud:2006kr, Hasebe:2012mz, Hasebe-2014-1, Hasebe-2014-2,  Balli:2014pqa, Hasebe-2017-1, Ishiki-Matsumoto-Muraki-2018, Adachi-Ishiki-Matsumoto-Saito-2020, Adachi-Ishiki-Kanno-2022}. Based on the idea of utilizing the Landau models to generate  quantum matrix geometries \cite{Hasebe-2023-1, Hasebe-2016},  fuzzy matrix geometries of two-, three-, and four-spheres have been revealed in arbitrary LLs  \cite{Hasebe-2016, Hasebe-2017, Hasebe-2020, Hasebe-2021}.  This method has been successfully applied to derive new solutions for the Matrix model \cite{Hasebe-2023-1} and to construct deformed matrix geometries \cite{Hasebe-2025}. These results indeed demonstrate the effectiveness of the Landau models for generating matrix geometries and  strongly suggest that Matrix model geometries are  realized in  Landau models.  
It is worthwhile to investigate whether this approach can be equally  applicable to the  derivation of supermatrix geometries. 
Introducing NCG in  superspace is known to   (partially) break the SUSY structure in general \cite{Seiberg-2003}, however, fuzzy supermanifolds elude this unfavorable property.  
Importantly, the graded Lie algebraic construction of SUSY NCG preserves the full SUSY of the original supermanifold. For instance, as the supersphere possesses the $\mathcal{N}=1$ SUSY, the fuzzy supersphere maintains the same $\mathcal{N}=1$ SUSY. The preservation of (super)symmetry is a remarkable feature of fuzzy spaces. Although fuzzy spaces are regarded as ``discretized" geometries, they  retain the continuous symmetries of the original manifolds. This stands in stark contrast to lattice models, where continuous spatial symmetries are  reduced to discrete  lattice symmetries. This unique property  is considered to be a key factor behind the recent development of the numerical simulations of  3D Ising criticality  using fuzzy-sphere regularization \cite{Zhu-Han-Huffman-Hofmann-He-2022, Hu-He-Zhu-2023}.  
Furthermore, very recent studies \cite{Zhou-Gaiotto-He-2025, Tang-Voinea-Hu-et-al-2025} suggest a close relationship between the fuzzy supersphere and the 3D SUSY Ising theory. The fuzzy supersphere and the fuzzy super Landau model are of increasing importance. However, an entire picture that encompasses higher super  LLs   has yet to be established.

In this paper, we investigate the super Landau model on a supersphere \cite{Hasebe-Kimura-2005} to explore a comprehensive understanding of the super LL structures and  associated matrix geometries. By incorporating  higher LL structures, we unveil that the Howe duality constitutes the fundamental  algebraic structure of the system. We show that the supermonopole harmonics form an orthonormal basis and  propose their consistent probabilistic interpretation by projecting them onto the body-sphere along with an appropriate definition of inner product.  Using the newly defined inner product, the supermatrix coordinates are evaluated for arbitrary Landau levels. 
This presents the first explicit derivation of the supermatrix geometry from a Landau model.   
 We also discuss the significance of  Howe duality in the context of quantum matrix geometry. 
 The theta correspondence  is shown to induce a geometric transformation between fuzzy objects in the framework of NCG. Viewing  Howe duality as a duality between  internal and external spaces, we argue that it is a general feature of  coset-type Landau models and   may underlie the geometric structure of  quantum geometries in Matrix models.

This paper is organized as follows. In Sec.\ref{sec:gradhopf}, we review the basics of the super Landau model.  Super-spinor derivative operators are constructed in Sec.\ref{sec:effecop}. In Sec.\ref{sec:supermonoeha}, we derive the supermonopole harmonics for integer LLs and explore their  probabilistic interpretation.  
This analysis is extended in Sec.\ref{sec:supermonoehahalf} for half-integer LLs,  where we redefine the  inner product to handle negative-norm states.    
 The super D-matrix is introduced in Sec.\ref{sec:superdmat}, where the underlying super Howe duality of the system is  revealed.  We derive the supermatrix geometry in 
 Sec.\ref{sec:supermatrix}, providing a geometric realization of the theta correspondence as a transformation of fuzzy manifolds.    Sec.\ref{sec:summary} is devoted to a summary and discussions.

\section{Review of the super Landau model}\label{sec:gradhopf}

  We review the basic structure of the super Landau model on a supersphere based on \cite{Hasebe-Kimura-2005}, adding some more useful  information. The correspondence between the original Landau model and its SUSY counterpart is summarized in Table \ref{table:qhesqhe}.

\begin{table}
\center 
\begin{tabular}{|c|c|c|}\hline
        &  The original setup &  The SUSY setup    \\ \hline  
         Fibre-bundle        &  Principal fibre bundle     &  Principal superfibre bundle      \\ \hline  
      Topological  map           &  Hopf map     &  Graded Hopf map      \\ \hline  
 Symmetry group          &  $SU(2)$     &  $UOSp(1|2)$       \\ \hline  
 Gauge group          &  $U(1)$     &  $U(1)$       \\ \hline  
 Base manifold           &  $S^2$     &  $S^{2|2}$        \\ \hline   
  Magnetic source           & Dirac Monopole     &  Supermonopole   \\ \hline 
  Landau level eigenstates           &  Monopole harmonics     &  Supermonopole harmonics        \\ \hline 
   Non-commutative geometry            &  Fuzzy sphere     &  Fuzzy supersphere       \\ \hline 
    \end{tabular}       
\caption{ The basic mathematical  setups for the original Landau model and the super Landau model. 
}
\label{table:qhesqhe}
\end{table}

\subsection{The graded Hopf map and supersphere}

The graded Hopf map \cite{LandiMarmo1987, Bartocci1990,Landi2001} is defined by the following mapping  
\be
S^{3|2} ~~\overset{S^1}{\longrightarrow}~~S^{2|2}. \label{gradedhopfab}
\ee
As a coset, $S^{2|2}$ is expressed as
\be
S^{2|2}~\simeq ~S^{3|2}/S^1 ~\simeq~UOSp(1|2)/U(1),  
\ee
whose bosonic part realizes the original Hopf map,  
$S^{2}~\simeq ~S^{3}/S^1 ~\simeq~SU(2)/U(1)$. 
With two complex Grassmann even quantities, $u$ and $v$, and a complex Grassmann odd quantity $\eta$, arbitrary $UOSp(1|2)$\footnote{While the notation $OSp(1|2)$ is frequently used in place of $UOSp(1|2)$ in the literature, we strictly distinguish between these in the present paper for preciseness. The bosonic subgroup of the $UOSp(1|2)$ is the compact $SU(2)$ group, while that of the $OSp(1|2)$ is the non-compact $SU(1,1)$ group.} group element is represented as \cite{Berezin-Tolstoy-1981}
\be
g=
\begin{pmatrix}
u & -v & \eta \\
v^* & u^* & \eta^* \\
-u \eta^* + v^*\eta & u^*\eta +v\eta^* & 1- \eta^*\eta
\end{pmatrix}. \label{guosp12m}
\ee
The condition $\text{sdet}(g) =1$ and $g^{\ddagger}=g^{-1}$.\footnote{
See Appendix \ref{sec:berez} for the definition of the superdeterminant and the super adjoint is given by 
\be
\begin{pmatrix}
A & C \\
D & B 
\end{pmatrix}^{\ddagger}:=\begin{pmatrix}
A^{\dagger} & -D^{\dagger} \\
C^{\dagger} & B^{\dagger} 
\end{pmatrix}.
\ee
} 
imposes a constraint on the parameters as 
\be
u^*u+v^*v -\eta^*\eta=1, \label{constrpsi}
\ee
which geometrically signifies  $UOSp(1|2)\simeq S^{3|2}$. 
With a three-component spinor $\psi$ (the super Hopf spinor) subject to the normalization condition $\psi^{\ddagger} \psi=\psi^t\psi^*=1$, the graded Hopf map (\ref{gradedhopfab}) is explicitly given by  
\be
\psi =\begin{pmatrix}
u \\
v\\
\eta
\end{pmatrix} ~~\longrightarrow~~\frac{x_i}{r} =2\psi^{t}l_i \psi^*, ~~~\frac{\theta_{\alpha}}{r} =2\psi^{t}l_{\alpha} \psi^*, \label{Hopfmapsusy}
\ee
where $r$ signifies a constant parameter and 
\be
\psi^{\ddagger} :=
\begin{pmatrix}
u^* &
v^* &
-\eta^*
\end{pmatrix}. 
\ee
Note the minus sign in front of $\eta^*$.\footnote{
Here, $*$ denotes the pseudo-conjugation which acts on the Grassmann odd quantities as  
\be
(\eta_1\eta_2)^* =\eta^*_1\eta_2^*, ~~~(\eta^*)^*=-\eta. 
\ee
Then, $\theta_{\alpha}$ (\ref{Hopfmapsusy}) satisfy 
\be
{\theta_{\alpha}}^*=C_{\alpha\beta}\theta_{\beta}, 
\ee
which implies that 
$\theta_{\alpha}$ form an $SU(2)$ (pseudo-)Majorana spinor. } 
The symbols $l_i$ and $l_{\alpha}$ denote the $UOSp(1|2)$ superspin matrices: 
\be
l_i =\frac{1}{2}\begin{pmatrix}
\sigma_i & 0 \\
0 & 0 
\end{pmatrix}, ~~~l_{\theta_1} =\frac{1}{2}\begin{pmatrix}
0 & 0 & 1  \\
0 & 0  & 0 \\
0 & 1 & 0  
\end{pmatrix}, ~~~l_{\theta_2} =\frac{1}{2}\begin{pmatrix}
0 & 0 & 0  \\
0 & 0  & 1 \\
-1 & 0 & 0  
\end{pmatrix}. \label{l1/2mat}
\ee
 The $uosp(1|2)$ algebra is summarized in Appendix \ref{appen;uso12mat}. 
  The super-coordinates (\ref{Hopfmapsusy}), the bosonic coordinates $x_i$  and the fermionic coordinates $\theta_{\alpha}$,  automatically satisfy the condition for $S^{2|2}$: 
\be
x_ix_i +C_{\alpha\beta}\theta_{\alpha}\theta_{\beta} =(\psi^{t} \psi^* )^2=r^2, \label{consxthe}
\ee
where $C$ denotes the charge conjugation matrix of the $SU(2)$: 
\be
C=i\sigma_2. 
\ee
In detail, the super-coordinates  (\ref{Hopfmapsusy}) are  given by  
\begin{align}
&\frac{x_1}{r} =uv^* +vu^*, ~~~\frac{x_2}{r}=-iuv^* +ivu^*, ~~~\frac{x_3}{r}=uu^*-vv^*, \nn\\
&\frac{\theta_1}{r}=v^*\eta +u\eta^*, ~~\frac{\theta_2}{r} =- u^*\eta +v\eta^* . \label{gramap}
\end{align}
Note that the super-coordinates are constructed from  the super Hopf spinor, suggesting that  the Hopf spinor is more fundamental than the coordinates. 
The supersphere (\ref{consxthe}) is invariant under the $\mathcal{N}=1$ SUSY transformation: 
\be
\delta_{\xi} x_i =\theta_{\alpha}(\sigma_i C)_{\alpha\beta}\xi_{\beta}, ~~~~\delta_{\xi}\theta_{\alpha} =x_i  (\sigma_i)_{\beta\alpha}\xi_\beta , \label{n1susyssp}
\ee
where $\xi_{\alpha}$ are Grassmann odd pseudo-Majorana  parameters, $\xi^*_{\alpha}=C_{\alpha\beta}\xi_{\beta}$.   
 
\subsection{Super Hopf spinor and the supermonopole gauge field}

The original Hopf spinor can be expressed as\footnote{The Hopf spinor regular on the southern hemisphere can be taken as 
\be
\phi'=\frac{1}{\sqrt{2r(r-y_3)}}\begin{pmatrix}
y_1+iy_2 \\
r-y_3
\end{pmatrix}. 
\ee
The corresponding gauge field is given by 
\be
\mathcal{A}'=dy_i \mathcal{A}'_i=2ig\phi'^{\dagger}d\phi'=g\frac{1}{r(r-y_3)}\epsilon_{ij3}y_j dy_i. 
\ee
The gauge transformation relating the $\mathcal{A}'$ and $\mathcal{A}$ is 
\be
\mathcal{A}' = \mathcal{A}-iu^{*}du, \label{gaudaa}
\ee
where 
\be
u =e^{-2ig \varphi}~~~~(\tan(\varphi) :=\frac{y_2}{y_1}). 
\ee
} 
\be
\phi(y_i)=\begin{pmatrix}
\mu \\
\nu 
\end{pmatrix}=\frac{1}{\sqrt{2r(r+y_3)}}
\begin{pmatrix}
r+y_3 \\
y_1-iy_2
\end{pmatrix}\cdot e^{i\chi}, \label{hopfspinor}
\ee
where  $y_i$ denote the ordinary c-number coordinates on $S^2$:  
\be
y_iy_i=r^2.
\ee
The $U(1)$ phase $e^{i\chi}$ signifies  the $S^1$-fibre on $S^{2}$. 
The corresponding gauge field is 
\be
\mathcal{A}=dy_i \mathcal{A}_i=2ig\phi^{\dagger}d\phi=-g\frac{1}{r(r+y_3)}\epsilon_{ij3}y_j dy_i. \label{mathau}
\ee
We took $e^{i\chi}=1$ in (\ref{mathau}).  
Similarly, the super Hopf spinor can be represented  as  
\be
\psi(x_i, \theta_{\alpha}) =\begin{pmatrix}
u \\
v\\
\eta
\end{pmatrix}=\frac{1}{\sqrt{2r^3(r+x_3)}}
\begin{pmatrix}
(r+x_3)(r-\frac{1}{4(r+x_3)}\theta C\theta) \\
(x_1-ix_2 )(r+\frac{1}{4(r+x_3)}\theta C\theta) \\
(x_1-ix_2)\theta_1  -(r+x_3) \theta_2 
\end{pmatrix}\cdot e^{i\chi}. \label{shopf}
\ee
The $U(1)$ phase factor $e^{i\chi}$ denotes  the $S^1$-fibre on $S^{2|2}$. 
Note that  the $x_i$ are Grassmann even quantities but $\it{not}$ ordinary c-numbers, as $x_i$ ``contain'' a Grassmann quantity $\theta C\theta$: indeed from (\ref{consxthe}), 
\be
x_i = y_i\sqrt{1-\frac{1}{r^2}\theta C\theta}. 
\ee
In the context of supermanifolds, the $c$-numbers $y_i$  are called the coordinates of the body, whereas the $\theta_{\alpha}$ are referred to as the soul coordinates. When the $\theta_{\alpha}$ are dropped, the bosonic coordinates $x_i$ reduce to the body coordinates $y_i$. 
Note that the $\eta$ (\ref{shopf}) is  an $SU(2)$ singlet 
\be
\eta=\frac{1}{r}C_{\alpha\beta}\theta_{\alpha}\psi_{\beta}=\frac{1}{r}(v \theta_1-u \theta_2),
\ee
and then 
\be
\eta^*\eta=-\frac{1}{2r^2}\theta C\theta. 
\ee
The super Hopf spinor (\ref{shopf}) is related to the original Hopf spinor (\ref{hopfspinor})   as 
\be
\psi =(1-\frac{1}{4r^2}\theta C\theta) \begin{pmatrix}
\mu \\
\nu \\
\frac{1}{r}(\nu\theta_1 -\mu\theta_2)
\end{pmatrix} .
\ee
The pseudo-conjugation of $\psi$ is given by 
\be
\psi^*  =\begin{pmatrix}
u^* \\
v^*\\
\eta^*
\end{pmatrix}=\frac{1}{\sqrt{2r^3(r+x_3)}}
\begin{pmatrix}
(r+x_3)(r-\frac{1}{4(r+x_3)}\theta C\theta) \\
(x_1+ix_2 )(r+\frac{1}{4(r+x_3)}\theta C\theta) \\
 (r+x_3)\theta_1+ (x_1+ix_2)\theta_2
\end{pmatrix}\cdot e^{-i\chi}. \label{cshopf}
\ee
The  supermonopole gauge field is derived as 
\be
A=2ig \psi^{\ddagger}d\psi =-2ig {\psi^*}^{\ddagger}d\psi^*, \label{gaugederi}
\ee
where $g$ signifies the magnetic charge of the supermonopole, $g=0, \pm 1/2, \pm 1,  \pm 3/2, \cdots $. (We took $e^{i\chi}=1$ in (\ref{gaugederi}).) 
In differential form, (\ref{gaugederi}) is given by  
\be
A =dx_i A_i +d\theta_{\alpha}A_{\alpha}, 
\ee
where  
\be
A_i =-g\frac{1}{r+x_3}\epsilon_{ij3}x_j \biggl(1+\frac{2r+x_3}{2r^2(r+x_3)}\theta C\theta\biggr),~~~~~
A_{\alpha} =-ig\frac{1}{r^3}(\theta \sigma_i C)_{\alpha}x_i. \label{supermogaubf}
\ee
Note that the Dirac string singularity appears only in  $A_i$, which is identical to the standard monopole gauge field:\footnote{ 
In the southern hemisphere, the regular super Hopf spinor is given by 
\be
\psi'=\frac{1}{\sqrt{2r^3(r-x_3)}}
\begin{pmatrix}
(x_i+ix_2) (r+\frac{1}{4r(r+x_3)}\theta C\theta) \\
(r-x_3) (r-\frac{1}{4r(r+x_3)}\theta C\theta) \\
(r-x_3)\theta_1 -(x_i+ix_2)\theta_2 
\end{pmatrix}.
\ee
The corresponding gauge field $A'=2ig {\psi'}^{\ddagger}d\psi'=dx_i A'_i +d\theta_{\alpha}A'_{\alpha}$ is  
\be
A'_i =g\frac{1}{r^3}\epsilon_{ij3}x_j \biggl(1+\frac{2r-x_3}{2r^2(r-x_3)}\theta C\theta\biggr), ~~~~A'_{\alpha} =-ig\frac{1}{r^3}(\theta \sigma_i C)_{\alpha}x_i. 
\ee
The bosonic part satisfies $dx_i A'_i =dy_i \mathcal{A}'_i$ and the fermionic gauge field $A'_{\alpha}$ is identical to $A_{\alpha}$. 
}
\be
dx_i A_i =dy_i \mathcal{A}_i. \label{aeqa}
\ee
Since the Dirac singularity reflects the non-trivial bundle topology,  the ``twist''  of the bosonic part of the superfibre bundle (\ref{aeqa}) is accounted for by  that of the original monopole bundle  on the body-sphere.  The gauge transformation between the supermonopole gauge fields on the norther and southern hemispheres is exactly the same as the original one (\ref{gaudaa}): 
\be
A'=A-iu^*du, 
\ee
where $u$ denotes the transition function 
\be
u: =e^{-2ig \varphi} =\biggl(\frac{y_1-iy_2}{\sqrt{r(r-y_3)}}\biggr)^{2g} = \biggl(\frac{x_1-ix_2}{\sqrt{r(r-x_3)}}\biggr)^{2g}(1+\frac{g}{r^2-{x_3}^2}\theta C\theta). \label{deftransfu}
\ee
We then have 
\be
-iu^*du =\frac{2g}{r^2-{y_3}^2} \epsilon_{ij3}y_j dy_i=\frac{2g}{r^2-{x_3}^2} \epsilon_{ij3}x_j dx_i (1+\frac{1}{r^2-{x_3}^2}\theta C\theta).
\ee
Note that the $-iu^*du$ does not include the  fermion coordinate 1-form $d\theta$.  The twist of the bundle is evaluated by the mapping from the equator  to the transition function. Since  the transition functions of the super and original Hopf maps  coincide (\ref{deftransfu}),   
the twist of the $S^1$-fibre of  the supermonopole can  reasonably be identified with that of the original monopole.   The winding number from the $S^1$-equator to the transition function $u\in U(1)$ is calculated as  
\be
i\frac{1}{2\pi}\int_0^{2\pi}u^*\frac{\partial u}{\partial \chi}d\chi  = 2g, \label{windingno}
\ee
which is equal to the Chern number of the monopole: 
\be
c_1 =\frac{1}{2\pi}\int_{S^2} f =2g, \label{chern1susy}
\ee
where  
\be
f=g\frac{1}{r^3}\epsilon_{ijk}y_k dy_i\wedge dy_j. 
\ee
The above discussion implies that the Chern number of the supermonopole is also identified as (\ref{chern1susy}). 
From (\ref{supermogaubf}),  the field strengths of the supermonopole are obtained as  
\be
F=dA=-\frac{1}{2}dx_j\wedge dx_i F_{ij}-d\theta_{\alpha}\wedge dx_i F_{i\alpha} -\frac{1}{2}d\theta_{\beta}\wedge d\theta_{\alpha}F_{\alpha\beta}, 
\ee 
where  
\begin{align}
F_{ij}&=-F_{ji}=\partial_i A_j -\partial_j A_i=g\frac{1}{r^3}\epsilon_{ijk}x_k (1+\frac{3}{2r^2}\theta C\theta   ), \nn\\
F_{i\alpha} &=-F_{\alpha i} =\partial_i A_{\alpha} -\partial_{\alpha} A_i=-ig\frac{1}{r^3}(\delta_{ij}-\frac{3}{r^2}x_ix_j )(\theta \sigma_j C)_{\alpha}, \nn\\
F_{\alpha\beta} &=F_{\beta\alpha}=\partial_{\alpha} A_\beta +\partial_{\beta} A_\alpha =-2i g \frac{1}{r^3}x_i (\sigma_i C)_{\alpha\beta}(1+\frac{3}{2r^2}\theta C\theta). \label{superfieldstrength}
\end{align}
Throughout this paper,  Grassmann derivatives such as  $\partial_{\alpha}$ are understood to act from the left.

\subsection{Super angular momentum operators}

The covariant derivatives are introduced as 
\be
D_i =\partial_i +iA_i, ~~~D_{\alpha} =\partial_{\alpha} +iA_\alpha, 
\ee
and we construct the super angular momentum operators as 
\be
L_i =\Lambda_i +r^2 B_i, ~~~~~~
L_{\alpha} = \Lambda_{\alpha}+r^2 B_{\alpha}. \label{llds}
\ee
Here, $\Lambda_i$ and $\Lambda_{\alpha}$ denote the covariant super angular momentum operators 
\be
\Lambda_i =-i\epsilon_{ijk}x_j D_k +\frac{1}{2}\theta_{\alpha}(\sigma_{i})_{\alpha\beta}D_{\beta}, ~~~~~\Lambda_{\alpha}= \frac{1}{2}x_i(C\sigma_i)_{\alpha\beta}D_{\beta} -\frac{1}{2}\theta_{\beta}(\sigma_i)_{\beta\alpha}D_{i}, 
\ee
and $B_i$ and $B_{\alpha}$ stand for the supermonopole magnetic field:  
\be
B_i = g \frac{1}{ r^3}x_i, ~~~~~B_{\alpha}= g \frac{1}{r^3}\theta_{\alpha}. \label{superbs}
\ee
These form a super-vector multiplet of the $UOSp(1|2)$ group, implying that $B_i$ and $B_{\alpha}$ are  superpartners. The supermagnetic field (\ref{superbs}) is radially distributed in $\mathbb{R}^{3|2}$, which indicates that  the supermagnetic monopole is located at the origin. 
The magnitude of the super magnetic field is 
\be
B=\sqrt{B_iB_i+C_{\alpha\beta}B_{\alpha}B_{\beta}} =\frac{|g|}{r^2},
\ee
and the magnetic charge of the supermonopole is evaluated from the (super) Gauss law:  
\be
\frac{\text{sgn}(g)}{2\pi}\int_{S^{2|2}}d\Omega_{2|2} B=\frac{\text{sgn}(g)}{2\pi}\int_{S^2}d\Omega_2 B=2g,  \label{integs22b}
\ee
where $d\Omega_{2|2}$ denotes the area element of $S^{2|2}$ (see Appendix \ref{sec:berez})
\be
\int_{S^{2|2}}d\Omega_{2|2} =\int_{S^2}d\Omega_2 \int d\theta_1 d\theta_2 (1-\frac{1}{2}\theta C\theta). \label{areas22}
\ee
As shown in (\ref{integs22b}),  the integral over the super magnetic field on a super-sphere is reduced to that of the magnetic field over the body-sphere, and the supermonopole charge is equal to the Chern number (\ref{chern1susy}).  This is consistent with the previous observation that the Chern number of the supermonopole is equal to that of the original monopole.

A bit  of calculations shows that  (\ref{llds}) satisfy the $uosp(1|2)$ algebra: 
\be
[L_i, L_j] =i\epsilon_{ijk}L_k, ~~~[L_i, L_{\alpha}] =\frac{1}{2}(\sigma_i)_{\beta\alpha}L_\beta, ~~~\{L_{\alpha}, L_{\beta}\} =\frac{1}{2}(C\sigma_i)_{\alpha\beta}L_i.
\ee
From the first relation, we see that $L_i$ constitute the  $su(2)$ algebra and  generate $SU(2)$ rotations in the superspace $\mathbb{R}^{3|2}$.   The second equation implies that the supercharges  $L_{\alpha}$ transform as an $SU(2)$ spinor and carry angular momentum 1/2. The last equation shows that $L_{\alpha}$ are fermionic operators, and  the two successive operations of the fermionic generators yield the  $SU(2)$ operations. 
 The set of $L_i$ and $L_{\alpha}$ amounts to the $UOSp(1|2)$ super-rotations in  $\mathbb{R}^{3|2}$. It is worth mentioning  a distinction between the present model and standard SUSY field theories based on the super Poincar\'e symmetry. While in the latter, the bosonic and fermionic states with different $1/2$ spin are related by the super transformation,  in the present model the  $\it{orbital}$ angular momentum, rather than intrinsic spin,  is involved: the supercharges $L_{\alpha}$ relate bosonic and fermionic states whose orbital angular momenta differ by 1/2. The super charges act as fermionic ladder operators of the angular momentum.

The super angular momentum operators (\ref{llds}) are expressed as 
\begin{subequations}
\begin{align}
&L_x=L_x^{0}+g\frac{x_1}{r+x_3}\biggl(1+\frac{1}{2r(r+x_3)}\theta C\theta\biggr), ~~L_y=L_y^{0}+g\frac{x_2}{r+x_3}\biggl(1+\frac{1}{2r(r+x_3)}\theta C\theta\biggr), \\
&L_z=L_z^{0}+g,\\
&L_{\theta_1} =L_{\theta_1}^{0} +\frac{g}{2r}(\theta_1+\frac{x_1+ix_2}{r+x_3}\theta_2), ~~~L_{\theta_2} =L_{\theta_2}^{0} +\frac{g}{2r}(\theta_2-\frac{x_1-ix_2}{r+x_3}\theta_1), 
\end{align}\label{exanof}
\end{subequations}
where $L^{0}_i$ and $L^{0}_\alpha$ denote the free super angular momentum operators: 
\be
L^{0}_i =-i\epsilon_{ijk}x_j {\partial}_k +\frac{1}{2}\theta_{\alpha}(\sigma_{i})_{\alpha\beta}\partial_{\beta},~~~
L^{0}_{\alpha} = \frac{1}{2}x_i(C\sigma_i)_{\alpha\beta}\partial_{\beta} -\frac{1}{2}\theta_{\beta}(\sigma_i)_{\beta\alpha}\partial_i.
\ee

The super rotations for the super-coordinates are represented as  
\be[L_i, x_j] =i\epsilon_{ijk}x_k, ~~~~~[L_i, \theta_{\alpha}] = \frac{1}{2}(\sigma_i)_{\beta\alpha}\theta_{\beta},~~~[L_\alpha, x_i] =-\frac{1}{2}(\sigma_i)_{\beta\alpha}\theta_{\beta}, ~~~\{L_{\alpha}, \theta_{\beta}\} =\frac{1}{2}(C\sigma_i)_{\alpha\beta}x_i. \label{uosp12coord}
\ee
The last two equations demonstrate  SUSY transformations of the super-coordinates.  
Obviously, the supersphere condition, $x_ix_i+C_{\alpha\beta}\theta_{\alpha}\theta_{\beta} =r^2$, is invariant under the $UOSp(1|2)$ SUSY transformation, implying that the supersphere respects the $\mathcal{N}=1$ SUSY generated by two pseudo-Majorana supercharges, $L_{\theta_1}$ and $L_{\theta_2}$.

\subsection{Super Landau Hamiltonian and super Landau levels}

\begin{figure}[tbph]
\center
\includegraphics*[width=160mm]{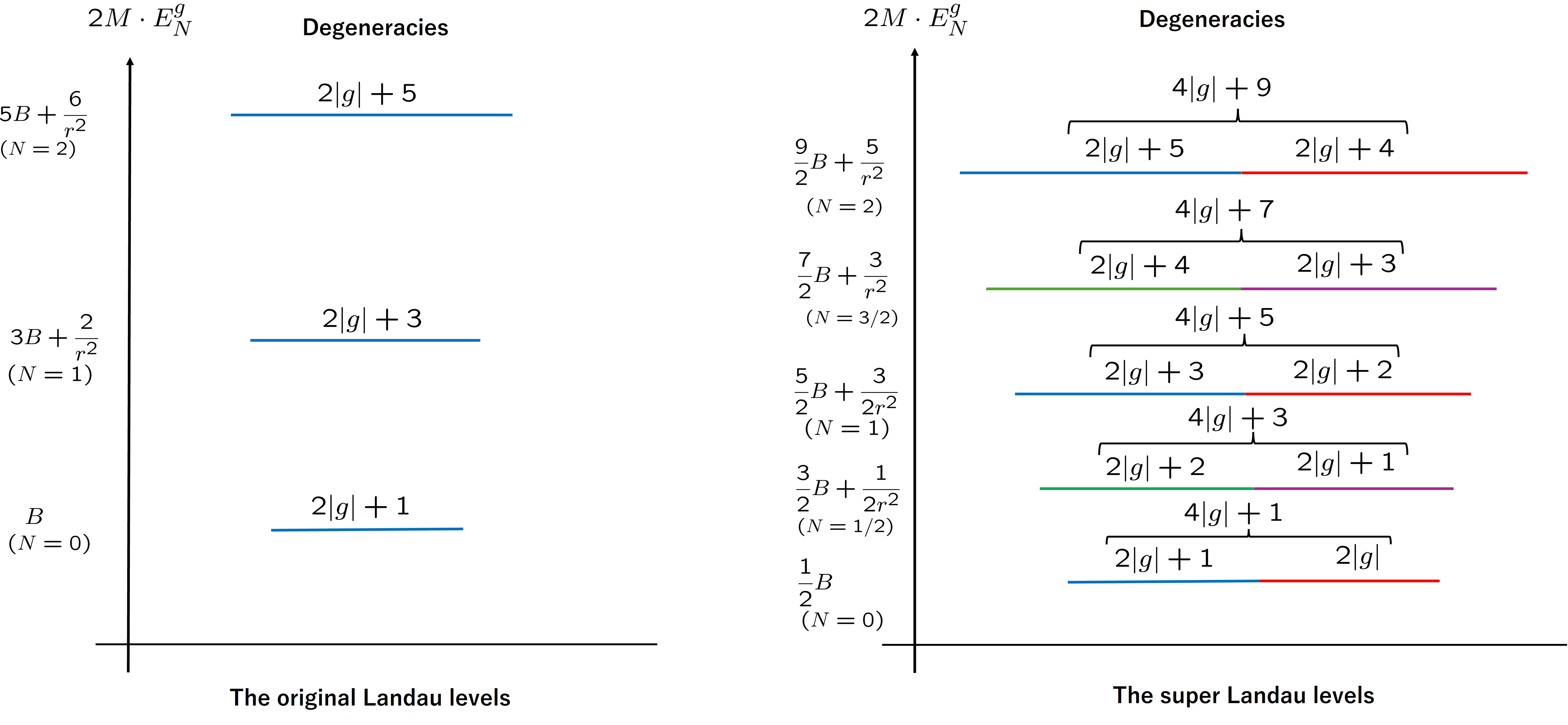}
\caption{Left: The original LLs (\ref{origlls}). Right: The super Landau levels (\ref{enesusyll}). Blue/red lines denote bosonic/fermionic states in integer Landau levels, while green/purple lines denote bosonic/fermionic states in half-integer Landau levels. Both of the LLs and the degeneracies are (almost) doubled in the SUSY system.  }
\label{SUSYLL.fig}
\end{figure}

Let us  consider  super quantum mechanics on a supersphere in the supermonopole background:      
\be
H=\frac{1}{2Mr^2} (\Lambda_i \Lambda_i +C_{\alpha\beta}\Lambda_{\alpha}\Lambda_{\beta}). \label{superllmodel}
\ee
Using the orthogonality relation  
\be
x_i\Lambda_i +C_{\alpha\beta}\theta_{\alpha}\Lambda_{\beta} = \Lambda_i x_i +C_{\alpha\beta}\Lambda_{\alpha}\theta_{\beta}=0, 
\ee
the Hamiltonian (\ref{superllmodel}) can be rewritten as  
\be
H=\frac{1}{2Mr^2}(L_i L_i +C_{\alpha\beta}L_{\alpha}L_{\beta})-\frac{1}{2Mr^2}g^2, 
\label{superlandauhal}
\ee
which is essentially the $UOSp(1|2)$ Casimir operator. 
Since the SUSY Hamiltonian is invariant under the transformations generated by $L_i$ and $L_{\alpha}$, the system possesses  $UOSp(1|2)$ super-rotational invariance. At the same time, $L_i$ and $L_{\alpha}$ constitute conserved quantities, as  they commute with the Hamiltonian. 
The standard results from $UOSp(1|2)$ representation theory (see Appendix \ref{appen;uso12mat}) determine the energy eigenvalues and their degeneracies:  
\begin{subequations}
\begin{align}
\text{Casimir eigenvalues}~~&:~~L_i L_i +C_{\alpha\beta}L_{\alpha}L_{\beta}=l(l+\frac{1}{2}), \\
\text{Degeneracies}~~&:~~4l+1. 
\end{align}
\end{subequations}
The superspin index  $l$ generally takes  integer or half-integer values.  In  particular, for the present system, $l$ is given by
\be
l=N+|g| \label{lng}
\ee
where $N$ denotes the super Landau level index 
\be
N=0, ~{1}/{2},~ 1,~ {3}/{2},~ 2,~ \cdots. 
\ee
The Landau levels include  both integer and half-integer values due to SUSY.  
The energy eigenvalues are then  given by 
\be
E^g_N=\frac{1}{2Mr^2} (N(N+\frac{1}{2})+(2N+\frac{1}{2})|g|), \label{enesusyll}
\ee
with degeneracies 
\be
4|g|+4N+1=4|g|+1, 4|g|+3, 4|g|+5, \cdots. 
\ee
Under the $SU(2)$ group of $UOSp(1|2)$, the super-multiplet in the $N$th LL decomposes into  two $SU(2)$ irreducible representations:  
\be
4|g|+4N+1 =(2|g|+2N+1)\oplus (2|g|+2N),  
\ee
where $(2|g|+2N+1)$ and $(2|g|+2N)$  correspond to bosonic and fermionic states, respectively. 
The LL energies (\ref{enesusyll}) can be expressed as   the sum of  contributions from the (super)spherical harmonics  and the planar super Landau levels:  
\be
E^g_N
 ={E}^0_N +\tilde{E}^B_N 
\ee
with  
\begin{subequations}
\begin{align}
&E_N^0 =\frac{1}{2Mr^2}N(N+\frac{1}{2}), \label{superspheene}\\
&\tilde{E}_N^B = (N+\frac{1}{4})\frac{B}{M}. \label{superll} 
\end{align}\label{superllener}
\end{subequations}
The formulas  (\ref{enesusyll}) is quite similar to the LLs on Haldane's sphere \cite{Haldane-1983},\footnote{
The original LLs on Haldane's sphere are 
\be
E^g_N= 
\frac{1}{2Mr^2}(N(N+1)+ 
 (2N+1)|g|),~~~~~~(N=0,1,2,\cdots)\label{origlls}
\ee
with degeneracies
\be
2l+1=2|g|+2N+1=2|g|+1, 2|g|+3, 2|g|+5, \cdots. 
\ee
} with the exception that  the constants differ slightly  and the SUSY Landau levels include half-integer values  (see Fig.\ref{SUSYLL.fig} also).  In particular, the LLL energy is given by 
\be
E_{N=0}^g=\tilde{E}_{N=0}^B=\frac{B}{4M}, \label{lllenergsup}
\ee
which is half of the original LLL energy due to the cancellation between the bosonic and fermionic degrees of freedom.

\section{Super-spinor derivative operators}\label{sec:effecop}

 Spinor derivative operators are  quite useful for analyzing the Landau model on a two-sphere \cite{Haldane-1983, Karabali-Nair-2002, Greiter-2011, Hasebe-2016}. In this section, we introduce super-spinor derivative operators that play the role of  effective super-angular momentum operators and  edth operators.   We denote the $L_i$ and $L_{\alpha}$ (\ref{llds}) as  $L^{g}_A$ $(A=x_1, x_2, x_3, \theta_1, \theta_2)$ with monopole charge $g$.

\subsection{Super rotations and quantum  numbers }\label{sec:superrot}

Let us first discuss  the super-rotations of the super-Hopf spinor. Using Eqs. (\ref{shopf}) and (\ref{exanof}),  we have 
\be
L_A^{1/2} \psi^{t} =\psi^t l_A,
\ee
while, for  its complex conjugate (\ref{cshopf}),
\be
L_A^{-1/2} {\psi^*}^{t} ={\psi^*}^t \tilde{l}_A.
\ee
Here, $l_{A}$ and $\tilde{l}_A$ respectively denote the super-spin matrices (\ref{l1/2mat}) and their complex representation  matrices:
\be
\tilde{l}_i=-l_i^*, ~~~\tilde{l}_{\alpha}=C_{\alpha\beta}l_{\beta},
\ee
which satisfy 
\be
[\tilde{l}_i, \tilde{l}_j]=i\epsilon_{ijk}\tilde{l}_k, ~~~[\tilde{l}_i, \tilde{l}_{\alpha}]=\frac{1}{2}(\sigma_{i})_{\beta\alpha}\tilde{l}_{\beta}, ~~\{\tilde{l}_{\alpha}, \tilde{l}_{\beta}\}=\frac{1}{2}(C\sigma_i)_{\alpha\beta}\tilde{l}_i. 
\ee
The super-Hopf spinor and its complex conjugate thus realize a $UOSp(1|2)$ $l=1/2$ fundamental representation, respectively. 
These two representation matrices are related as 
\be
\tilde{l}_A  =\mathcal{R}^t l_A \mathcal{R}, 
\ee
where $\mathcal{R}$ signifies  the $UOSp(1|2)$ charge conjugation matrix: 
\be
\mathcal{R}=\begin{pmatrix}
0 & -1 & 0 \\
1 & 0 & 0 \\
0 & 0 & -1 
\end{pmatrix},
\ee
which satisfies 
\be
\mathcal{R}^{-1} =\mathcal{R}^t. 
\ee
 The quantum numbers of the components of the super Hopf spinor (and its conjugate), along with  the super-coordinates, are summarized in Table \ref{table:correspDNYM}.

\begin{table}
\center 
\begin{tabular}{|c||c|c|c|c|}\hline
        &  superspin $l$ &   spin $s$ & magnetic  quantum number $m$ & supermonopole charge $g$    \\ \hline \hline
        $u$   ($u^*$)            &  1/2     &  1/2 & +1/2 ($-1/2$) & +1/2 ($-1/2$)       \\ \hline  
 $v$   ($v^*$)            &  1/2     &  1/2 & $-1/2$ (+1/2) & +1/2 ($-1/2$)       \\ \hline  
 $\eta$   ($\eta^*$)            &  1/2     &  0 & 0 (0) & +1/2 ($-1/2$)       \\ \hline   
 $x+iy$               &  1     &  1 & 1 & 0       \\ \hline 
    $x_3$               &  1     &  1 & 0  & 0       \\ \hline 
 $x-iy$               &  1     &  1 & $-1$  & 0       \\ \hline 
 $\theta_1$               &  1     &  1/2 & +1/2 & 0       \\ \hline 
  $\theta_2$               &  1     &  1/2 & $-1/2$ & 0       \\ \hline 
    \end{tabular}       
\caption{ Quantum numbers of the components of the super Hopf spinor   and  the super-coordinates.  
}
\label{table:correspDNYM}
\end{table}

\subsection{Effective super angular momentum operators}

An arbitrary LL is specified by the quantum numbers, $(N, g)$, or equivalently, $(l, g)$, 
while $\psi_a = u, v, \eta$ and $\psi_a^* = u^*, v^*, \eta^*$ carry   the same super-spin index of $l=1/2$, and  $g=1/2$ and $g=-1/2$, respectively (Table \ref{table:correspDNYM}). Taking into account the matching of the quantum numbers,  the $N$th LL eigenstates are given by (a linear combination of) the following homogeneous polynomials:
\be
\Psi^{(n_1, n_2)}:= \overbrace{\psi_{a_1} \cdots \psi_{a_{n_1}}}^{n_1} ~\overbrace{{\psi_{b_1}^*} \cdots {\psi_{b_{n_2}}^*}}^{n_2},  \label{lleigpoly}
\ee
where $n_1$ and $n_2$ are non-negative integers subject to\footnote{As we will see in Sec.\ref{sec:supermonoehahalf}, for half-integer LLs, the first relation of (\ref{condll})  is modified to  
\be
n_1+n_2=2l+1. 
\ee
} 
\be
n_1+n_2=2l, ~~~~n_1-n_2=2g. \label{condll}
\ee
Importantly, the action of $L_A^{g}$  on these homogeneous functions is effectively represented by that of $L_A^{\pm 1/2}$ on each component:\footnote{While $L_A$, $\psi_a$, and $\psi^*_b$ are assumed to be Grassmann-even, a formula analogous to (\ref{acangupoly})  holds  for the Grassmann-odd cases.}
\be
L^{g}_A \Psi^{(n_1, n_2)} =\sum_{i=1}^{n_1}\psi_{a_1}\cdots (L_A^{1/2}\psi_{a_i}) \cdots \psi_{a_{n_1}} \psi^*_{b_1}\cdots \psi^*_{b_{n_2}} + \sum_{i=1}^{n_2}\psi_{a_1}\cdots  \psi_{a_{n_1}} \psi^*_{b_1}\cdots (L_A^{-1/2}\psi^*_{b_i}) \cdots \psi^*_{b_{n_2}} . \label{acangupoly}
\ee
Under the constraint (\ref{condll}), the action of $L_A^{g}$ on the homogeneous function reduces to the one-particle action. 
From the results in Sec. \ref{sec:superrot}, these effective operators are identified as 
\be
L_A =\psi^t l_A \frac{\partial}{\partial \psi} +\psi^{\dagger} \tilde{l}_A \frac{\partial}{\partial \psi^*}, \label{laopeeffe}
\ee
which obviously satisfy the $uosp(1|2)$ algebra. Their explicit forms are given by 
\begin{align}
&L_+=L_x+iL_y = u\frac{\partial}{\partial v} -v^*\frac{\partial}{\partial u^*},~~L_-=L_x-iL_y =  v\frac{\partial}{\partial u}-u^*\frac{\partial}{\partial v^*},\nn\\
&L_z =\frac{1}{2} (u\frac{\partial}{\partial u} -v\frac{\partial}{\partial v})
-\frac{1}{2}(u^*\frac{\partial}{\partial u^*} -v^*\frac{\partial}{\partial v^*}), \nn\\         
&L_{\theta_1} = \frac{1}{2}(u\frac{\partial}{\partial \eta}+\eta\frac{\partial}{\partial v})+\frac{1}{2}(v^*\frac{\partial}{\partial \eta^*}-\eta^*\frac{\partial}{\partial u^*}),~~L_{\theta_2} = \frac{1}{2}(v\frac{\partial}{\partial \eta}-\eta\frac{\partial}{\partial u})-\frac{1}{2}(u^*\frac{\partial}{\partial \eta^*}+\eta^*\frac{\partial}{\partial v^*}).  \label{ladderopex}
\end{align}
When we introduce the complex representation 
\be
\tilde{\psi}:= \mathcal{R} \psi^* =\begin{pmatrix}
-v^* \\
u^* \\
-\eta^*
\end{pmatrix} , \label{complexhopfsp}
\ee
$L_A$ (\ref{laopeeffe}) can also be represented as 
\be
L_A =\psi^t l_A \frac{\partial}{\partial \psi} +\tilde{\psi}^{t} l_A \frac{\partial}{\partial \tilde{\psi}}. \label{effecang}
\ee
Since these are bilinear differential forms of the super Hopf spinor and its complex conjugate, the degrees $n_1$ and $n_2$ are conserved,  which is consistent with the fact that $g$ and $l$ are fixed in each LL. In practical calculations associated with the LL eigenstates, the $L_A$ (\ref{effecang}) are far more efficient than the original coordinate-derivative operators (\ref{exanof}).

We also introduce an operator $L$ whose eigenvalues signify (the half of) the total degree, $n_1+n_2$:\footnote{
In particular in the LLL, $L_A$  reduce to \cite{Hasebe-2005-1}
\be
L_A =\psi^t l_A\frac{\partial}{\partial \psi}, 
\ee
so the $UOSp(1|2)$ Casimir is expressed as 
\be
L_iL_i+C_{\alpha\beta}L_{\alpha}L_{\beta} =L(L+\frac{1}{2}), \label{llcllcs}
\ee
with  
\be
L=\frac{1}{2}\psi^t \frac{\partial}{\partial\psi}. \label{opll}
\ee
The relation (\ref{llcllcs}) does not  hold, in general, for  $L_A$ (\ref{effecang}) and $L$ (\ref{generallopeff}).  
}   
\be
L=\frac{1}{2}\psi^t \frac{\partial}{\partial \psi} +\frac{1}{2}\tilde{\psi}^{t} \frac{\partial}{\partial \tilde{\psi}}=\frac{1}{2}\psi^t \frac{\partial}{\partial \psi} +\frac{1}{2}\psi^{\dagger} \frac{\partial}{\partial \psi^*},  \label{generallopeff}
\ee
which  commutes $L_A$s:  
\be
[L, L_A] =0. 
\ee
 As demonstrated in Secs.\ref{subsubse:act1} and \ref{subsubse:act2},  the eigenvalue of $L$ coincides with the super-spin index $l$ (\ref{lng}) for integer LLs, while it equals $l + 1/2$ for half-integer LLs.  
 
\subsection{Effective edth operators}\label{subsec:edthop}

While the edth operators were originally introduced by Newman and Penrose for the analysis of gravitational radiation \cite{Newman-Penrose-1966}, these operators 
were also  effective in the analysis of the Landau model \cite{Hasebe-2016}. Here, we extend this formalism to the SUSY case.

In contrast to the super-angular momentum operators, which do not mix the super-Hopf spinor and its complex representation, the effective edth operators are constructed to interchange these representations:  
\be
\psi ~~\longleftrightarrow ~~\tilde{\psi} =\mathcal{R} \psi^* . 
\ee
This interchange  is realized by\footnote{$R_-$ and $R_+$ act as 
\be
R_-\psi^t =\tilde{\psi}^t, ~~~~R_+{\tilde{\psi}}^t ={\psi}^t.
\ee
}
\be
R_- =\tilde{\psi}^t \frac{\partial}{\partial \psi}=\psi^{\dagger} \mathcal{R}^t  \frac{\partial}{\partial \psi},~~~~~R_+ ={\psi}^t \frac{\partial}{\partial \tilde{\psi}}=\psi^t \mathcal{R} \frac{\partial}{\partial \psi^*}
\ee
or, more explicitly,  
\be
R_- =u^*\frac{\partial}{\partial v} -v^*\frac{\partial}{\partial u}-\eta^*\frac{\partial}{\partial \eta}, ~~~~R_+ =-u\frac{\partial}{\partial v^*} +v\frac{\partial}{\partial u^*}-\eta\frac{\partial}{\partial \eta^*}. \label{rpmexp}
\ee 
These provide a natural SUSY extension of the spinor-derivative operators for Haldane's sphere \cite{Karabali-Nair-2002, Greiter-2011}.\footnote{To the best of the author’s knowledge, the effective edth operators were first employed in  \cite{Karabali-Nair-2002}, where they are realized as right actions of the D-matrix.}    
Since the super-Hopf spinor and its complex representation carry super-monopole charges of $1/2$ and $-1/2$, respectively, $R_{\pm}$ shift the super-monopole charge by $\Delta g = \pm 1$. The $R_{\pm}$ thus act as ladder operators for monopole charge. Note that they do not change the super-spin $l$, as they are bilinear super-spinor differential operators. Since $l = N + |g|$ is fixed,  the $R_{\pm}$ shifts the  LL index $N$ by $\mp 1$ corresponding to the shift of $g$ by $\pm 1$. Then, $R_{\pm}$ can also be viewed as lowering/raising operators for the Landau levels (that simultaneously shift the monopole charge).

The commutation relation between $R_+$ and $R_-$ yields a new operator,  
\be
R_3:= \frac{1}{2}[R_+,  R_-]=\frac{1}{2}\biggl(\psi^t\frac{\partial}{\partial \psi} -\tilde{\psi}^{t}\frac{\partial}{\partial \tilde{\psi}}\biggr)= \frac{1}{2}\biggl(\psi^t\frac{\partial}{\partial \psi} -\psi^{\dagger}\frac{\partial}{\partial \psi^*}\biggr) .
\ee
Together with $R_3$, $R_i$ satisfy the $SU(2)$ algebra: 
\be
[R_i, R_j] =i\epsilon_{ijk}R_k, 
\ee
where 
\be
R_1:=\frac{1}{2}(R_++R_-), ~~~R_2:=-i\frac{1}{2}(R_+-R_-).
\ee
It is straightforward  to verify that $L_i$ commute with both $L_A$  and $L$: 
\be
[R_i, L_A]= [R_i, L]=0. 
\ee
Although $R_i$ and $L_A$ are commutative, the $SU(2)$ transformations generated by $R_i$ does not bring further degeneracies to  Landau levels. Instead, $R_{\pm}$ act as ladder operators between Landau levels. In other words, $L_A$ represent the symmetry group generators, while $R_i$ are spectrum-generating operators. One may wonder whether there exist fermionic counterparts for $R_i$, analogous to $L_{\alpha}$ for $L_i$.  We will revisit this question in Sec. \ref{sec:superdmat} along with the discussion of the super Howe duality.

\section{Supermonopole harmonics in the integer Landau levels}\label{sec:supermonoeha}

In this section, we explicitly construct the supermonopole harmonics by using the super-spinor derivative operators. We also discuss  the probabilistic interpretation of these states.

\subsection{The original monopole harmonics}\label{subsec:orimono}

We begin with a brief review of the original monopole harmonics. In the Dirac gauge, the monopole harmonics are given by \cite{Wu-Yang-1976}:\footnote{We can ``interchange'' $m$ and $g$ in  (\ref{monohar}) as 
\begin{align}
\Phi^{g}_{l,m}(y_i)& =\sqrt{\frac{2l+1}{4\pi}\frac{(l+g)!(l-g)!}{(l+m)!(l-m)!}}(\frac{r+y_3}{2r})^{\frac{g+m}{2}}(\frac{r-y_3}{2r})^{\frac{g-m}{2}}P_{l-g}^{(g-m, g+m)}(\frac{y_3}{r}) ~(\frac{y_1+iy_2}{\sqrt{{r}^2-{y_3}^2}})^{m-g}  \nn\\
&  =  (-1)^{l-g} \sqrt{\frac{2l+1}{4\pi}\frac{(l+g)!(l-g)!}{(l+m)!(l-m)!}} \sum_{n=\text{max}(0, -m-g)}^{\text{min}(l-g, l-m)}(-1)^n 
\begin{pmatrix}
l+m \\
g+m+n
\end{pmatrix}\begin{pmatrix}
l-m \\
n 
\end{pmatrix}
\mu^{g+m+n}\nu^{l-m-n}{\mu^*}^{n}{\nu^*}^{l-g-n}.  \label{monoharex}
\end{align}
} 
\begin{align}
&\Phi^{g}_{l,m}(y_i) =(-1)^{m-g}\sqrt{\frac{2l+1}{4\pi}\frac{(l+m)!(l-m)!}{(l+g)!(l-g)!}}(\frac{r+y_3}{2r})^{\frac{m+g}{2}}(\frac{r-y_3}{2r})^{\frac{m-g}{2}}P_{l-m}^{(m-g, m+g)}(\frac{y_3}{r}) ~(\frac{y_1+iy_2}{\sqrt{{r}^2-{y_3}^2}})^{m-g}  \nn\\
&~~~~~~~~~~~~~~~~~~~~~~~~~~~~~~~~~(g,m =l, l-1, l-2, \cdots, -l),
\label{monohar}
\end{align}
where $P_{l+m}^{(\alpha, \beta)}$ denote the Jacobi polynomials:  
\be
P_N^{(\alpha, \beta)}(\cos\vartheta) = \sum_{n=0}^N (-1)^n\begin{pmatrix}
N +\alpha \\
n +\alpha
\end{pmatrix}
\begin{pmatrix}
N +\beta \\
n 
\end{pmatrix}(\cos\frac{\vartheta}{2})^{{2(N-n)}}(\sin\frac{\vartheta}{2})^{2n}.
\ee
Here, we parameterized $y_i$ as 
\be
y_1=r\sin\vartheta\cos\varphi, ~~y_2=r\sin\vartheta\sin\varphi, ~~y_3=r\cos\vartheta. \label{ysparangle}
\ee
The monopole harmonics satisfy  
\be
\Phi_{l,m}^g(\vartheta, \varphi)^*=(-1)^{g-m}\Phi_{l,-m}^{-g}(\vartheta, \varphi) 
\ee
and 
\be
|\Phi^{g}_{l,m} (y_i)| =|\Phi^{g}_{l,m} (y_3)| = |\Phi^{g}_{l,-m} (-y_3)|. \label{absphimy}
\ee
Their orthonormality relation on $S^2$ is expressed as  
\be
\int_{S^2}d\Omega_2 ~{\Phi^{g}_{l', m'}(y_i)}^*~\Phi_{l,m}^g(y_i) =\delta_{l' l}\delta_{m'm}~~~~~(d\Omega_2=\sin\vartheta d\vartheta d\varphi).
\ee
It is useful to express the monopole harmonics in terms of the Hopf spinor components (\ref{hopfspinor}):
\be
\Phi_{l, m}^g (y_i) = \sqrt{\frac{2l+1}{4\pi} \frac{(l+m)!(l-m)!}{(l+g)!(l-g)!}}~\sum_{n=\text{max}(0, m-g)}^{\text{min}(l-g, l+m)}(-1)^n
\begin{pmatrix}
l+g \\
g-m+n
\end{pmatrix}
\begin{pmatrix}
l-g \\
n 
\end{pmatrix}  \mu^{l+m-n}\nu^{g-m+n}{\mu^*}^{l-g-n}{\nu^*}^n.
\ee
Low-dimensional examples are the  components of the Hopf spinor: 
\begin{align}
&\Phi^{1/2}_{1/2,1/2}(y_i)=\frac{1}{\sqrt{2\pi}}\mu,~~~\Phi^{1/2}_{1/2,-1/2}(y_i)=\frac{1}{\sqrt{2\pi}} \nu,\nn\\
&\Phi^{-1/2}_{1/2,1/2}(y_i)=-\frac{1}{\sqrt{2\pi}}\nu^*, ~~~~\Phi^{-1/2}_{1/2,-1/2}(y_i)=\frac{1}{\sqrt{2\pi}}\mu^*, \label{munuphi}
\end{align}
and the vector coordinates:   
\be
\Phi^{0}_{1, \pm 1}(y_i)=\mp \sqrt{\frac{3}{8\pi}}\frac{1}{r}(y_1\pm i y_2) , ~~~~\Phi^{0}_{1,0}(y_i)=\sqrt{\frac{3}{4\pi}} \frac{1}{r}y_3  .
\ee

\subsection{Supermonopole harmonics}

In the integer LL $(N=l-|g|)$, there are  $(4l+1)$ degenerate eigenstates consisting of  
\be
\Psi_{l,m}^g(x_i, \theta_{\alpha}) = \{Y_{l,m_1}^g(x_i, \theta_{\alpha}) ,~\mathcal{Y}_{l,m_2}^g(x_i,  \theta_{\alpha})\}, \label{intllsupermono}
\ee
where  
\be
g=\overbrace{l, l-1, l-2, \cdots, -l}^{2l+1},~~~~m_1=\overbrace{l, l-1, l-2, \cdots, -l}^{2l+1},~~~m_2= \overbrace{l-{1}/{2}, l-{3}/{2}, l-{5}/{2}, \cdots,   -l+{1}/{2}}^{2l}. 
\ee
Here, $Y_{l,m_1}^g(x_i, \theta_{\alpha})$ and $\mathcal{Y}_{l,m_2}^g(x_i, \theta_{\alpha})$ represent the bosonic and fermionic sectors of the supermonopole harmonics, respectively.

\subsubsection{Bosonic monopole harmonics}

It is straightforward to extend the discussion in Sec.\ref{subsec:orimono}  to derive the bosonic sector of the supermonopole harmonics. The highest-weight state of the supermultiplet ($m=l$) may be given by  
\be
u^{l+g}{v^*}^{l-g}. \label{intnhighest}
\ee
We can easily confirm that (\ref{intnhighest}) is indeed annihilated by the fermionic raising operator $L_{\theta_1}$ (\ref{ladderopex}).  By repeatedly applying the lowering operator $L_-$ to (\ref{intnhighest}), we can derive the remaining monopole harmonics of the bosonic sector. These are identical in form to the standard monopole harmonics, provided that the original Hopf spinor is replaced by the bosonic part of the super-Hopf spinor: 
\begin{align}
Y^{g}_{l,m}(x_i, \theta_{\alpha}) &=\sqrt{\frac{1}{4\pi} \frac{(l+m)!(l-m)!}{(l+g)!(l-g)!}}~\sum_{n=\text{max}(0, m-g)}^{\text{min}(l-g, l+m)}(-1)^n
\begin{pmatrix}
l+g \\
g-m+n
\end{pmatrix}
\begin{pmatrix}
l-g \\
n 
\end{pmatrix}  u^{l+m-n}v^{g-m+n}{u^*}^{l-g-n}{v^*}^n \nn\\
&~~~~~~~~~~~~~~~~~~~~~~~~~~~~~~~~~(g,m =l, l-1, l-2, \cdots, -l). \label{bosonmonohar}
\end{align}
Here, the normalization factor is chosen to match the integral over the supersphere (\ref{norsim1}).    Since $Y^g_{l,m}$ are homogeneous functions of  the bosonic components of the super Hopf spinor, (\ref{bosonmonohar}) can be factorized into a body part (involving $y_i$) and a soul part (involving 
$\theta_{\alpha}$). 
The supermonopole harmonics are then related to the original monopole harmonics as  
\be
Y^{g}_{l,m}(x_i, \theta_{\alpha}) =\frac{1}{\sqrt{2l+1}}(1-\frac{l}{2r^2}\theta C \theta) \cdot \Phi_{l, m}^g (y_i). 
\label{relorimobos}
\ee
From this equation, the properties of the bosonic supermonopole harmonics are readily deduced from those of the original monopole harmonics. For instance, the pseudo-complex conjugation is  
\be
Y^{g}_{l,m}(x_i, \theta_{\alpha})^*=(-1)^{g-m} ~Y^{-g}_{l,-m}(x_i, \theta_{\alpha}).
\ee

\subsubsection{Fermionic monopole harmonics}

The fermionic monopole harmonics can be derived by applying the fermionic lowering operator $L_{\theta_2}$ to the bosonic monopole harmonics. According to (\ref{ruleladder}), a fermionic state is obtained from the bosonic state through the formula  $|l, m\rangle_\text{F} = -\frac{2}{\sqrt{l+m+\frac{1}{2}}} L_{\theta_2} |l, m+\frac{1}{2}\rangle_{\text{B}}$. Substituting  $|l,m\rangle_{\text{B}} = Y^{g}_{l,m}$ (\ref{relorimobos}) and  $L_{\theta_2} = \frac{1}{\sqrt{l+m+\frac{1}{2}}} (\eta\frac{\partial}{\partial u} + \eta^*\frac{\partial}{\partial v^*})$ (\ref{ladderopex}) into this formula, we obtain 
\begin{align}
\mathcal{Y}^{g}_{l, m}(x_i, \theta_\alpha)& =\sqrt{\frac{1}{4\pi} \frac{(l+m-\frac{1}{2})!(l-m-\frac{1}{2})!}{(l+g)!(l-g)!}}~\sum_{n=(0, m-g+\frac{1}{2})}^{\text{min}(l-g, l+m+\frac{1}{2})} (-1)^n 
\begin{pmatrix}
l+g \\
g-m+n-\frac{1}{2}
\end{pmatrix}
\begin{pmatrix}
l-g \\
n 
\end{pmatrix}\nn\\
&~~~~\cdot u^{l+m-n-\frac{1}{2}}v^{g-m+n-\frac{1}{2}}{u^*}^{l-g-n}{v^*}^{n-1} \biggl((l-n+m+\frac{1}{2})v^* \eta  +n u\eta^*\biggr).\label{fermimonohar}
\end{align}
With the bosonic monopole harmonics,   (\ref{fermimonohar}) takes a remarkably compact form:
\begin{align}
\mathcal{Y}^{g}_{l, m}(x_i, \theta_\alpha)&= \sqrt{l+g}~Y_{l-\frac{1}{2}, m}^{g-\frac{1}{2}}(x_i, \theta_{\alpha}) \cdot \eta -\sqrt{l-g}~Y_{l-\frac{1}{2},m}^{g+\frac{1}{2}}(x_i, \theta_{\alpha}) \cdot \eta^*  \nn\\
&~(m= l-1/2, l-1, l-3/2, \cdots, -l+1/2,~~~~g=l, l-1, l-2, \cdots, -l ), \label{genefermmon1}
\end{align}
which,  in terms of the original monopole harmonics, is expressed as\footnote{
With  the Clebsch-Gordan coefficients, (\ref{genefermmon}) can be written as 
\be
\mathcal{Y}^{g}_{l, m}(x_i, \theta_\alpha)=\langle l-\frac{1}{2}, g-\frac{1}{2}; \frac{1}{2}, \frac{1}{2}|l, g\rangle~\Phi^{g-\frac{1}{2}}_{l-\frac{1}{2},m} (y_i)\cdot \eta - \langle l-\frac{1}{2}, g+\frac{1}{2}; \frac{1}{2}, -\frac{1}{2}|l, g\rangle~\Phi^{g+\frac{1}{2}}_{l-\frac{1}{2},m} (y_i)\cdot \eta^*. 
\ee
}  
\begin{align} 
\mathcal{Y}^{g}_{l, m}(x_i, \theta_\alpha)&=\sqrt{\frac{l+g}{2l}}~\Phi^{g-\frac{1}{2}}_{l-\frac{1}{2},m} (y_i)\cdot \eta - \sqrt{\frac{l-g}{2l}}~\Phi^{g+\frac{1}{2}}_{l-\frac{1}{2},m} (y_i)\cdot \eta^*. \label{genefermmon}
\end{align}
This concise expression is particularly efficient in deriving the supermatrix geometry in Sec.\ref{sec:supermatrix}.  
The pseudo-complex conjugation  is given by 
\be
\mathcal{Y}^{g}_{l,m}(x_i, \theta_{\alpha})^*=(-1)^{g-m+\frac{1}{2}} ~\mathcal{Y}^{-g}_{l,-m}(x_i, \theta_{\alpha}), 
\ee
In terms of the body and  soul coordinates, (\ref{genefermmon}) can also be represented as  
\be
\mathcal{Y}^{g}_{l, m}(x_i, \theta_\alpha)
={\mathcal{V}}_{l, m-\frac{1}{2}}^g(y_i)~ \frac{\theta_1}{r} -{\mathcal{U}}_{l, m+\frac{1}{2}}^g(y_i)~\frac{\theta_2}{r}, \label{fermoextheta}
\ee
where  
\begin{subequations}
\begin{align}
&{\mathcal{U}}_{l, m+\frac{1}{2}}^g(y_i) :=  \frac{g}{l}\sqrt{\frac{l+m+\frac{1}{2}}{2l+1}}~\Phi_{l, m+\frac{1}{2}}^g(y_i) +\frac{1}{l}\sqrt{\frac{(l+g)(l-g)(l-m-\frac{1}{2})}{2l-1}} ~\Phi_{l-1,m+\frac{1}{2}}^g  (y_i) , \\
&{\mathcal{V}}_{l, m-\frac{1}{2}}^g(y_i) :=\frac{ g}{l}\sqrt{\frac{l-m+\frac{1}{2}}{2l+1}}~\Phi_{l, m-\frac{1}{2}}^g(y_i) -\frac{1}{l}\sqrt{\frac{(l+g)(l-g)(l+m-\frac{1}{2})}{2l-1}} ~\Phi_{l-1,m-\frac{1}{2}}^g  (y_i)  .
\end{align}\label{compfermimo}
\end{subequations}
In the derivation of (\ref{fermoextheta}), we used (\ref{prodmunumono}). The pseudo-conjugations of $\mathcal{U}_{l, m+\frac{1}{2}}^{g}$ and $\mathcal{V}_{l,  m-\frac{1}{2}}^g$ are respectively given by 
\be
{\mathcal{U}}_{l, m+\frac{1}{2}}^g(y_i)^*=(-1)^{g-m+\frac{1}{2}}~{\mathcal{V}}^{-g}_{l, -m-\frac{1}{2}}(y_i), ~~~~~~{\mathcal{V}}_{l, m-\frac{1}{2}}^g(y_i)^*=-(-1)^{g-m+\frac{1}{2}}~{\mathcal{U}}^{-g}_{l, -m+\frac{1}{2}}(y_i).
\ee

\subsubsection{Action of the operators}\label{subsubse:act1}

It is straightforward to show  that the  super-spinor derivative operators act on the supermonopole harmonics as 
\begin{subequations}
\begin{align}
&R_{\pm} Y_{l,m}^g =\sqrt{(l\mp g)(l\pm g+1)}~Y^{g\pm 1}_{l, m}, ~~~~~~~~R_z Y_{l,m}^g=g Y_{l,m}^g,  \\
&R_{\pm} \mathcal{Y}_{l,m}^g =\sqrt{(l\mp g )(l\pm g+1)}~\mathcal{Y}^{g\pm 1}_{l, m}, ~~~~~~~~R_z \mathcal{Y}_{l,m}^g=g \mathcal{Y}_{l,m}^g, 
\end{align}\label{boslleth}
\end{subequations}
\begin{subequations}
\begin{align}
&L_{\pm} Y_{l,m}^g =\sqrt{(l\mp m)(l\pm m+1)}~Y^g_{l, m\pm 1}, ~~~L_z Y_{l,m}^g=m Y_{l,m}^g,  \\
&L_{\pm} \mathcal{Y}_{l,m}^g =\sqrt{(l\mp m -\frac{1}{2})(l\pm m+\frac{1}{2})}~\mathcal{Y}^g_{l, m\pm 1}, ~~~L_z \mathcal{Y}_{l,m}^g=m \mathcal{Y}_{l,m}^g, \\
&L_{\theta_1} Y_{l,m}^g =\frac{1}{2}\sqrt{l-m} ~\mathcal{Y}^{g}_{l, m+\frac{1}{2}}, ~~~~~~~~~~~~~~L_{\theta_2} Y_{l,m}^g =-\frac{1}{2}\sqrt{l+m}~ \mathcal{Y}^{g}_{l, m-\frac{1}{2}}, \\
&L_{\theta_1} \mathcal{Y}_{l,m}^g =\frac{1}{2}\sqrt{l+m+\frac{1}{2}}~ {Y}^{g}_{l, m+\frac{1}{2}}, ~~~~~~~~L_{\theta_2} ~\mathcal{Y}_{l,m}^g =\frac{1}{2}\sqrt{l-m+\frac{1}{2}}~ {Y}^{g}_{l, m-\frac{1}{2}}, 
\end{align}\label{actllsumo1}
\end{subequations}
and 
\be
L Y_{l,m}^g =l Y_{l, m}^g,~~~~~L\mathcal{Y}_{l,m}^g =l \mathcal{Y}_{l, m}^g. \label{bosllL}
\ee
Equation (\ref{actllsumo1}) coincides with (\ref{ruleladder}), indicating that the $\{Y_{l,m_1}^g, ~\mathcal{Y}_{l, m_2}^g\}$ form a $UOSp(1|2)$ irreducible representation.

\subsection{Probabilistic interpretation }

As the supermonopole harmonics are defined on a supermanifold, they depend on Grassmann variables, so their physical interpretation is far from clear. Here, we propose a probabilistic interpretation 
by projecting them onto the underlying body-sphere.

Using the Berezin integral on a supersphere (\ref{areas22}), we can demonstrate that the supermonopole harmonics form an orthonormal basis:\footnote{For the integrals of Grassmann numbers, 
\begin{align}
&\int d\theta_1 d\theta_2 (1-\frac{1}{2}\theta C\theta)~{Y^{g}_{l', m'}(x_i, \theta_{\alpha})}^*~ Y^g_{l, m}(x_i, \theta_{\alpha})= {\Phi_{l', m'}^{g}}(y_i)^*~{\Phi^{g}_{l,m'}(y_i)}, \nn\\
&\int d\theta_1 d\theta_2 (1-\frac{1}{2}\theta C\theta)~{\mathcal{Y}^{g}_{l', m'}(x_i, \theta_{\alpha})}^*~ \mathcal{Y}^g_{l, m}(x_i, \theta_{\alpha})
= \frac{l+g}{2l}~ \Phi^{g-\frac{1}{2}}_{l'-\frac{1}{2}, m'}(y_i)^*~\Phi^{g-\frac{1}{2}}_{l-\frac{1}{2}, m}(y_i)+ \frac{l-g}{2l}~\Phi^{g+\frac{1}{2}}_{l'-\frac{1}{2}, m'}(y_i)^*~\Phi^{g+\frac{1}{2}}_{l-\frac{1}{2}, m}(y_i), \nn\\
&\int d\theta_1 d\theta_2 (1-\frac{1}{2}\theta C\theta)~{\mathcal{Y}^{g}_{l', m'}(x_i, \theta_{\alpha})}^*~ {Y}^g_{l, m}(x_i, \theta_{\alpha})
=0.
\end{align}
}  
\begin{subequations}
\begin{align}
&\int_{S^{2|2}}d\Omega_{2|2} ~{Y^{g}_{l', m'}}(x_i, \theta_{\alpha})^*~ Y^g_{l, m}(x_i, \theta_{\alpha}) = \delta_{l' l}\delta_{m' m}, \label{norsim1}\\
&\int_{S^{2|2}}d\Omega_{2|2} ~{\mathcal{Y}^{g}_{l', m'}(x_i, \theta_{\alpha})}^* ~\mathcal{Y}^g_{l, m}(x_i, \theta_{\alpha}) = \delta_{l' l}\delta_{m' m}, \\
&\int_{S^{2|2}}d\Omega_{2|2} ~{\mathcal{Y}^{g}_{l, m'}(x_i, \theta_{\alpha})}^* ~{Y}^g_{l, m} (x_i, \theta_{\alpha})=0.
\end{align}
\end{subequations}
Since the Grassmann quantities are the primary obstacles for the probabilistic interpretation, we project
them out by mapping the absolute square of the super states onto the body-sphere:
\be
\rho(y_i):= \int d\theta_1 d\theta_2 ~(1-\frac{1}{2}\theta C\theta) ~{\Psi(x_i, \theta_{\alpha})}^*~\Psi(x_i, \theta_{\alpha}).
\ee
Here, $\int d\theta_1 d\theta_2 ~(1-\frac{1}{2}\theta C\theta)$ denotes the Grassmann part of the integral over a supersphere (\ref{areas22}).  
With this definition, the probability densities for $Y_{l,m}^g$ are obtained as 
\be
\rho_{ l, m}^g (y_i) =\int d\theta_1 d\theta_2 ~(1-\frac{1}{2}\theta C\theta) ~{Y_{l,m}^{g}(x_i, \theta_{\alpha})}^*~Y_{l,m}^g (x_i, \theta_{\alpha})=|\Phi_{l m }^g(y_i)|^2, 
\ee
which are nothing but the probability densities for the original monopole harmonics. 
Meanwhile, the probability densities for $\mathcal{Y}_{l,m}^g$ are given by 
\be
\varrho_{ l, m}^g(y_i) =\int d\theta_1 d\theta_2 ~(1-\frac{1}{2}\theta C\theta) ~{\mathcal{Y}_{l,m}^{g}(x_i, \theta_{\alpha})}^*~\mathcal{Y}_{l,m}^g (x_i, \theta_{\alpha})=|{\mathcal{U}}_{l, m+\frac{1}{2}}^g (y_i)|^2 + |{\mathcal{V}}_{l, m-\frac{1}{2}}^g (y_i)|^2, \label{profermint}
\ee
which represent  the sum of the absolute squares of the amplitudes, ${\mathcal{U}}_{l, m+\frac{1}{2}}^g$ and ${\mathcal{V}}_{l, m-\frac{1}{2}}^g$, and are thereby  manifestly positive semi-definite. 
From (\ref{genefermmon}), we have another expression of (\ref{profermint})
\be
\varrho_{ l, m}^g(y_i) 
=\frac{l+g}{2l}\rho_{ l-\frac{1}{2}, m}^{g-\frac{1}{2}}(y_i) +\frac{l-g}{2l}\rho_{ l-\frac{1}{2}, m}^{g+\frac{1}{2}}(y_i).
\ee

Let us recall that $|\Phi_{l,m}^{g}(y_i)|$ depends only on $y_3$ (\ref{absphimy}),  so the probability densities also depend solely on $y_3$ and  are axially symmetric about the $y_3$-axis. Furthermore, from (\ref{absphimy}),  the probability densities satisfy the following property:
\be
\rho_{ l, -m}^{g}(y_3)=  \rho_{l, m}^{g}(-y_3),~~~~ \varrho_{ l, -m}^{g}(y_3)=  \varrho_{l, m}^{g}(-y_3).
\ee
In addition from $|\Phi_{l,-m}^{-g}(y_i)|= |\Phi_{l,m}^{g}(y_i)|$, it follows that 
\be
\rho_{l, -m}^{-g}(y_3)=  \rho_{l, m}^{g}(y_3), ~~~~\varrho_{l, -m}^{-g}(y_3)=  \varrho_{ l, m}^{g}(y_3).
\ee
Namely,  the probability densities are invariant under the simultaneous reversal of the signs of $m$ and $g$. Physically, this implies  that the probability density of the supermonopole harmonics with a given magnetic quantum number $m$ in an antimonopole background is identical to that with the opposite quantum number $-m$ in a monopole background. 
For a better understanding, we depict 
the probability densities for the $l=2$ LLL  supermonopole harmonics $(g=l)$ and the $l=2$  superspherical harmonics ($g=0$) 
 in Fig.\ref{l2pro.fig}.\footnote{See Appendix \ref{sec:specic} for more about the LLL eigenstates and the superspherical harmonics.}   

\begin{figure}[tbph]
\center
\includegraphics*[width=140mm]{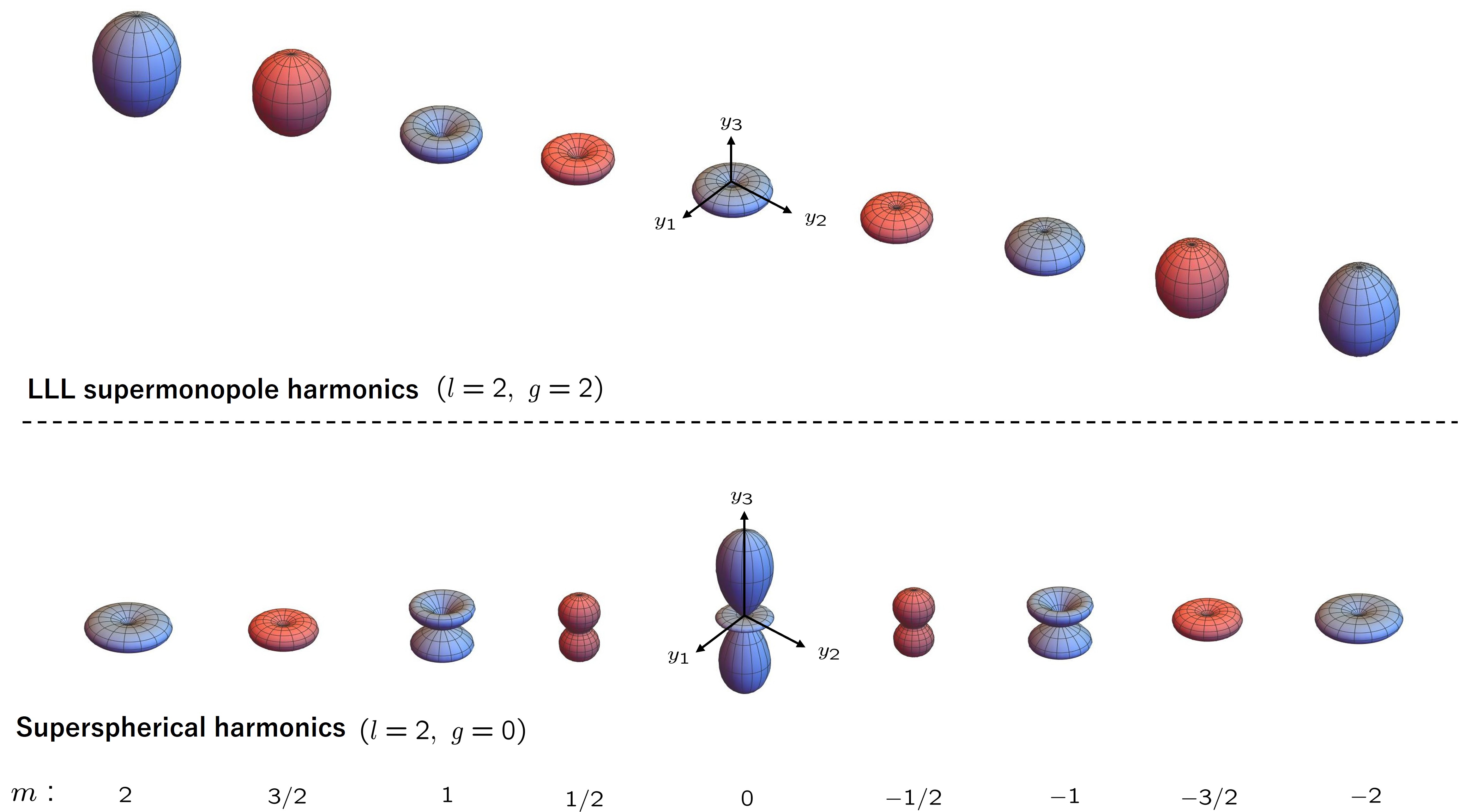}
\caption{The probability densities for the LLL supermonopole harmonics and the superspherical harmonics. }
\label{l2pro.fig}
\end{figure}

\section{Supermonopole harmonics in the half-integer Landau levels}\label{sec:supermonoehahalf}

In this section, we derive the supermonopole harmonics in the half-integer LLs and discuss their probabilistic interpretation. 

\subsection{Supermonopole harmonics}

We denote $(4l+1)$-fold degenerate  supermonopole harmonics in half-integer LLs  as 
\be
\Psi_{l,m}^g(x_i, \theta_{\alpha}) = 
\begin{pmatrix}
Z_{l,m_1}^g(x_i, \theta_{\alpha}) \nn\\
\mathcal{Z}_{l,m_2}^g(x_i,  \theta_{\alpha})
\end{pmatrix}, \label{halfllsupermono}
\ee
where $Z_{l,m_1}^g$ and $\mathcal{Z}_{l,m_2}^g$ represent the bosonic and fermionic sectors, respectively. The ranges of the quantum numbers are given by   
\be
g=\overbrace{l-1/2, l-3/2, l-5/2, \cdots, -l+1/2}^{2l},~~~~m_1=\overbrace{l, l-1, l-2, \cdots, -l}^{2l+1},~~~m_2= \overbrace{l-{1}/{2}, l-{3}/{2}, l-{5}/{2}, \cdots,   -l+{1}/{2}}^{2l}. 
\ee
Since the LL index $N(=l-|g|)$ is a half-integer, the difference between $g$ and $l$ must  also be a half-integer. Consequently, the monopole charge $g$ takes $2l$ distinct values, in contrast to the $2l+1$ values in the integer LL case.

\subsubsection{Highest weight states of the half-integer Landau levels}

Let us first construct  the highest weight states for the $1/2$th Landau level from  those of the $N=0$ and $N=1$ LLs with the same superspin $l(=g>0)$, 
\be
Y^{g}_{g, g}(x_i, \theta_{\alpha})=\frac{1}{\sqrt{4\pi}}u^{2g},~~~~Y_{g, g}^{g-1}(x_i, \theta_{\alpha})=-\sqrt{\frac{2g}{4\pi}}u^{2g-1}v^*.
\ee
We then introduce  the following Grassmann odd state:  
\be
{Z}_{l', m'}^{g'} ~\propto ~Y^{g}_{g, g}(x_i, \theta_{\alpha})~\eta^* +\alpha 
~Y^{g-1}_{g, g}(x_, \theta_\alpha)~\eta , \label{zyy}
\ee
where $\alpha$ is a constant to be determined. Let us recall that $\eta$ and $\eta^*$ are $SU(2)$ singlets, while  $Y^{g}_{g, g}$ and $Y^{g-1}_{g, g}$ are  $SU(2)$ highest weight states with spin $s=g$; thereby, ${Z}_{l', m'}^{g'}$ is also an $SU(2)$ highest weight state (satisfying $L_+ Z^{g'}_{l', m'}=0$) with spin  $s=g$. By considering the matching of the quantum numbers between the left- and right-hand sides of (\ref{zyy}), we can identify the quantum numbers of the ${Z}_{l', m'}^{g'} $ as\footnote{$\eta$ and $\eta^*$ carry the supermonopole charge $\pm 1/2$, respectively.} 
\be
g'=g-\frac{1}{2}, ~~~m'=g. 
\ee
The coefficient $\alpha$ is determined by requiring that $Z^{g'}_{l', m'}$ be the highest weight state of a supermultiplet: 
\be
L_{\theta_1} Z^{g-1/2}_{g, g}=0. 
\ee
This yields     
$\alpha =\frac{1}{\sqrt{2g}}$,  
and so 
\be
{Z}_{l', g}^{g-1/2} ~\propto Y^{g}_{g, g}(x_i, \theta_{\alpha})~\eta^* +\frac{1}{\sqrt{2g}} 
~Y^{g-1}_{g, g}(x_, \theta_\alpha)~\eta =\frac{1}{\sqrt{4\pi}}u^{2g}\eta^* - \frac{1}{\sqrt{4\pi}}
~u^{2g-1}v^*\eta . \label{high1/2s}
\ee
Now ${Z}_{l', g}^{g-1/2}$ is promoted to the highest weight state of the $UOSp(1|2)$ supermultiplet with    superspin 
\be
l'=g. 
\ee
The LL to which (\ref{high1/2s}) belongs  is then determined as 
\be
N=l'-g'=1/2. 
\ee
Other eigenstates in the $1/2$th LL can be obtained by applying  $L_{\theta_2}$ (and $L_{-}$) to (\ref{high1/2s}). 

Furthermore, by acting $R_-$ on ${Z}^{g-1/2}_{g, g}$, we can construct other highest weight states with the same superspin $l=g$ but with different monopole charges:   
\be
{Z}^{g-n-1/2}_{g, g} ~\propto ~R_-^{n} {Z}^{g-1/2}_{g, g}. \label{anohiwe}
\ee
Since $L$s and $R$s  commute,  ${Z}^{g-n-1/2}_{g, g}$ obviously satisfies the highest weight state conditions, $L_{\theta_1}{Z}^{g-n-1/2}_{g, g} =0$.  Therefore, ${Z}^{g-n-1/2}_{g, g}$  realize the highest weight states of  the following half-integer LLs: 
\be
N=g -(g-n-1/2) =n+{1}/{2}. \label{halfnrep}
\ee
As before, by applying $L_{\theta_2}$ (and $L_{-}$) to (\ref{anohiwe}), we can explicitly derive  the  degenerate eigenstates for the half-integer LLs (\ref{halfnrep}).

The half-integer LL eigenstates are represented by homogeneous functions of degree $2l+1$ (in contrast to degree $2l$  for integer LLs) with  $\text{max}(|g|) = l - 1/2$  and $\text{max}(|m|) = l$.

\subsubsection{Bosonic monopole harmonics}

Following the above procedure, we obtain 
 the (normalized) bosonic monopole harmonics  as  the following Grassmann $\it{odd}$ homogeneous polynomials:  
\begin{align}
{Z}^g_{l, m}(x_i, \theta_{\alpha})& =\sqrt{l-g+\frac{1}{2}}~Y_{l,m}^{g-\frac{1}{2}}(x_i, \theta_{\alpha}) \cdot \eta +\sqrt{l+g+\frac{1}{2}}~Y_{l,m}^{g+\frac{1}{2}}(x_i, \theta_{\alpha}) \cdot \eta^* \nn\\
&=\sqrt{\frac{l-g+\frac{1}{2}}{2l+1}}~\Phi_{l,m}^{g-\frac{1}{2}}(y_i) \cdot \eta +\sqrt{\frac{l+g+\frac{1}{2}}{2l+1}}~\Phi_{l,m}^{g+\frac{1}{2}}(y_i) \cdot \eta^*\nn\\
&~~~~~~~~~(m=l, l-1, l-2, \cdots, -l,~~~~~g=l-\frac{1}{2}, l-\frac{3}{2}, \cdots, -(l-\frac{1}{2})).  \label{bosogoddpol}
\end{align}
By employing $\theta_1$ and $\theta_2$ instead of $\eta$ and $\eta^*$, one can rewrite (\ref{bosogoddpol}) as  
\begin{align}
{Z}^g_{l,m} (x_i, \theta_{\alpha})& =\frac{1}{l+\frac{1}{2}}\biggl(  \sqrt{\frac{(l+g+\frac{1}{2})(l-g+\frac{1}{2})(l-m+1)}{2(l+1)} }~\Phi_{l+\frac{1}{2}, m-\frac{1}{2}}^g(y_i) +g\sqrt{\frac{l+m}{2l}} ~\Phi^g_{l-\frac{1}{2}, m-\frac{1}{2}}(y_i)   \biggr)\frac{\theta_1}{r} \nn\\
&-\frac{1}{l+\frac{1}{2}}\biggl(  \sqrt{\frac{(l+g+\frac{1}{2})(l-g+\frac{1}{2})(l+m+1)}{2(l+1)} }~\Phi_{l+\frac{1}{2}, m+\frac{1}{2}}^g(y_i) -g\sqrt{\frac{l-m}{2l}} ~\Phi^g_{l-\frac{1}{2}, m+\frac{1}{2}}(y_i)   \biggr)\frac{\theta_2}{r}, 
\end{align}
where  (\ref{prodmunumono}) and (\ref{munuphi}) were used. The complex conjugation is given by 
\be
{Z^g_{l,m}}(x_i, \theta_{\alpha})^* =(-1)^{g-m-\frac{1}{2}} Z_{l,-m}^{-g}(x_i, \theta_{\alpha}). 
\ee

\subsubsection{Fermionic monopole harmonics}

As in the integer LL case, the  fermionic monopole harmonics can be derived using the formula $\mathcal{Z}_{l, m}^g = -\frac{2}{\sqrt{l+m+\frac{1}{2}}} L_{\theta_2}Z_{l, m+\frac{1}{2}}^g$: 
\begin{align}
\mathcal{Z}^g_{l, m}(x_i, \theta_{\alpha})& =\overbrace{(u^*u +v^*v-(2l+1) \eta^*\eta)}^{=1-2l \eta^
*\eta} ~Y_{l-\frac{1}{2}, m}^g(x_i, \theta_{\alpha})\nn\\
&~~~~~~~~~(m=l-\frac{1}{2}, l-\frac{3}{2}, \cdots, -(l-\frac{1}{2}),~~~~~g=l-\frac{1}{2}, l-\frac{3}{2}, \cdots, -(l-\frac{1}{2})), \label{ferhalinnnmo}
\end{align}
which are Grassmann $\it{even}$ polynomial functions.   
In terms of  $\theta_1$ and $\theta_2$, (\ref{ferhalinnnmo}) is represented as  
\be
\mathcal{Z}^g_{l,m} (x_i, \theta_{\alpha})=\frac{1}{\sqrt{2l}} (1+\frac{2l+1}{4r^2}\theta C\theta)~\Phi_{l-\frac{1}{2}, m}^g(y_i).
\ee
The complex conjugation is 
\be
{\mathcal{Z}^g_{l,m}}(x_i, \theta_{\alpha})^* =(-1)^{g-m} \mathcal{Z}_{l,-m}^{-g}(x_i, \theta_{\alpha}). 
\ee
Note that for the half-integer LL eigenstates, the fermion parity and the Grassmann parity are opposite:  $Z^g_{l,m}$ are Grassmann odd, while  $\mathcal{Z}^g_{l,m}$ are Grassmann even. 

Just as the monopole harmonics (with a given $g$) form a complete set for functions on a sphere \cite{Wu-Yang-1976}, the supermonopole harmonics realize a complete basis on  a supersphere.  
Any function on a supersphere can be expanded in terms of the fermion coordinates and represented as the sum of the four terms:  
\be
f(x_i, \theta_{\alpha}) =f_0(y_i)+f_1(y_i)\theta_1 +f_2(y_i)\theta_2 +f_3(y_i)\theta_1 \theta_2. 
\ee
The supermonopole harmonics $Y_{l,m}^g$ and $\mathcal{Z}^{l,m}_g$ can span the first 
and  last terms,  while $\mathcal{Y}_{l,m}^g$ and ${Z}_{l,m}^g$  the second and  third terms.  Thus, the supermonopole harmonics (with a given $g$) realize a complete set on the supersphere. 

\subsubsection{Action of the operators}\label{subsubse:act2}

The super-spinor derivative operators act  as
\begin{subequations}
\begin{align}
&R_{\pm} Z_{l,m}^g =\sqrt{(l\mp g-\frac{1}{2})(l\pm g+\frac{1}{2})}~Z^{g\pm 1}_{l, m}, ~~~~~~~~R_z Z_{l,m}^g=g Z_{l,m}^g,  \\
&R_{\pm} \mathcal{Z}_{l,m}^g =\sqrt{(l\mp g -\frac{1}{2})(l\pm g+\frac{1}{2})}~\mathcal{Z}^{g\pm 1}_{l, m}, ~~~~~~~~R_z \mathcal{Z}_{l,m}^g=g \mathcal{Z}_{l,m}^g, 
\end{align}\label{femilleth}
\end{subequations}
\begin{subequations}
\begin{align}
&L_{\pm} Z_{l,m}^g =\sqrt{(l\mp m)(l\pm m+1)}~Z^g_{l, m\pm 1}, ~~~L_z Z_{l,m}^g=m Z_{l,m}^g,  \\
&L_{\pm} \mathcal{Z}_{l,m}^g =\sqrt{(l\mp m -\frac{1}{2})(l\pm m+\frac{1}{2})}~\mathcal{Z}^g_{l, m\pm 1}, ~~~L_z \mathcal{Z}_{l,m}^g=m \mathcal{Z}_{l,m}^g, \\
&L_{\theta_1} Z_{l,m}^g =\frac{1}{2}\sqrt{l-m} ~\mathcal{Z}^{g}_{l, m+\frac{1}{2}}, ~~~~~~~~~~~~~~L_{\theta_2} Z_{l,m}^g =-\frac{1}{2}\sqrt{l+m}~ \mathcal{Z}^{g}_{l, m-\frac{1}{2}}, \label{lalphaz}\\
&L_{\theta_1} \mathcal{Z}_{l,m}^g =\frac{1}{2}\sqrt{l+m+\frac{1}{2}}~ {Z}^{g}_{l, m+\frac{1}{2}}, ~~~~~~~~L_{\theta_2} ~\mathcal{Z}_{l,m}^g =\frac{1}{2}\sqrt{l-m+\frac{1}{2}}~ {Z}^{g}_{l, m-\frac{1}{2}}, 
\end{align}\label{actllsumo}
\end{subequations}
and 
\be
L Z_{l,m}^g =(l+\frac{1}{2}) Z_{l, m}^g,~~~~~L\mathcal{Z}_{l,m}^g =(l+\frac{1}{2}) \mathcal{Z}_{l, m}^g. \label{femillL}
\ee
The relations (\ref{actllsumo}) are equal to those for the integer LLs (\ref{actllsumo1}), implying that the $\{Z_{l, m_1}^{g},~\mathcal{Z}_{l, m_2}^g\}$  form another $UOSp(1|2)$ multiplet. 
The relations (\ref{femilleth}) and (\ref{femillL}) are distinct from the integer LL results,  (\ref{boslleth}) and (\ref{bosllL}). These differences are consistent with the previous observations:  (\ref{femilleth}) implies  $\text{max}(|g|) = l - 1/2$, and (\ref{femillL}) implies that   the half-integer supermonopole harmonics have degree $2l+1$.

\subsection{Ghosts and a new definition of the inner product}

The integrals of the supermonopole harmonics over a supersphere are evaluated as 
\begin{subequations}
\begin{align}
&\int_{S^{2|2}}d\Omega_{2|2} ~{Z^{g}_{l', m'}}(x_i, \theta_{\alpha})^*~ Z^g_{l, m}(x_i, \theta_{\alpha}) = \delta_{l' l}\delta_{m' m}, \\
&\int_{S^{2|2}}d\Omega_{2|2} ~{\mathcal{Z}^{g}_{l', m'}(x_i, \theta_{\alpha})}^* ~\mathcal{Z}^g_{l, m}(x_i, \theta_{\alpha}) =- \delta_{l' l}\delta_{m' m}, \label{mathzmathz}\\
&\int_{S^{2|2}}d\Omega_{2|2} ~{\mathcal{Z}^{g}_{l, m'}(x_i, \theta_{\alpha})}^* ~{Z}^g_{l, m} (x_i, \theta_{\alpha})=0, 
\end{align}
\end{subequations}
and 
\begin{align}
&~~~\int_{S^{2|2}}d\Omega_{2|2} ~{\mathcal{Z}^{g}_{l, m'}(x_i, \theta_{\alpha})}^* ~{Y}^g_{l, m} (x_i, \theta_{\alpha})=\int_{S^{2|2}}d\Omega_{2|2} ~{\mathcal{Z}^{g}_{l, m'}(x_i, \theta_{\alpha})}^* ~\mathcal{Y}^g_{l, m} (x_i, \theta_{\alpha})\nn\\
&=\int_{S^{2|2}}d\Omega_{2|2} ~{{Z}^{g}_{l, m'}(x_i, \theta_{\alpha})}^* ~{Y}^g_{l, m} (x_i, \theta_{\alpha})=\int_{S^{2|2}}d\Omega_{2|2} ~{{Z}^{g}_{l, m'}(x_i, \theta_{\alpha})}^* ~\mathcal{Y}^g_{l, m} (x_i, \theta_{\alpha})=0.
\end{align}
The supermonopole harmonics thus form an orthonormal basis, except for the minus sign on the right-hand side of (\ref{mathzmathz}). This implies that the fermionic monopole harmonics $\mathcal{Z}_{l,m}^g$ are negative norm states, $i.e.$, ghosts. It is known that 
 representations of the $UOSp(1|2)$ group are pseudo-orthogonal \cite{Scheunert-Nahm-Rittenberg-1977, Daumens-1993}  and super Landau models contain   ghost states \cite{Hasebe-2005-2, Ivanov-2007, Curtright-Ivanov-Mezincescu-Townsend-2007}. To handle these ghosts,  there are basically three  approaches.  The first  is simply to discard the half-integer Landau levels as unphysical and   decouple the physical sector from  ghosts. As the whole Hilbert space is spanned by  the integer and the half-integer Landau level eigenstates,  almost half of the  Hilbert space has to be abandoned in this approach. The second is to utilize only the  bosonic monopole harmonics in the half-integer LLs, but discarding the fermionic sectors containing  ghosts. In this remedy,   the SUSY structure in the half-integer LLs has to be abandoned, leaving only the $SU(2)$ symmetry.      
The last one is to modify the definition of the inner product to retain the ghost states as  physical states  \cite{Curtright-Ivanov-Mezincescu-Townsend-2007}.   All of the fermionic monopole harmonics in the half-integer Landau levels are ghosts, but by just flipping the sign of their norms simultaneously, we can transform them positive norm states without ruining the  SUSY structure. This is equivalent to modifying the  metric of the super Hilbert space  \cite{Ivanov-2007}. In this paper, we adopt the last.  
We add a minus sign  for the definition of bra of  $\mathcal{Z}$:  
\be
\langle \mathcal{Z}_{l,m}^g| :=-{\mathcal{Z}_{l,m}^{g}}^*, \label{braspec}
\ee
while the bras for other supermonopole harmonics are as usual. We can represent this new definition of the bras by introducing the Scasimir: 
\be
S=-2C_{\alpha\beta}L_{\alpha}L_{\beta} +\frac{1}{4},
\ee
whose eigenvalues are $\pm (l+\frac{1}{4})$ for bosonic and fermionic eigenstates, respectively, so  the eigenvalues correspond to the fermion parity (see Appendix \ref{subsec:scas} for details). 
We then define  the  bras of the supermonopole harmonics in a unified fashion:  
\be
\langle \Psi|:=(\text{sgn}(S)^{2N}\Psi(x_i, \theta_{\alpha}))^*. \label{genedebra}
\ee
Here, $N$ denotes the LL of $\Psi(x_i, \theta_{\alpha})$.\footnote{With the Scasimir, $N$ can be represented as 
\be
N=|S|-|g|-\frac{1}{4}. 
\ee
}
For instance, if $\Psi=\mathcal{Z}$, where $N$ is a half-integer and $\text{sgn}(S)=-1$,   then (\ref{genedebra}) reduces to (\ref{braspec}). 
Accordingly,  the inner product  is modified as 
\be
 \langle \Psi|\Psi'\rangle = \int_{S^{2|2}}d\Omega_{2|2}  ~(\text{sgn}(S)^{2N}\Psi(x_i, \theta_{\alpha}))^*~\Psi'(x_i, \theta_{\alpha}) , \label{newinnerin}
\ee
and matrix element of an operator $O$ is given by 
\be
 \langle \Psi|O|\Psi'\rangle = \int_{S^{2|2}}d\Omega_{2|2}  ~(\text{sgn}(S)^{2N}\Psi(x_i, \theta_{\alpha}))^*~ O~\Psi'(x_i, \theta_{\alpha}) . \label{newinner}
\ee
The correct matrix elements of $L_{\alpha}$ (\ref{lilalpha}) can be  reproduced from  (\ref{lalphaz}) under the new formula  (\ref{newinner}), which supports the validity of this definition.  Henceforth, we use these new definitions when evaluating the inner products and matrix elements. For instance, 
 the probability densities  are calculated as 
\begin{subequations}
\begin{align}
&\rho_{l,m}^g(y_i) =\int d\theta_1 d\theta_2 ~(1-\frac{1}{2}\theta C\theta) ~{Z}^{g}_{l,m}(x_i, \theta_{\alpha})^*~Z^g_{l,m}(x_i, \theta_{\alpha}) =\frac{l+g+\frac{1}{2}}{2l+1}|\Phi^{g+\frac{1}{2}}_{l,m}(y_i)|^2 + \frac{l-g+\frac{1}{2}}{2l+1}|\Phi^{g-\frac{1}{2}}_{l,m}(y_i)|^2, \\ 
&\varrho_{l,m}^g(y_i) =-\int d\theta_1 d\theta_2 ~(1-\frac{1}{2}\theta C\theta)~ \mathcal{Z}^{g}_{l,m}(x_i, \theta_{\alpha})^*~\mathcal{Z}^g_{l,m}(x_i, \theta_{\alpha}) =|\Phi^{g}_{l-\frac{1}{2},m}(y_i)|^2 .
\end{align}
\end{subequations}

\section{Super D-matrix and Howe duality}\label{sec:superdmat}

In this section, we introduce the super D-matrix of the $UOSp(1|2)$ group and discuss  a systematic realization of the supermonopole harmonics from the perspective of  super Howe duality. 

\subsection{D-matrix and Howe duality}

The Euler angle representation of Wigner's D-matrix is given by 
\be
D_s(\alpha, \beta, \gamma) = e^{-i\alpha S_z^{(s)}}e^{-i\beta S_y^{(s)}}e^{-i\gamma S_z^{(s)}}, \label{oridwigmat}
\ee
where $S_i^{(s)}$ denote the $SU(2)$ spin matrices with spin  $s$, and thereby $D_s$ denotes a $(2s+1)\times (2s+1)$ matrix. 
It is known that the monopole harmonics are realized as the matrix elements of the D-matrix:\footnote{
With Wigner's small D-matrix 
\be
d_{s, g, m}(\beta) =(e^{i\beta S_y^{(s)}})_{g,m}, 
\ee
the monopole harmonics (\ref{monohawig}) are represented as 
\be
\Phi^{g}_{s,m}(y_i)  =\sqrt{\frac{2l+1}{4\pi}}~d_{s,g,m}(\vartheta)~e^{i(m-g)\varphi}. 
\ee
} 
\cite{Wu-Yang-1976, Dray1985, Dray1986}
\be
\Phi^{g}_{s,m}(y_i) =\sqrt{\frac{2l+1}{4\pi}}~D_{s}(\varphi, -\vartheta, -\varphi)_{g ,m}  , \label{monohawig}
\ee
where $\vartheta$ and $\varphi$ signify the angle coordinates in $\mathbb{R}^3$  (\ref{ysparangle}).  The $\mathcal{D}_s$ consists of $(2s+1)$ rows comprising the LLL and higher LL eigenstates: its $g$th row realizes the $N=s-|g|$th LL eigenstates as components \cite{Hasebe-2016, Hasebe-2023-1}. (This is a consequence of the Peter-Weyl theorem.)   
The left and right actions of the D-matrix independently induce transformations between rows and columns, respectively \cite{Karabali-Nair-2002, Greiter-2011, Hasebe-2016}. In other words, for a fixed $g$, the $(2s+1)$ elements in a given row form an irreducible representation under the right $SU(2)$ action, while  for a fixed $m$, the $(2s+1)$ elements in a given column constitute another  irreducible representation under the left $SU(2)$ action. This  interchangeability manifests the $(SU(2), SU(2))$ self-dual Howe duality \cite{Howe1979seriesAI}\footnote{The $(SU(2),SU(2))$ duality is a particular realization of Howe duality in the Landau model. Howe duality, however, is a mathematical structure that also arises in certain other quantum mechanical systems. See \cite{Rowe:2012ym, Basile:2020gqi} as pedagogical accounts of Howe duality  for physicists. } of the D-matrix, which realizes a duality between its rows and columns, or equivalently,  between  the quantum numbers $g$ and $m$. The Hilbert space $\mathcal{H}$ of the Landau model then admits a multiplicity-free  decomposition:
\begin{equation}
\mathcal{H} \cong \bigoplus_{s=0}^{\infty} \left( V_{SU(2)_L}^{(s)} \otimes V_{SU(2)_R}^{(s)} \right), \label{howesu2}
\end{equation}
where $V_{SU(2)_{L/R}}^{(s)}$ denotes the vector space on which each $SU(2)$ group with a common spin $s$ acts.\footnote{Physically, this duality corresponds to a transformation between monopole harmonics and spin-coherent states \cite{Hasebe-2024-1}.} The one-to-one mapping between $V_{SU(2)_{L}}^{(s)}$ and $V_{SU(2)_{R}}^{(s)}$ is known as the theta correspondence.   It should be emphasized that the existence of the Howe duality  is only  realized through the inclusion of higher Landau levels.

\subsection{Super D-matrix and supermonopole harmonics}\label{subsec:superdmat}

An arbitrary group element of $UOSp(1|2)$ is given by \cite{Berezin-Tolstoy-1981} 
\be
g=F\cdot B, 
\ee
where $F$ and $B$ are defined as 
\begin{subequations}
\begin{align}
&F(\eta, \eta^*):=e^{2\eta L_{\theta_1}+2\eta^*L_{\theta_2}} =\bs{1}+2\eta L_{\theta_1}+2\eta^*L_{\theta_2}-2\eta^*\eta (L_{\theta_1}L_{\theta_2} -L_{\theta_2}L_{\theta_1}),  \\
&B(\chi, \vartheta, \varphi):= e^{-i\chi L_z}e^{-i\vartheta L_y}e^{-i\varphi L_z}. 
\end{align}
\ese
Here, $L_i$ and $L_{\alpha}$ are the $UOSp(1|2)$ representation matrices rather than differential operators. For an irreducible representation with superspin $l$, the super D-matrix is expressed as 
\be
\mathcal{D}_l(\alpha,\beta,\gamma,\eta,\eta^*)= F_l(\eta, \eta^*) \cdot B_l(\alpha, \beta, \gamma) , \label{superdmat}
\ee
where 
\be
F_l(\eta, \eta^*)=\begin{pmatrix}
(1+l \eta^*\eta) \bs{1}_{2l+1} & \eta \tau^{(l)}_{\theta_1} +\eta^*\tau^{(l)}_{\theta_2} \\
\eta {\tau^{(l)}_{\theta_2}}^t-\eta^*{\tau_{\theta_1}^{(l)}}^t & (1-(l+\frac{1}{2})\eta^*\eta)\bs{1}_{2l}
\end{pmatrix}, \label{fm}
\ee
and 
\be
B_l(\alpha, \beta, \gamma)
=\begin{pmatrix}
D_{l}(\alpha, \beta, \gamma)    & 0 \\
0 & D_{l-\frac{1}{2}}(\alpha, \beta, \gamma)  
\end{pmatrix}. \label{genebosonpar}
\ee
The block diagonal components of $B_l$ are the standard D-matrices (\ref{oridwigmat}).   See  Appendix \ref{appen;uso12mat} for the explicit forms of $\tau_{\alpha}^{(l)}$ and  $F_l(\eta, \eta^*)$.  In the following, we take the Dirac gauge in which (\ref{genebosonpar}) is given by 
\be
B_l(\varphi, \vartheta, -\varphi)=e^{-i\varphi L_z^{(l)}} e^{-i\vartheta L_y^{(l)}}e^{i\varphi L_z^{(l)}}=e^{i\vartheta (  L^{(l)}_x \sin\varphi-L^{(l)}_y\cos\varphi)},
\ee
where $L_{i=x,y,x}^{(l)}$ represent the bosonic generators of the $uosp(1|2)$ matrices  (\ref{lilalpha}). 
For low-dimensional representations, (\ref{superdmat}) are given by  
\bse
\begin{align}
&\mathcal{D}_{1/2}=\begin{pmatrix}
u & -v & \eta \\
v^* & u^* & \eta^* \\
\hdashline
-u \eta^*  + v^* \eta  &  u^* \eta +v\eta^* &  1-\eta^*\eta 
\end{pmatrix}, \label{gl12}\\
&\mathcal{D}_{1}= \begin{pmatrix}
u^2 & -\sqrt{2}~uv & v^2 & \sqrt{2}~u\eta & -\sqrt{2}~ v \eta \\
\sqrt{2}~ uv^* & u^* u-v^* v & -\sqrt{2}~u^*v & u \eta^* +v^*\eta & u^*\eta-v\eta^* \\
{v^*}^2 & \sqrt{2}~u^*v^* & {u^*}^2 & \sqrt{2}~ v^*\eta^* & \sqrt{2}~ u^*\eta^* \\
\hdashline
-\sqrt{2}~(u^2\eta^*-uv^*\eta) & (u^*u-v^*v)\eta +2uv \eta^* & -\sqrt{2}~(v^2\eta^* +u^*v\eta) & (1-2\eta^*\eta)u & -(1-2\eta^*\eta)v \\
\sqrt{2}~({v^*}^2\eta-uv^*\eta^*) & -(u^*u-v^*v)\eta^*+2u^* v^* \eta  & \sqrt{2}~({u^*}^2\eta +u^*v \eta^*) & (1-2\eta^*\eta)v^* & (1-2\eta^*\eta)u^* 
\end{pmatrix}. \label{gl1}
\end{align}
\ese
Matrix $\mathcal{D}_{1/2}$ is exactly equal to (\ref{guosp12m}).   
The first two rows of $\mathcal{D}_{1/2}$ denote the  $g=1/2$ and $g=-1/2$ LLL eigenstates, (\ref{lll12}) and   (\ref{lll-12}).  The last row stands for the $l=1/2$ superspherical harmonics (\ref{l1/2superspheh}). 
Similarly, the first and third  rows of $\mathcal{D}_{1}$ represent the $g=1$ and $g=-1$ LLL eigenstates,  (\ref{lll1}) and  (\ref{lll-1}). The second row is the $l=1$ superspherical harmonics  (\ref{l1spusp}). The last two rows respectively correspond to the  $g=1/2$ and $g=-1/2$ lowest  half-integer LL eigenstates, (\ref{halfinlll+}) and (\ref{halfinlll-}).   
In general, the supermonopole harmonics are realized as the matrix elements of the super D-matrix:
\be
\Psi_{l, m}^g (x_i, \theta_{\alpha})= \frac{1}{\sqrt{4\pi}} \mathcal{D}_{l}(\varphi, \vartheta, -\varphi, \eta, \eta^*)_{g,m}.  ~~~(\text{up to sign})
\ee
As illustrated in Fig.\ref{nlrep.fig}, the integer LL supermonopole harmonics with $l=N+|g|$ appear as the rows of the  upper $(2l+1)\times (4l+1)$ rectangular matrix of $\mathcal{D}_l$,   while  the half-integer LL supermonopole harmonics  with $l=N+|g|$ appear as the rows of the lower $(2l)\times (4l+1)$ rectangular matrix of $\mathcal{D}_l$.  
\begin{figure}[tbph]
\center
\includegraphics*[width=145mm]{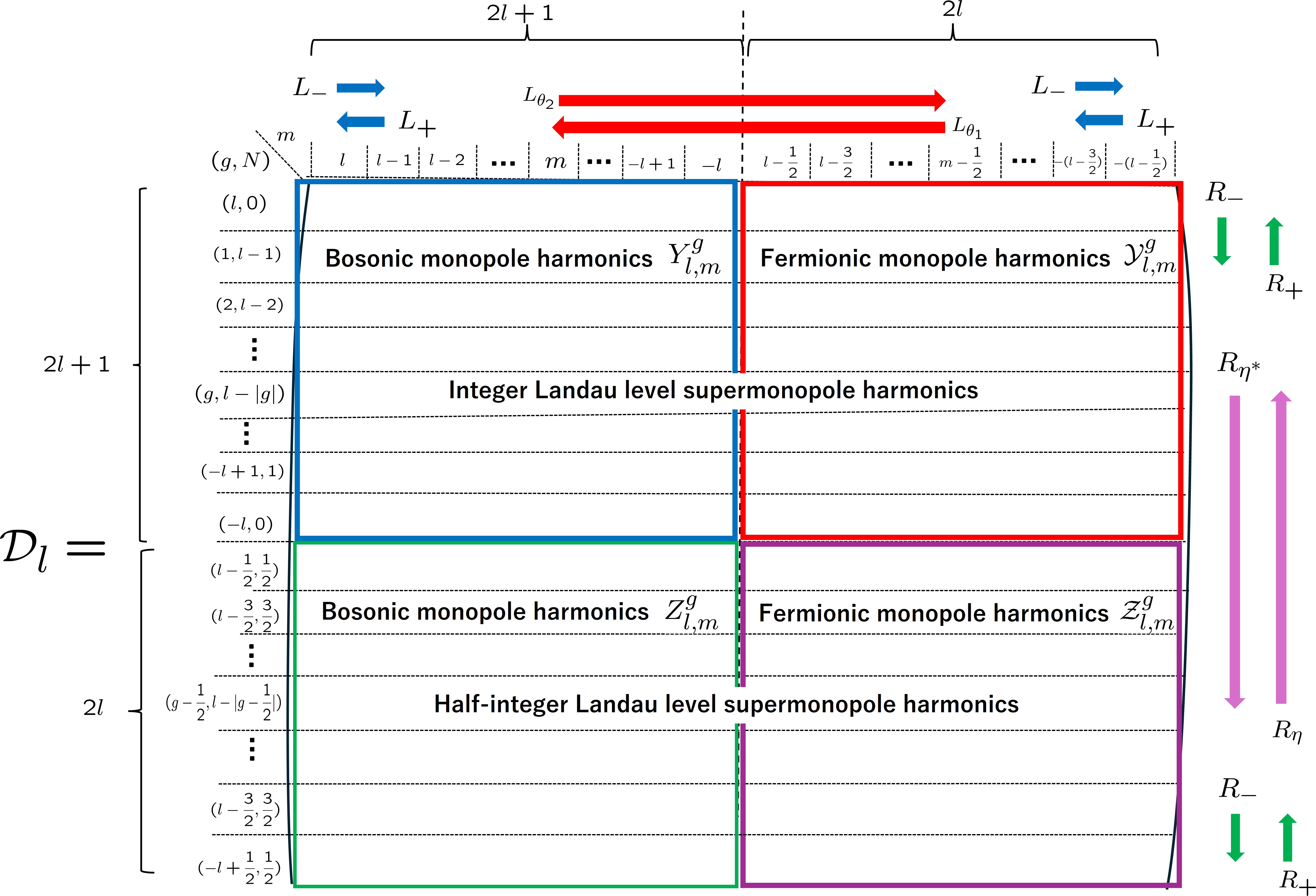}
\caption{Structure of the super $D$-matrix ($l=N+|g|$). The rows in the blue (red) rectangles correspond to bosonic (fermionic) monopole harmonics in the integer Landau levels, while those in the green (purple) rectangles realize the bosonic (fermionic) monopole harmonics in the  half-integer Landau levels. The colored arrows indicate the actions of the ladder operators of the two independent $uosp(1|2)$ algebras. }
\label{nlrep.fig}
\end{figure}

\subsection{Super Howe duality}\label{subsec:superhowe}

As in the case of the standard D-matrix, the super D-matrix undergoes two independent transformations, namely the left and right $UOSp(1|2)$ super-rotations.  
This suggests
that the $(4l+1)$ matrix elements in a given column (with varying $g$) constitute a left $UOSp(1|2)$ multiplet,
and those in a given row (with varying $m$) form a right $UOSp(1|2)$ multiplet. While the $SU(2)$ ladder
operators $R_{\pm}$ (\ref{rpmexp}) partially realize the left $UOSp(1|2)$ transformations, they only connect the components within each of the
two $SU(2)$ sectors in a column,  so they fail to relate integer and half-integer LLs. 
To bridge this gap, we construct the fermionic ladder operators that shift the monopole charge $g$ by $\pm 1/2$.

Let us recall the lowest-dimensional super D-matrix (\ref{gl12}), in which   
the fundamental  representations for $g=0$ and $g=\pm 1/2$ are realized as its rows. We then construct the fermionic ladder operators interchanging these states: 
\begin{subequations}
\begin{align}
&R_{\eta}:=\frac{1}{2}\Phi^t \mathcal{R} \frac{\partial}{\partial \psi^*} =\frac{1}{2} ((-u\eta^* +v^*\eta)\frac{\partial}{\partial v^*} - (u^*\eta+v\eta^* )\frac{\partial}{\partial u^*}-(1-\eta^*\eta)\frac{\partial}{\partial \eta^*}), \\
&R_{\eta^*}:=-\frac{1}{2}\Phi^t  \frac{\partial}{\partial \psi} =\frac{1}{2} ((-u\eta^* +v^*\eta)\frac{\partial}{\partial u} - (u^*\eta+v\eta^* )\frac{\partial}{\partial v}+(1-\eta^*\eta)\frac{\partial}{\partial \eta}), 
\end{align}\label{retaetab}
\end{subequations}
where $\Phi$ denotes the $l=1/2$ superspherical harmonics: 
\be
\Phi:=
\begin{pmatrix}
u\eta^* -v^*\eta \\
u^*\eta + v\eta^*\\
-(1-\eta^*\eta)
\end{pmatrix}. 
\ee
Notably, $R_{\alpha=\eta, \eta^*}$  are  quadratic-order polynomial differential operators unlike the other super-spinor bilinear derivative operators. It is straightforward to verify that $R_{\alpha}$ and $R_i$ satisfy the $uosp(1|2)$ algebra:  
\begin{align}
&[R_i, R_j] =i\epsilon_{ijk}R_k, ~~~~[R_i, R_{\alpha}]=\frac{1}{2}(\sigma_i)_{\beta\alpha}R_{\beta}, ~~~\{R_{\alpha}, R_{\beta}\} =\frac{1}{2}(C\sigma_{i})_{\alpha\beta} R_i \nn\\ &~~~~~~~~~~~~~(i,j,k=1,2,3,~~\alpha,\beta=\eta, \eta^*).
\end{align}
Furthermore, the generators $L_A$ and $R_A$ are shown to be super-commutative:  
\be
[L_i, R_j]=[L_i, R_{\alpha}]=[L_{\alpha}, R_{\alpha}]=\{L_{\alpha}, R_{\beta}\}=0, 
\ee
where we used the Grassmann properties of $\eta$ and $\eta^*$ along with  (\ref{constrpsi}).   
They  act as the fermionic ladder operators for the monopole charge:  
\begin{subequations}
\begin{align}
&R_{\eta} Y^g_{l,m} =\frac{1}{2}\sqrt{l-g}~ Z^{g+\frac{1}{2}}_{l,m} ,~~~~~~~~~~R_{\eta^*} Y^g_{l,m} =-\frac{1}{2}\sqrt{l+g} ~Z^{g-\frac{1}{2}}_{l,m} ,\\
&R_{\eta} Z^g_{l,m} =\frac{1}{2}\sqrt{l+g+\frac{1}{2}}~ Y^{g+\frac{1}{2}}_{l,m} ,~~~~R_{\eta^*} Z^g_{l,m} =\frac{1}{2}\sqrt{l-g+\frac{1}{2}} ~Y^{g-\frac{1}{2}}_{l,m} , \label{actrz}
\end{align}
\end{subequations}
and 
\begin{subequations}
\begin{align}
&R_{\eta} \mathcal{Y}^g_{l,m} =-\frac{1}{2}\sqrt{l-g}~ \mathcal{Z}^{g+\frac{1}{2}}_{l,m} ,~~~~~~~~R_{\eta^*} \mathcal{Y}^g_{l,m} =\frac{1}{2}\sqrt{l+g} ~\mathcal{Z}^{g-\frac{1}{2}}_{l,m} ,\\
&R_{\eta} \mathcal{Z}^g_{l,m} =-\frac{1}{2}\sqrt{l+g+\frac{1}{2}}~ \mathcal{Y}^{g+\frac{1}{2}}_{l,m} ,~~~R_{\eta^*} \mathcal{Z}^g_{l,m} =-\frac{1}{2}\sqrt{l-g+\frac{1}{2}} ~\mathcal{Y}^{g-\frac{1}{2}}_{l,m} . \label{actrzcal}
\end{align}
\end{subequations}
The  $R_{\alpha}$ operators  thus shift the supermonopole charge by $\pm 1/2$, and they interchange the integer and half-integer Landau level eigenstates.\footnote{
Careful readers may wonder about the consistency of the degrees in (\ref{actrz}) and (\ref{actrzcal}). The left-hand sides involve the quadratic-order polynomial differential operators acting on degree-$(2l+1)$ variables, which seemingly suggests a total degree  $2l+2$. However,  the nilpotency of the Grassmann variables reduces the actual degree to $2l$, which indeed matches the degrees of $Y_l$ and $\mathcal{Y}_{l}$ on the right-hand sides. 
}  We  identified two independent $uosp(1|2)$ algebras acting on the rows and columns indices, with $R_A$ and $L_A$ serving as their respective ladder operators (see Fig.~\ref{nlrep.fig}). The original $(SU(2), SU(2))$ Howe duality is now extended to the $(UOSp(1|2), UOSp(1|2))$ duality for the super D-matrix. 

The super Hilbert space  is then decomposed as 
\begin{equation}
\mathcal{H} \cong \bigoplus_{l=0}^{\infty} \left( V_{UOSp(1|2)_L}^{(l)} \otimes V_{UOSp(1|2)_R}^{(l)} \right), \label{howeosp12}
\end{equation}
where $V_{UOSp(1|2)_{L/R}}^{(l)}$ denotes the vector space on which each $UOSp(1|2)$ group with  a common superspin  $l$ acts. More specifically,  $V_{UOSp(1|2)_{L}}^{(l)}$ and $V_{UOSp(1|2)_{R}}^{(l)}$  represent the space of supermonopole harmonics generated by $L_A$ and  the $UOSp(1|2)$ irreducible representation space generated by $R_A$, respectively. The theta correspondence realizes a one-to-one map between $V_{UOSp(1|2)_{L}}^{(l)}$ and $V_{UOSp(1|2)_{R}}^{(l)}$ in the present SUSY system.

\section{Super Quantum Matrix Geometry}\label{sec:supermatrix}

There are basically three methods for deducing NCG from Landau models. The former two, summarized in Appendix \ref{append:lllgeo}, are intuitive  but restricted to the LLL. On the other hand, the third approach, known as the level projection method, is  rigorous  for deriving precise matrix coordinates  and applicable not only to the LLL but also to higher LLs \cite{ Hasebe-2023-1, Hasebe-2016}.

\subsection{Fuzzy sphere geometry of the body-sphere}\label{subsec:fuzzsphe}

As the body-part of the present SUSY model realizes Haldane's sphere, its effective geometry  is naturally given by a  fuzzy sphere geometry. Within each LL, the matrix coordinates of the body-sphere are evaluated as  
\be
(Y_{i})_{m'm} =\langle \Phi_{s m'}^{g}|y_i|\Phi_{s m}^g\rangle =\int_{S^2} d\Omega_2 {\Phi_{s m'}^{g}}^*~y_i~\Phi_{s m}^g.
\ee
From (\ref{s2invec}), the matrix elements are readily obtained as \cite{Hasebe-2016} 
\begin{align}
&\langle \Phi_{s m'}^g |y_1\pm iy_2|\Phi_{s m}^g\rangle =r\frac{g}{s(s+1)}\sqrt{s(s+1) \mp m(m\pm 1)}  ~\delta_{m', m \pm 1}, \nn\\
&\langle \Phi_{s m'}^g |y_3|\Phi_{s m}^g\rangle =r\frac{g}{s(s+1)}m ~\delta_{m', m }, 
\end{align}
which implies 
\be
Y_i =\alpha{S}^{(s)}_i, \label{fuzzymats2}
\ee
with the NC scale factor 
\be
\alpha =r\frac{g}{s(s+1) }. \label{scalfacbody}
\ee
Note that the $SU(2)$ Casimir eigenvalue $s(s+1)$ appears as the denominator of $\alpha$. 
The matrix coordinates (\ref{fuzzymats2})  constitute the NCG of the fuzzy sphere:   
\be
[Y_i, Y_j] =i\alpha \epsilon_{ijk}Y_k, ~~~~~~
Y_iY_i =\alpha^2 s  (s+1)\bs{1}_{2s+1}= r^2\frac{g^2}{s(s+1)}\bs{1}_{2s+1}. 
\ee

\subsection{Fuzzy supersphere geometry }

We extend the level projection method in Sec.\ref{subsec:fuzzsphe} to derive the supermatrix geometry.

\subsubsection{Fuzzy supersphere geometry in the integer Landau levels}\label{subsec:matrcoorintegerll}

Using the integer LL eigenstates  (\ref{intllsupermono}); $|\Psi^g_{lm}\rangle=(|Y_{lm_1}^g\rangle ~~~|\mathcal{Y}_{lm_2}^g\rangle)^t$, 
we evaluate the matrix elements of the supersphere coordinates as 
\begin{subequations}
\begin{align}
&(X_i)_{m' m} =\langle \Psi_{lm'}^{g}| x_i |\Psi_{lm}^g \rangle =\int_{S^{2|2}} d\Omega_{2|2} ~
{\Psi^{g}_{l m'}}^{*} x_i 
\Psi^{g}_{l m}, \\
&(\Theta_{\alpha})_{m' m} =\langle \Psi_{lm'}^g| \theta_\alpha |\Psi_{lm}^g \rangle =\int_{S^{2|2}} d\Omega_{2|2} ~
{\Psi^{(g)}_{l, m'}}^{*} \theta_\alpha 
\Psi^{(g)}_{l, m},
\end{align}
\end{subequations}
where $\int_{S^{2|2}}d\Omega_{2|2}$  is defined by (\ref{areas22}). From the  $SU(2)$ selection rules,  it is easy to see that they take the following block-structured forms: 
\be
X_i=\begin{pmatrix}
\langle Y| x_i |Y\rangle & 0 \\
0 & \langle \mathcal{Y} |x_i |\mathcal{Y}\rangle
\end{pmatrix},~~~~\Theta_\alpha=\begin{pmatrix}
0_{2l+1} & \langle Y | \theta_\alpha |\mathcal{Y}\rangle  \\
 \langle \mathcal{Y} |\theta_\alpha |{Y}\rangle  & 0_{2l} 
\end{pmatrix}.
\ee
The remaining matrix elements can be evaluated using either the standard monopole harmonics formulas, (\ref{prodmunumono}) and (\ref{s2invec}), or the newly derived supermonopole harmonics formula (\ref{threenewforsusy}). Here we adopt the latter. By applying (\ref{sf1}) and (\ref{sf2}) together with (\ref{l1spusp}), we  readily obtain
\begin{align}
&\langle Y_{l, m_1'}^g |x_1\pm ix_2|Y_{l, m_1}^g\rangle =r\frac{g}{l(l+\frac{1}{2})}\sqrt{l(l+1) \mp m_1(m_1\pm 1)}  ~\delta_{m_1', m_1 \pm 1}, \nn\\
&\langle Y_{l, m_1'}^g |x_3|Y_{l, m_1}^g\rangle =r\frac{g}{l(l+\frac{1}{2})}m_1 ~\delta_{m_1', m_1 }, 
\end{align}
and, from (\ref{sf3}) and (\ref{sf4}), 
\begin{align}
&\langle \mathcal{Y}_{l, m_2'}^g |x_1\pm ix_2|\mathcal{Y}_{l, m_2}^g\rangle =r\frac{g}{l(l+\frac{1}{2})}\sqrt{(l-\frac{1}{2})(l+\frac{1}{2}) \mp m'_2(m_2\pm 1)}  ~\delta_{m_2', m_2 \pm 1},\nn\\
&\langle \mathcal{Y}_{l, m_2'}^g |x_3|\mathcal{Y}_{l, m_2}^g\rangle =r\frac{g}{l(l+\frac{1}{2})}m_2 ~\delta_{m_2', m_2 }. 
\end{align}
Similarly, using  (\ref{sf5}) and (\ref{sf6}) with  (\ref{l1spusp}), the fermionic components are evaluated as     
\begin{align}
&\langle {Y}_{l, m_1'}^g|\theta_{1,2}|\mathcal{Y}_{l, m_2}^g\rangle =r\frac{g}{2l(l+\frac{1}{2})}\sqrt{l\pm m_2 +\frac{1}{2}}  ~\delta_{m_1', m_2 \pm \frac{1}{2}}, \nn\\
&\langle \mathcal{Y}_{l, m_2'}^g |\theta_{1,2}|{Y}_{l, m_1}^g\rangle =\pm r\frac{g}{2l(l+\frac{1}{2})}\sqrt{l\mp m_1 }  ~\delta_{m_2', m_1 \pm \frac{1}{2}}. 
\end{align}
The supermatrix coordinates are then identified as 
\be
X_i =\alpha{L}^{(l)}_i ~~~\Theta_{\alpha}=\alpha {L}^{(l)}_{\alpha}, \label{fuzzysuperspheinteger}
\ee
with the NC scale factor 
\be
\alpha =r\frac{g}{l(l+\frac{1}{2})}, \label{susncs}
\ee
where  $L_i^{(l)}$ and $L_{\alpha}^{(l)}$ are  the $uosp(1|2)$ matrices with superspin $l$ (\ref{lilalpha}). These are $(4l+1) \times (4l+1)$ matrices, reflecting the total degeneracy of the given Landau level.  
Analogous to the $SU(2)$ case (\ref{scalfacbody}),  the $UOSp(1|2)$ Casimir eigenvalue $l(l+1/2)$ appears as the denominator of the NC scale factor (\ref{susncs}).  
In the LLL ($l=|g|$), (\ref{susncs})  reduces to  
\be
\alpha ~\rightarrow~\frac{r}{g+\frac{1}{2}\text{sgn}(g)},  
\ee
which  (in the limit $|g|\rightarrow \infty$) coincides with the NC scale factor  obtained in 
Appendix \ref{append:lllgeo}.  It is also noted that, for a fixed $l$,  $\alpha$ (\ref{susncs}) is proportional to $g$, which will be important in  Sec.\ref{subsec:fuzzyimplfss}.  
With the identification (\ref{fuzzysuperspheinteger}), one may find that  the fuzzy supersphere geometry is  realized in an arbitrary super Landau level: 
\be
[X_i, X_j] =i\alpha \epsilon_{ijk}X_k, ~~~[X_i, \Theta_{\alpha}] = \frac{\alpha}{2}(\sigma_i)_{\beta\alpha}\Theta_{\beta}, ~~~\{\Theta_{\alpha}, \Theta_{\beta}\} = \frac{\alpha}{2}(C\sigma_i )_{\alpha\beta} X_i, \label{fuzzysuperspheralge}
\ee 
The first relation signifies the NC algebra of the bosonic coordinates, while the last denotes the NAC geometry of the fermionic sector. The second relation implies that these bosonic and fermionic coordinates are entangled, forming a closed SUSY NC algebra as a whole. The supermatrix coordinates satisfy 
\be
X_iX_i +C_{\alpha\beta}\Theta_{\alpha}\Theta_{\beta} =\alpha^2 l  (l+\frac{1}{2})\bs{1}_{4l+1}= r^2\frac{g^2}{l(l+\frac{1}{2})}\bs{1}_{4l+1}. \label{condfuzzssphere}
\ee
The right-hand side (the $UOSp(1|2)$ Casimir) is interpreted as  the square of the  radius of the fuzzy supersphere: 
\be
R_{\text{F}} : = r\frac{|g|}{\sqrt{l(l+\frac{1}{2})}}. \label{deffurad}
\ee
With higher LLs included, there are two independent parameters: $g$ and $l$, or  equivalently, $g$ and $N(=l-|g|$). In either of the limits $|g| \rightarrow \infty$ or  $N \rightarrow \infty$, the NC scale factor (\ref{susncs}) vanishes, indicating that the fuzzy supersphere reduces to a continuous supersphere. However, the NC scale factor decreases more rapidly in the  $N \rightarrow \infty$ limit than  in the  $|g| \rightarrow \infty$ limit. Consequently, the radius $R_{\text{F}}$ (\ref{deffurad}) also vanishes  as  $N \rightarrow \infty$, whereas it converges to the finite classical radius $r$ of the classical sphere as $|g| \rightarrow \infty$.

\subsubsection{Fuzzy supersphere geometry in the half-integer Landau levels}

In a  manner analogous to the analysis of the integer LLs, we deduce the supermatrix geometry from the half-integer LL eigenstates   (\ref{halfllsupermono}); $|\Psi_{lm}^g \rangle =(|Z_{lm_1}^g \rangle ~~|\mathcal{Z}_{l m_2}^g\rangle)^t$.  
From the $SU(2)$ selection rules,  we find  the matrix coordinates again take the following form:
\be
X_i=\begin{pmatrix}
\langle Z|x_i|Z\rangle & 0 \\
0 & \langle \mathcal{Z}|x_i|\mathcal{Z}\rangle
\end{pmatrix}, ~~~\Theta_\alpha=\begin{pmatrix}
0_{2l+1} & \langle Z|\theta_\alpha|\mathcal{Z}\rangle  \\
 \langle \mathcal{Z}|\theta_{\alpha}|{Z}\rangle & 0_{2l}
\end{pmatrix}.
\ee
Let us recall that the inner products (\ref{newinner}) associated with $\langle \mathcal{Z}|$ are given by 
\be
\langle \mathcal{Z}|x_i|{Z}\rangle =-\int_{S^{2|2}}d\Omega_{2|2} \mathcal{Z}^* x_i Z,~~~~\langle \mathcal{Z}|\theta_{\alpha}|{Z}\rangle =-\int_{S^{2|2}}d\Omega_{2|2} \mathcal{Z}^* \theta_{\alpha} Z. 
\ee
By proceeding in a similar manner to Sec.\ref{subsec:matrcoorintegerll}, we derive the  matrix coordinates as 
\be
X_i =\alpha L_i^{(l)}, ~~~~\Theta_{\alpha} =\alpha L_{\alpha}^{(l)}. 
\ee
These  are identical to those obtained for the integer LLs (\ref{fuzzysuperspheinteger}). We  have thus  confirmed that the fuzzy supersphere geometries  emerge  also in the half-integer Landau levels under  the new definition of the  inner products for the ghosts.

\subsection{Implications of the SUSY NC algebra}\label{subsec:implfss}

The matrix coordinates $X_i$ and $\Theta_{\alpha}$  act also as the generators of $UOSp(1|2)$ transformations. The SUSY NC algebra (\ref{fuzzysuperspheralge}) represents the super-rotational   covariance of the fuzzy supersphere, serving as the NC realization  of the covariance of the classical supersphere (\ref{uosp12coord}). The fuzzy supersphere condition (\ref{condfuzzssphere}) remains manifestly invariant under the SUSY transformations generated by $\Theta_{1}$ and $\Theta_2$, indicating that the $\mathcal{N}=1$ SUSY remains exact  after fuzzification. More specifically, under the $\mathcal{N}=1$  SUSY transformation identical to that of the classical supersphere (\ref{n1susyssp})
\be
\delta_{\xi} X_i =\Theta_{\alpha}(\sigma_i C)_{\alpha\beta}\xi_{\beta}, ~~~~\delta_{\xi}\Theta_{\alpha} =X_i  (\sigma_i)_{\beta\alpha}\xi_\beta , \label{n1susysspff}
\ee
the fuzzy supersphere condition (\ref{consxthe}) is invariant, and the SUSY NC algebra (\ref{fuzzysuperspheralge}) is also covariantly preserved. Thus, the NCG does not break the SUSY structure of the original classical system.   
This exact preservation of $\mathcal{N}=1$ SUSY is a key feature of the fuzzy geometries  constructed from  super Lie algebras, as mentioned in the Introduction. 

In terms of the supermonopole field strengths (\ref{superfieldstrength}), the SUSY NC algebra (\ref{fuzzysuperspheralge}) can be expressed as 
\be
[X_i, X_j] =i\frac{r^4}{l(l+\frac{1}{2})}F_{ij} + \cdots, ~~~~[X_i, \Theta_{\alpha}] = i\frac{r^4}{2l(l+\frac{1}{2})}C_{\alpha\beta}F_{i\beta} +\cdots , ~~~~\{\Theta_{\alpha}, \Theta_{\beta}\} = i\frac{r^4}{4l(l+\frac{1}{2})}C_{\alpha\gamma}C_{\beta\delta}F_{\gamma\delta}+\cdots ,   \label{sncgfield}
\ee
where $\cdots$ denotes higher-order corrections. Equations (\ref{sncgfield}) clearly show that the existence of SUSY field strengths is crucial for the realization of the SUSY non-commutative geometry.

Since the three bosonic matrix coordinates are equivalent under the $SU(2)$ symmetry, we may, without loss of generality, diagonalize the third component of $X_i$  (\ref{lilalpha}): 
\be
X_3 =\alpha 
\left(\begin{array}{c:c}
S_z^{(l)} & 0 \\
\hdashline 
0 & S_z^{(l-1/2)}
\end{array}
\right)=\alpha \cdot \text{diag}(l, l-1, l-2,~\cdots, -l: l-{1}/{2}, l-{3}/{2}, \cdots, -l+{1}/{2}). 
\ee
The bosonic degrees of freedom are represented by the eigenvalues of the upper-left block $S_z^{(l)}$, while the fermionic degrees of freedom correspond to those of the lower-right block $S_z^{(l-1/2)}$. 
The bosonic and fermionic components constitute the quantum elements of the fuzzy supersphere (see Fig.\ref{fuzzys.fig}). 
The radius of the latitude corresponding to the quantum element at $X_3=\alpha \cdot m$ $(m=l, l-1/2, l-1, \cdots, -l)$ is given by
\be
R_m=\sqrt{R_F^2-{X_3}^2} = r{\sqrt{\frac{1}{l(l+\frac{1}{2})}-(\frac{m}{l(l+\frac{1}{2})})^2}}\cdot |g|.  \label{defrm}
\ee
In other words, the supermonopole harmonics are eigenstates of 
$|\alpha| \sqrt{L_1 L_1+L_2 L_2+C_{\alpha\beta}L_{\alpha}L_{\beta}}$  
with  the eigenvalues (\ref{defrm}). 

\begin{figure}[tbph]
\center
\includegraphics*[width=130mm]{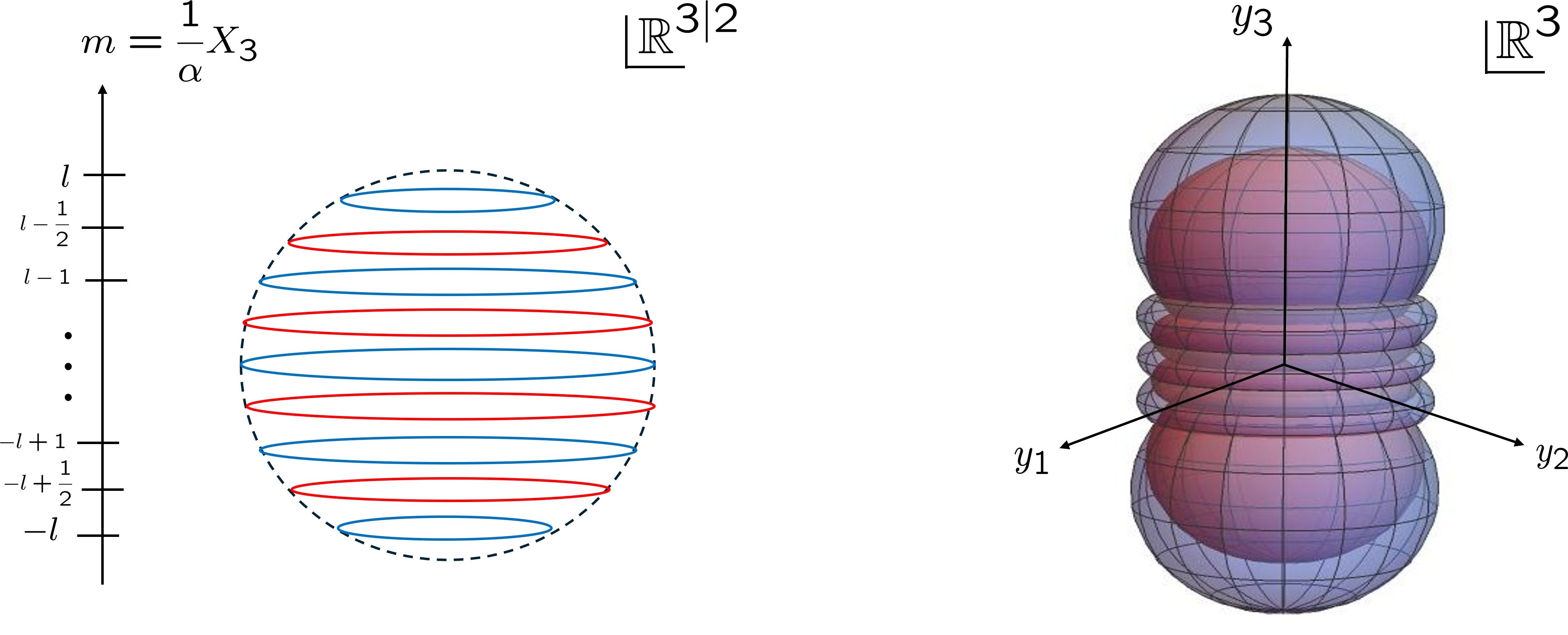}
\caption{Left: the fuzzy supersphere. The blue and red latitudes respectively correspond to the bosonic and fermionic degrees of freedom. Right: The probability densities of the $l=2$ LLL eigenstates from Fig.\ref{l2pro.fig} (also see (\ref{extmonolll})).  
The elements (= latitudes) in the left figure correspond to the supermonopole harmonics in the right figure.  }
\label{fuzzys.fig}
\end{figure}

\subsection{Fuzzy geometry transformation via the super Howe duality}\label{subsec:fuzzyimplfss}

Here, we  fix $l = N + |g|$ and vary $g$  from $l$ to $-l$ under this constraint.      
The fuzzy supersphere matrix coordinates  are given by the $(4l+1) \times (4l+1)$ $uosp(1|2)$ matrices  (\ref{fuzzysuperspheinteger}). Such fuzzy supermatrix geometries correspond to the columns of the super D-matrix $\mathcal{D}_{l}$, since each row of $\mathcal{D}_l$  describes the $(4l+1)$ component basis states of the super Landau level (Fig.\ref{nlrep.fig}).   More specifically, the rows of $\mathcal{D}_l$, labeled by $g$, represent fuzzy supersphere geometries of the same $(4l+1)\times(4l+1)$ matrix size, but with distinct NC scale factors $\alpha=g/(l(l+1/2))$, or equivalently, distinct radii (\ref{deffurad}). Then, as $g$ varies from $l$ to $-l$, we have $(4l+1)$ fuzzy supersphere geometries of different radii.  When $g$ is regarded as  ``time'', this represents  an evolution of the fuzzy supersphere of $(4l+1)$ latitudes as shown in Fig.\ref{evo.fig}.  
\begin{figure}[htbp]
  \hspace{-0.7cm}
 \includegraphics[width=160mm]{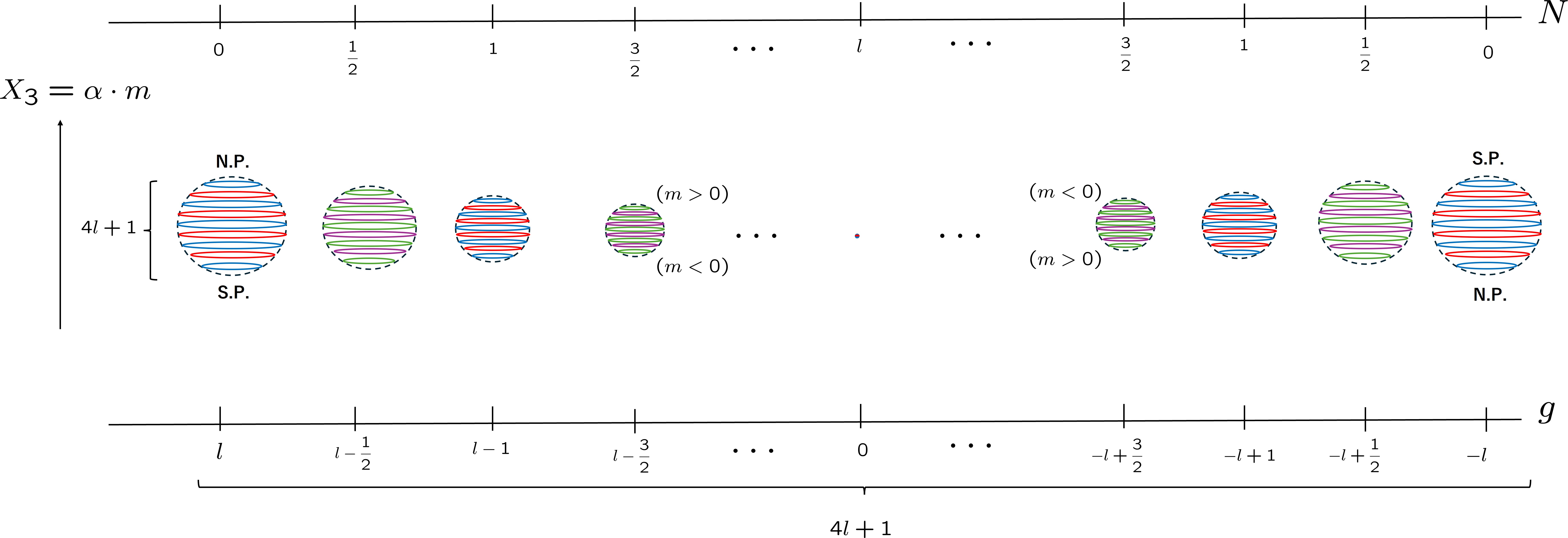}
  \caption{Evolution of the fuzzy supersphere geometry. As the supermonopole charge $g$ varies, the fuzzy superspheres associated with integer LLs (constituted from blue and red latitudes) and half-integer LLs (constituted from green and purple latitudes) appear alternately. The radius of the fuzzy supersphere shrinks as $|g|$ decreases, and upon crossing $g=0$, the northern ($m>0$) and southern ($m<0$) hemispheres are interchanged.}
  \label{evo.fig}
\end{figure}

Let us consider  the evolution of the quantum state with fixed $m$. This quantum state corresponds to the latitude with radius $R_m$ (\ref{defrm}) which depends linearly on $|g|$. As $g$ decreases from $l$ to $-l$ (with $\Delta g=1/2$), the $R_m$ sweeps out a (fuzzy) cone. In the super D matrix picture, this procedure realizes a sequence of  transitions between the columns of the integer and half-integer Landau levels.   
 The resulting fuzzy cone is then composed of $(4l+1)$ latitudes corresponding to the integer and half-integer LL eigenstates connected by the supercharges $R_\alpha$. We  refer to  this conic structure  as the  fuzzy supercone (of finite length).     
  The  slope  of the  ``surface'' of the fuzzy  supercone  is  given by\footnote{We assume the physical time to be $\frac{r}{l} \cdot g$. This definition accounts for the range of $g$ (from $-l$ to $l$) and the required dimension of length.} 
\be
c:=\frac{l}{r}\frac{\partial R_m}{\partial |g|} = {\sqrt{\frac{l}{l+\frac{1}{2}}-(\frac{m}{l+\frac{1}{2}})^2}} ~<~1. 
\ee
As $|m|$ decreases, $c$ increases and reaches the maximum value $c=1/\sqrt{1+1/(2l)}$ at $m=0$. 
With fixed $l$, $m$ varies from $l$ to $-l$, so there are $(4l+1)$ fuzzy supercones with distinct slopes (see Fig.\ref{flight.fig}). 
 Regarding $m$  as ``time'', we  see the evolution of the fuzzy supercone   in Fig.\ref{flight.fig}.  
\begin{figure}[htbp]
\includegraphics[width=160mm]{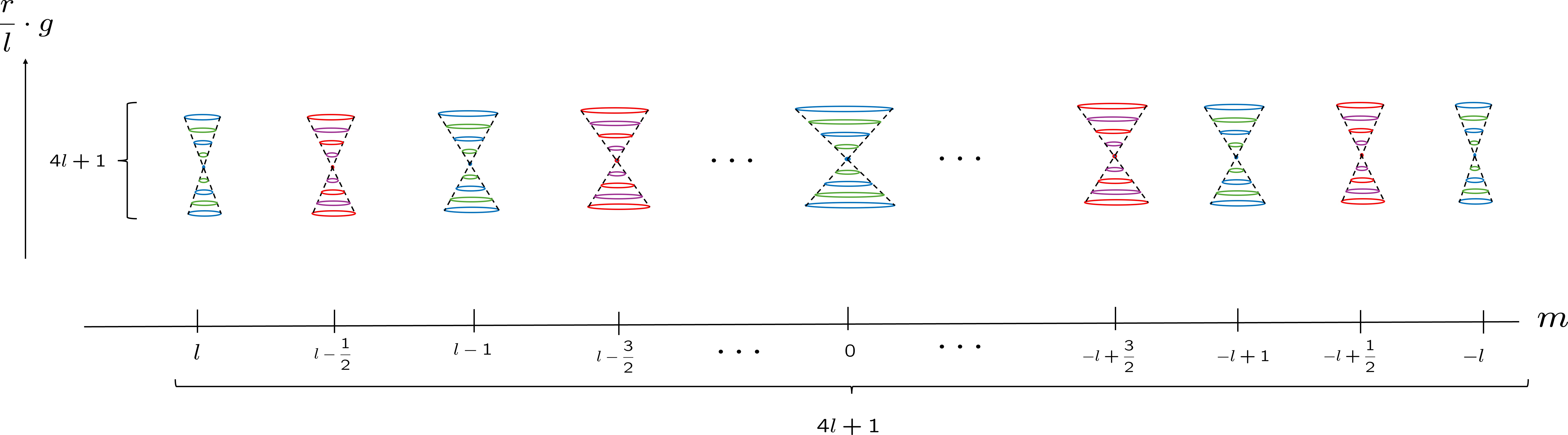}
  \caption{Evolution of the fuzzy supercone .  The slope of the fuzzy super cones decreases, as $m$ varies from $l$ to $0$, and subsequently increases  as $m$ moves from $0$ to $-l$.    }
  \label{flight.fig}
\end{figure}

The roles of $g$ and $m$ are interchanged between  the fuzzy supersphere and the fuzzy supercone.  Furthermore, the evolution of the fuzzy supercone (Fig.\ref{evo.fig}) is uniquely determined by that of the fuzzy  supersphere (Fig.\ref{flight.fig}), and vice versa. 
 Let us recall that the Howe duality represents the duality between the row ($g$) and column ($m$). Hence, the present geometric correspondence  demonstrates a geometric realization of the theta correspondence.  
  Some of the readers may be curious about the appearance of the fuzzy supercone instead of the fuzzy supersphere. 
 Since the present Howe duality is self-dual, $(UOSp(1|2), UOSp(1|2))$, the interchange of $m$ and $g$ should lead to  dual fuzzy superspheres instead of  fuzzy supercones. In this perspective, the finite-length fuzzy supercone is considered to be  a ``disguise''  of the dual fuzzy supersphere.\footnote{The physical origin of the conical geometry is, however, still unclear. 
The light cone describes the geometry  of  massless particle trajectories, and it is also known that  massless particle is closely related to charge-monopole system  \cite{Balachandran:1983oit, Goldberg-et-al-1967}.  There might be  some deeper reason behind the appearance of the conical geometry. }  
 In the language of the orbit method, each fuzzy supersphere indexed by  $g$ corresponds to a quantized  coadjoint orbit generated by one supergroup $UOSp(1|2)$. When we consider a family of such quantized  orbits labeled by $g$ from $l$ to $-l$, another family of  quantized orbits  emerges along the ``stacking direction'' of the original orbits, which manifests the  Howe duality in the framework of NCG.  
 More specifically, the stacked fuzzy superspheres yield a new nested fuzzy geometry, which can also be viewed as a family of the fuzzy supercones (see Fig.\ref{emer.fig}). These two superficially different descriptions are in fact equivalent and represent the same nested object. This provides an intuitive understanding of the Howe duality in the context of  NCG.

\begin{figure}[htbp]
  \center
\includegraphics[width=140mm]{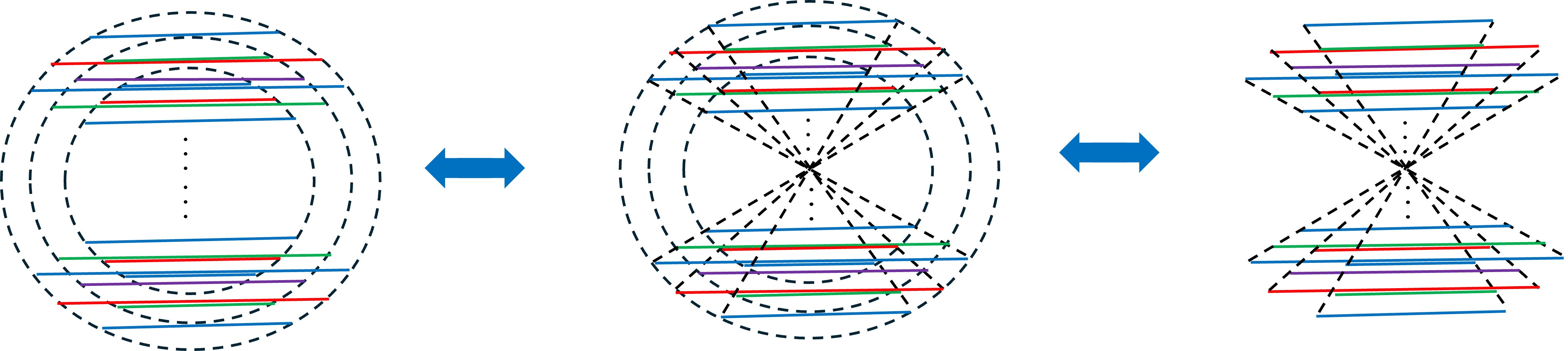}
  \caption{An intuitive picture of the emergence of fuzzy supercones.  The left  denotes the stacked fuzzy superspheres with their latitudes. From this nested structure, fuzzy supercone geometries naturally emerge as shown in the middle. The right depicts the dual description  as a family of the fuzzy supercones.}
  \label{emer.fig}
\end{figure}

\subsection{Howe duality as an underlying structure}

Finally, we discuss whether Howe duality appears in other Landau models and their associated matrix geometries.\footnote{In \cite{Bykov:2026tjt, Krivorol:2026egh}, the authors introduced a doubled space containing the Hilbert space of the Landau model, which naturally gives rise to a Howe duality. The relation between their Howe duality and the present one remains an open problem.}  Let us recall the discussions in Sec.\ref{subsec:superhowe}, where   $L_A$, which  change the external index $m$, act as the symmetry generators of the external space, while $R_A$ that change the internal index $g$, act as symmetry generators of the internal space. 
Since the $UOSp(1|2)_L$ and $UOSp(1|2)_R$ form a dual pair, the  Howe duality  can be interpreted as a  duality between  the internal  and  external spaces:   
\be
\text{Internal~(SUSY)~space} ~~~\overset{\text{(Super) Howe duality}}{\longleftrightarrow}~~~\text{ External~(SUSY)~space} 
\ee
In this interpretation, the theta correspondence realizes a bijection between the internal and external spaces (\ref{howeosp12}).
Specifically, in the present system, the $\mathcal{N}=1$ SUSY relating the bosonic and fermionic monopole harmonics in the external space $\it{necessarily}$ induces the corresponding $\mathcal{N}=1$ structure in the internal space, leading to the appearance of half-integer Landau levels as superpartners of the integer ones.   
The internal space structure is thus uniquely determined by that of the external space and vice versa.  

It should be emphasized that this internal-external space correspondence  is not a special feature in the present SUSY system, but rather a generic feature  of  systems with Howe duality.  In a general prescription  for constructing   fuzzy coset manifolds $G/H$ from Landau models, the gauge groups of the Landau models are identified with the stabilizer groups $H$ \cite{Hasebe-2023-1}.  While  $H$ originally represents a redundancy in the external  symmetry $G$  to define the  coset manifold,  it is reinterpreted in the Landau model as a redundancy of the  internal space. In other words,  the redundant external symmetry is equivalent to the internal gauge symmetry in the framework of coset-type Landau models, indicating the internal-external space correspondence. The D-matrix  provides a concrete non-linear realization of the coset, as demonstrated  in the previous studies of the Landau models on complex (projective) manifolds  \cite{Karabali-Nair-2002, Karabali:2004xq, Karabali-Nair-2006, Daoud:2006kr, Balli:2014pqa} and on higher dimensional spheres  \cite{Nair-Daemi-2004, Hasebe-2016, Hasebe-2017, Hasebe-2020, Hasebe-2021}.  
Since  Howe duality originates from  the commutativity of the left- and right-actions of the D-matrix,  it arises naturally in   coset-type Landau models   and their  associated matrix geometries. Thus,  Howe duality can be regarded as an underlying structure of these matrix geometries.  
 It has been shown that the analogous structure between internal and external spaces   gives rise to a  dimensional hierarchy   \cite{Hasebe-Kimura-2003, Hasebe-2014-1}. 
Such a duality  is closely related to the intrinsic structure of  Matrix models, as their internal and external spaces   are described by matrices of the same size arising from a common algebra \cite{Ho:2001as, Kimura:2003ab}. Since the internal and external spaces share a common origin, a large external space necessarily implies an extended internal space. 
This particular feature of the internal space gives rise to a higher-spin structure in Matrix models \cite{Hanada:2005vr, Steinacker:2016vgf, Sperling:2018xrm, Steinacker:2023zrb, Ho:2025htr}.

In the language of the orbit method, a coadjoint orbit generated by a group action corresponds to an irreducible representation of that group, while the theta correspondence determines a relation between two representation spaces. Therefore, it is natural to expect a unique correspondence between the two families of such orbits. Since the fuzzy geometries are represented  as quantized coadjoint orbits, 
the  theta correspondence between  fuzzy objects can be regarded  as a quantized  version of the relation between two orbit families, as exemplified by the geometric transformation in Sec.\ref{subsec:fuzzyimplfss}.  This observation also indicates the underlying role of Howe duality in fuzzy geometries.

\section{Summary and Discussions}\label{sec:summary}

We have performed a comprehensive analysis of the super-Landau model on a supersphere, encompassing both integer and half-integer higher LLs. By employing super-spinor derivative operators, we derived explicit forms of the supermonopole harmonics.   
In  the half-integer LLs, the fermionic monopole harmonics  are negative-norm states.  A consistent prescription for resolving  this probability problem  is proposed using a modified inner product and a projection onto the body-sphere. 
We revealed that the $(UOSp(1|2), UOSp(1|2))$ self-dual super Howe duality constitutes the underlying structure of the entire super Hilbert space based on the super D-matrix formalism. We then explicitly derived the matrix elements for the supersphere-coordinates in  arbitrary LLs, showing  that the fuzzy supersphere geometries arise in all LLs. 
We finally presented a geometric interpretation of the super Howe duality  as the theta correspondence between the fuzzy superspheres and fuzzy supercones. 

We have also discussed that the Howe duality  realizes an internal-external space duality.  
 This observation leads to  two intriguing possibilities. 
 The first  concerns  a new interpretation of quantum numbers associated with external-space symmetries. Since the external space and internal spaces are essentially equivalent, the external quantum numbers may be interpreted as internal ones, for instance  a possible topological interpretation of the magnetic quantum number. This issue will be discussed elsewhere.   
The second   pertains to the structure of Matrix model geometries. 
As discussed in the main text, the internal and external spaces in Matrix models share a common matrix origin, while  Howe duality naturally realizes this equivalence as an internal-external space duality. Furthermore, the previous studies, together with  the orbit method, also  suggest that Howe duality serves  as an underlying structure in fuzzy geometries.      Needless to mention, for Matrix models, matrices are of prime importance, while  we have observed that matrix geometries are obtained from quantum states or irreducible representations.   
This representation-theoretic point of view may offer new insights into a fundamental understanding  Matrix model geometries.   

Furthermore, though we highlighted Howe duality in the Landau models, the concept is not restricted to such models. 
Howe duality is a central concept in representation theory, providing a unified perspective that links   irreducible representations of two distinct groups as manifestations  of a  single underlying  representation. Since fuzzy manifolds are constructed from irreducible representations (and their algebraic structures),  the theta correspondence naturally relates two geometrically distinct fuzzy manifolds as distinct realizations of the same fuzzy object, namely  a family of quantized coadjoint orbits. 
Howe duality encompasses cases where one member of the two groups is compact and the other is non-compact, with different dimensions.  In such cases, we may have phenomenologically plausible matrix geometries that feature a compact internal space and a non-compact external space of different dimensions. The theta correspondence then accounts for  geometric transformations between fuzzy Riemannian (positive-definite) and pseudo-Riemannian (indefinite)  spaces.  Additionally, the see-saw structure of Howe duality allows  a single Hilbert space to accommodate multiple dual pairs. This   suggests   that a fuzzy space possesses multiple equivalent descriptions in terms of pairs of fuzzy spaces. Another remarkable feature of Howe duality is that irreducible representations of a larger group, which are difficult to construct directly, can be derived from those of a simpler group. In the context of NCG, this approach may be applied to obtain an intricate fuzzy manifold from a simpler one.
 
The application of supermonopole harmonics also offers an interesting direction to be pursued. Since  monopole harmonics are mathematically equivalent to spin-weighted spherical harmonics or Wigner’s D-matrices, the  supermonopole harmonics  will possess an equally broad utility. One promising application is about the quantum mechanics of massless superparticle. As spin-weighted spherical harmonics describe massless particle states, their SUSY counterpart  naturally describes massless superparticle states, which will be important in  studies of  supergravity and supertwistor theory. 

As mentioned in the Introduction, recent studies indicate the existence of a SUSY structure in the Moore-Read state. Specifically, in the zero-momentum limit, spin-3/2 gravitino-mode excitations are expected to appear along with the graviton-mode.  NC algebra  is  crucial for describing the magneto-roton along with its low energy limit,  the recently observed chiral graviton \cite{Liou-Haldane-Yang-Rezayi-2019, Liang-Liu-Yang-et-al-2024}. We then naturally anticipate that the SUSY NCG may  provide a possible geometric framework for the chiral gravitino-mode. It is rather stimulating to speculate  how such a sophisticated mathematics of  SUSY NCG  could manifest itself in real physics.

\section*{Acknowledgments}

I would like to thank Yusuke Kimura for the previous collaboration on which this work is based. 
 This work was supported by JSPS KAKENHI Grant No.21K03542.

\appendix

\section{The $uosp(1|2)$ algebra and relevant mathematics}\label{appen;uso12mat}

\subsection{The $uosp(1|2)$ algebra and irreducible representation}

The $uosp(1|2)$ algebra consists of three bosonic generators $L_i$ and two fermionic generators $L_{\alpha}$, which satisfy  
\be
[L_i, L_j] =i\epsilon_{ijk}L_k, ~~~[L_i, L_{\alpha}] =\frac{1}{2}(\sigma_i)_{\beta\alpha}L_\beta, ~~~\{L_{\alpha}, L_{\beta}\} =\frac{1}{2}(C\sigma_i)_{\alpha\beta}L_i, \label{uosp12algebra}
\ee
where $C:=i\sigma_2$ denotes the $SU(2)$ charge conjugation matrix. 
The first equation denotes the $su(2)$ algebra, and the second relation shows the $SU(2)$ spinor property of  the fermionic generators.  
The last equation  implies that two successive applications of the fermionic operators generate the bosonic angular momentum operations:  
\be
{L_{\theta_1}}^2 =\frac{1}{4}L_+, ~~~~{L_{\theta_2}}^2=-\frac{1}{4}L_- ~~~~~(L_{\pm}:= L_x\pm iL_y),
\ee
and 
\be
\{L_{\theta_1}, L_{\theta_2}\} = -\frac{1}{2}L_z. 
\ee
The $UOSp(1|2)$ Casimir is given by 
\be
\mathcal{C}=L_iL_i+C_{\alpha\beta}L_{\alpha}L_{\beta}. \label{casosp12}
\ee
Irreducible representations of the $UOSp(1|2)$ are indexed by the following superspin index 
\be
l=0,1/2, 1, 3/2, \cdots,  
\ee
and  the corresponding Casimir eigenvalue and the dimension of the representation are respectively given by 
\bse
\begin{align}
&l(l+1/2), \\
&4l+1.
\end{align}
\ese
Under the $SU(2)$  subgroup of  $UOSp(1|2)$,  the irreducible representation decomposes into  
\be
4l+1 =(2l+1)\oplus (2l). 
\ee
We refer to the $(2l+1)$-dimensional and $(2l)$-dimensional representations as the bosonic and fermionic states, denoted by $|l,m_1\rangle_B$ and $|l,m_2\rangle_F$, respectively. The magnetic quantum numbers range as $m_1 = l, l-1, \dots, -l$  and $m_2 = l-1/2, l-3/2, \dots, -l+1/2$. These states are assumed to satisfy
\be
\mathcal{C}|l,m\rangle_{B/F}=l(l+\frac{1}{2})|l, m\rangle_{B/F}.  
\ee
The action of the $UOSp(1|2)$ operators on these states is given by \cite{Pais-Rittenberg-1975, Scheunert-Nahm-Rittenberg-1977, Berezin-Tolstoy-1981} 
\begin{align}
&L_{\pm}|l, m\rangle_B =\sqrt{(l\mp m)(l\pm m+1)}~| l, m\pm 1\rangle_B,~~~~~~~~~~~~~~~L_{3}|l, m\rangle_B=m~| l, m\rangle_B,\nn \\
&L_{\pm}|l, m\rangle_F=\sqrt{(l\mp m-\frac{1}{2})(l\pm m+\frac{1}{2})}~| l, m\pm 1\rangle_F,~~~~~~~~~L_{3}|l, m\rangle_F=m~|l,  m\rangle_F,\nn\\
&L_{\theta_1}|l, m\rangle_B=\frac{1}{2}
\sqrt{l-m}~| l, m+\frac{1}{2}\rangle_F,~~~~~~~~~~~~~L_{\theta_2}|l, m\rangle_B=-\frac{1}{2}
\sqrt{l+m}~| l, m-\frac{1}{2}\rangle_F,\nn\\
&L_{\theta_1}|l, m\rangle_F=\frac{1}{2}
\sqrt{l+m+\frac{1}{2}}~| l, m+\frac{1}{2}\rangle_B,~~~~~~~L_{\theta_2}|l, m\rangle_F=\frac{1}{2}
\sqrt{l-m+\frac{1}{2}}~|l, m-\frac{1}{2}\rangle_B. \label{ruleladder}
\end{align}

\subsection{Scasimir and fermion parity}\label{subsec:scas}

The Scasimir \cite{q-alg/9605021} is introduced as 
\be
S=-2C_{\alpha\beta}L_{\alpha}L_{\beta}+\frac{1}{4}. 
\ee
This operator is a "square root" of the Casimir operator $\mathcal{C}$ (\ref{casosp12}) in the following sense:
\be
S^2=\mathcal{C}+\frac{1}{16}.
\ee 
Note that the Scasimir plays the role of a ``center''; it commutes with the bosonic generators but  $\it{anticommutes}$ with the fermionic generators: 
\be
[S, L_i] =\{S, L_{\alpha}\}=0. 
\ee
 The bosonic and fermionic states are the eigenstates of $S$ with eigenvalues of the same magnitude but opposite signs: 
\be
S|l,m\rangle_{B} = (l+\frac{1}{4})|l,m\rangle_{B},~~~~~S|l,m\rangle_{F} =-(l+\frac{1}{4})|l,m\rangle_{F}.
\ee
The sign of $S$ indicates the fermion parity. The fermion number $F$ and the fermion parity $(-1)^F$ are then respectively identified as 
\be
F=\frac{1}{2}(1-\text{sgn}(S)), ~~~~(-1)^F=\text{sgn}(S). 
\ee

\subsection{Supermatrix realization}

The $uosp(1|2)$  matrices for superspin $l$ are expressed as  
\be 
L^{(l)}_i  =\left(\begin{array}{c:c}
S_i^{(l)} & 0 \\
 \hdashline
0 & S_i^{(l-\frac{1}{2})} 
\end{array}
\right),~~~~L^{(l)}_{\alpha} = \frac{1}{2}\left(\begin{array}{c:c}
0_{2l+1} & \tau^{(l)}_{\alpha} \\
\hdashline 
C_{\alpha\beta}{\tau_{\beta}^{(l)}}^{t} & 0_{2l}
\end{array}\right). \label{lilalpha}
\ee
Here, $S_i^{(s)}$ denote the $SU(2)$ spin matrices of  spin  $s$: 
\be
S_i^{(s)}S_i^{(s)} =s(s+1) \bs{1}_{2s+1}. 
\ee
The $(2l+1) \times (2l)$ rectangular matrices $\tau^{(l)}_{\alpha}$ in $L_{\alpha}^{(l)}$ (\ref{lilalpha}) are defined as
\be
\tau_{\theta_1}^{(l)} := \begin{pmatrix}
\sqrt{2l} & 0 & 0 & 0 & 0 & 0   \\
0 & \sqrt{2l-1}  & 0 & 0 &  0 & 0   \\
0 & 0 & \sqrt{2l-2}   & 0 &  0 & 0   \\
0 & 0 & 0   & \ddots &  0 & 0  \\
0 & 0 & 0   & 0 &  \sqrt{2} & 0   \\
0 & 0 & 0   & 0 &  0 & \sqrt{1}   \\
0 & 0 & 0   & 0 &  0 & 0  
\end{pmatrix}, ~~
\tau_{\theta_2}^{(l)} := \begin{pmatrix}
0 & 0 & 0   & 0 &  0 & 0   \\
\sqrt{1} & 0 & 0 & 0 & 0 & 0   \\
0 & \sqrt{2}  & 0 & 0 &  0 & 0   \\
0 & 0 & \sqrt{3}   & 0 &  0 & 0  \\
0 & 0 & 0   & \ddots &  0 & 0  \\
0 & 0 & 0   & 0 &   \sqrt{2l-1} &  0  \\
0 & 0 & 0   & 0 &  0 & \sqrt{2l}   
\end{pmatrix}. 
\ee
The bosonic generators $L_i$ in (\ref{lilalpha}) are  the direct sum of $SU(2)$ spin generators whose spins differ by $1/2$. In the matrix representation, the Scasimir  is expressed as 
\be
S^{(l)}=(l+\frac{1}{4})\left(\begin{array}{c:c}
1_{2l+1} & 0 \\
\hdashline 
0 & -1_{2l}
\end{array}\right).
\ee

The  transformation matrix $F_l$ (\ref{fm}) is given by  
\begin{align}
&F_l(\eta, \eta^*)=e^{2\eta L_{\theta_1}^{(l)}+2\eta^* L_{\theta_2}^{(l)}}=\nn\\
&\hspace{-1.5cm}
\hspace{-0.5cm}\small{
\left(
\begin{array}{cccccc:ccccc}
 1+l\eta^*\eta & 0 &  0 & 0 & 0 & 0 & \sqrt{2l}~\eta  & 0 &  0 & 0 & 0 \\
  0 &  1+l\eta^*\eta & 0 & 0 & 0 & 0 & \eta^* &   \sqrt{2l-1}~\eta & 0 & 0 & 0 \\ 
  0 &  0 & 1+l\eta^*\eta & 0 & 0& 0 & 0 &  \sqrt{2}~\eta^* & \sqrt{2l-2}~\eta & 0 & 0\\
  0 & 0 &  0 & \ddots & 0 & 0 & 0 & 0 &  \sqrt{3}~\eta^* & \ddots & 0\\
  0 & 0 & 0  & 0 & 1+l\eta^*\eta & 0 & 0 & 0 & 0  & \ddots  & \eta \\
  0 & 0 & 0    & 0 & 0 & 1+l\eta^*\eta & 0 & 0 & 0    & 0 & \sqrt{2l}~\eta^*\\
 \hdashline
-\sqrt{2l}~\eta^* & \eta & 0 &    0  & 0 & 0 & 1-(l+\frac{1}{2})\eta^*\eta  & 0 &  0 & 0 & 0  \\
0 & -\sqrt{2l-1}~\eta^* & \sqrt{2}~\eta &    0  & 0 & 0 & 0 &  1-(l+\frac{1}{2})\eta^*\eta  & 0 & 0 & 0 \\
0 & 0 & -\sqrt{2l-2}~\eta^* &    \sqrt{3}~\eta  & 0 & 0 & 0 &  0 & 1-(l+\frac{1}{2})\eta^*\eta  & 0 & 0 \\
0 & 0 & 0 &    \ddots  & \ddots  & 0 & 0 & 0 &  0 & \ddots & 0 \\
0 & 0 & 0 &    0  & -\eta^* &  \sqrt{2l}~\eta & 0 & 0 & 0  & 0 & 1-(l+\frac{1}{2})\eta^*\eta
\end{array} 
\right) .}
\end{align}

\normalsize

\section{Berezinian and the area element of $S^{2|2}$}\label{sec:berez}

The Berezinian integral on  $S^{2|2}$ is derived as follows.   
We consider the coordinate transformation from the super-Euclidean coordinates in $\mathbb{R}^{3|2}$ to the superspherical coordinates: 
\begin{align}
&x_1 =r(1-\frac{1}{2r^2}\theta C\theta)\cos(\varphi)\sin(\vartheta), ~~x_2 =r(1-\frac{1}{2r^2}\theta C\theta)\sin(\varphi)\sin(\vartheta), ~~x_3 =r(1-\frac{1}{2r^2}\theta C\theta)\cos(\vartheta), \nn\\
&\theta'_1 =\theta_1, ~~~\theta'_2=\theta_2. \label{fromthetaxtoys}
\end{align}
Let us recall that $x_i = \left(1 - \frac{1}{2r^2} \theta C \theta \right) y_i$, in which $y_i$ represent the standard Cartesian coordinates in $\mathbb{R}^3$ (\ref{ysparangle}). 
The superdeterminant (Berezinian) associated with the coordinates transformation is given by 
\be
Ber=\text{sdet}\begin{pmatrix}
A & C \\
D & B
\end{pmatrix} =\frac{\text{det}(A-C\frac{1}{B}D)}{\text{det}(B)}. \label{berezj}
\ee
In the present case (\ref{fromthetaxtoys}), the blocks of the super-Jacobian matrix take the form of  
\be
A =\begin{pmatrix}
\frac{\partial \bs{x}}{\partial r} & \frac{\partial \bs{x}}{\partial \vartheta} & \frac{\partial \bs{x}}{\partial \varphi}
\end{pmatrix}=\begin{pmatrix}
(1+\frac{1}{2r^2}\theta C\theta) \frac{\partial \bs{y}}{\partial r} & (1-\frac{1}{2r^2}\theta C\theta) \frac{\partial \bs{y}}{\partial \vartheta} & (1-\frac{1}{2r^2}\theta C\theta) \frac{\partial \bs{y}}{\partial \varphi}
\end{pmatrix},
\ee
and 
\be
C_{i\alpha} =\frac{\partial x_i}{\partial \theta_{\alpha}}=-\frac{1}{r^2}x_iC_{\alpha\beta}\theta_{\beta},~~~D_{\alpha i} =\frac{\partial\theta_{\alpha}}{\partial x_i}=0,~~~B_{\alpha\beta}=\frac{\partial\theta'_{\alpha}}{\partial \theta_{\beta}}=\delta_{\alpha\beta}. 
\ee
Since $D=0$ and $\det(B)=1$, the Berezinian simplifies to $\det(A)$, which is evaluated as
\be
{Ber}=\text{det}(A)=(1+\frac{1}{2r^2}\theta C\theta)(1-\frac{1}{2r^2}\theta C\theta)^2 ~~\text{det} \begin{pmatrix}
\frac{\partial\bs{y}}{\partial r} &  \frac{\partial \bs{y}}{\partial \vartheta} & \frac{\partial \bs{y}}{\partial \varphi}
\end{pmatrix}=(1-\frac{1}{2r^2}\theta C\theta)~r^2\sin\vartheta . 
\ee
Consequently, the Berezin integral on $S^{2|2}$ is related to the integral on $S^2$  as 
\be
\int_{S^{2|2}} d\Omega_{2|2} =\int_{S^{2}} d\vartheta d\varphi d\theta_1 d\theta_2 ~Ber = \int_{S^2}\overbrace{d\vartheta d\varphi \sin\vartheta}^{=d\Omega_2} \int  d\theta_1 d\theta_2 (1-\frac{1}{2}\theta C\theta).
\ee
For an arbitrary function on $\mathbb{R}^{3|2}$ 
\be
f(x_i, \theta_{\alpha})=f_0(y_i)+f_1(y_i)\theta_1+ f_2(y_i)\theta_2 +f_3(y_i)\theta_1\theta_2, 
\ee
its integral over $S^{2|2}$ yields   
\be
\int_{S^{2|2}}d\Omega_{2|2}~ f(x_i, \theta_{\alpha}) =\int_{S^2}d\Omega_2 ~f_0(y_i)-\int_{S^2}d\Omega_2 ~f_3(y_i).
\ee
In particular, the integral of unity over $S^{2|2}$ is identical to that over $S^2$: 
\be
\int_{S^{2|2}} d\Omega_{2|2} ~1 =\int_{S^2} d\Omega_2 ~1. 
\ee

\section{Special LL eigenstates}\label{sec:specic}

We examine two special limits of $g$,  the LLL limit $|g| = \text{max}(|g|) = l$, and  the superspherical harmonics limit, $g = \text{min}(|g|) = 0$. In these limits,  the right-hand sides of (\ref{compfermimo}) reduce to the first and second terms, respectively.    We also discuss the lowest half-integer LL eigenstates.

\subsection{Lowest Landau level eigenstates ($|g|=\text{max}|g|=l$)}

In the LLL ($N=0$), the energy eigenvalues are given by (\ref{lllenergsup}). 
For $g>0$ , from (\ref{bosonmonohar}) and (\ref{genefermmon1}), the LLL eigenstates are obtained as  
\begin{subequations}
\begin{align}
Y^{g}_{g, m}(x_i, \theta_{\alpha})& =\sqrt{\frac{1}{4\pi} \frac{(2g)!}{(g+m)!(g-m)!}} ~u^{g+m}v^{g-m}, 
\\
\mathcal{Y}^{g}_{g, m}(x_i, \theta_\alpha)&=\sqrt{2g}~ Y^{g-\frac{1}{2}}_{g-\frac{1}{2}, m} \cdot \eta  
=\sqrt{\frac{1}{4\pi} \frac{(2g)!}{(g+m-\frac{1}{2})!(g-m-\frac{1}{2})!}}~  u^{g+m-\frac{1}{2}}v^{g-m-\frac{1}{2}} \eta. 
\end{align}
\end{subequations}
These are holomorphic functions of the super Hopf spinor components \cite{Hasebe-Kimura-2005}. For example, 
\begin{subequations}
\begin{align}
&(l, g)=(1/2,1/2)~:~~Y^{1/2}_{1/2, 1/2}=\frac{1}{\sqrt{4\pi}}u,~~Y^{1/2}_{1/2, -1/2}=\frac{1}{\sqrt{4\pi}}v,~~\mathcal{Y}^{1/2}_{1/2, 0}=\frac{1}{\sqrt{4\pi}}\eta,   \label{lll12} \\ 
&(l, g)=(1,1)~:~Y^{1}_{1, 1}=\frac{1}{\sqrt{4\pi}}u^2 ,~~Y^{1}_{1, 0}(x_i, \theta_{\alpha})=\frac{1}{\sqrt{2\pi}} uv ,~{Y}^{1}_{1,-1}=\frac{1}{\sqrt{4\pi}}v^2,~\mathcal{Y}^{1}_{1, 1/2}=\frac{1}{\sqrt{2\pi}} u\eta, ~\mathcal{Y}^{1}_{1, -1/2}=\frac{1}{\sqrt{2\pi}} v\eta. \label{lll1}
\end{align}
\end{subequations}
Equation (\ref{lll12}) corresponds to  the components of the super Hopf spinor themselves, whereas   (\ref{lll1}) appears in Sec. \ref{subsec:superdmat}. 
 
Meanwhile for $g<0$, the LLL eigenstates are 
\begin{align}
Y^{-|g|}_{|g|, m}(x_i, \theta_{\alpha})& =(-1)^{|g|+m}\sqrt{\frac{1}{4\pi} \frac{(2|g|)!}{(|g|+m)!(|g|-m)!}} ~{u^*}^{|g|-m}{v^*}^{|g|+m}, \nn\\
\mathcal{Y}^{-|g|}_{|g|, m}(x_i, \theta_\alpha)&=-\sqrt{2|g|}~ Y^{-|g|+\frac{1}{2}}_{|g|-\frac{1}{2}, m} \cdot \eta^*  
=(-1)^{|g|+m+\frac{1}{2}}\sqrt{\frac{1}{4\pi} \frac{(2|g|)!}{(|g|+m-\frac{1}{2})!(|g|-m-\frac{1}{2})!}}~  {u^*}^{|g|-m-\frac{1}{2}}{v^*}^{|g|+m-\frac{1}{2}} \eta^* , 
\end{align}
which are anti-holomorphic functions of the super Hopf spinor components. 
For instance, 
\begin{subequations}
\begin{align}
&(l, g)=(1/2,-1/2)~:~~Y^{-1/2}_{1/2, 1/2}=-\frac{1}{\sqrt{4\pi}}v^*,~~Y^{-1/2}_{1/2, -1/2}=\frac{1}{\sqrt{4\pi}}u^*,~~\mathcal{Y}^{-1/2}_{1/2, 0}=-\frac{1}{\sqrt{4\pi}}\eta^*,    \label{lll-12} \\ 
&(l, g)=(1, -1):Y^{-1}_{1, 1}=\frac{1}{\sqrt{4\pi}}{v^*}^2 ,~Y^{-1}_{1, 0}=-\frac{1}{\sqrt{2\pi}} u^*v^* ,~{Y}^{-1}_{1,-1}=\frac{1}{\sqrt{4\pi}}{v^*}^2,~\mathcal{Y}^{1}_{1, 1/2}=\frac{1}{\sqrt{2\pi}} v^*\eta^*, ~\mathcal{Y}^{1}_{1, -1/2}=-\frac{1}{\sqrt{2\pi}} u^*\eta^*. \label{lll-1}
\end{align}
\end{subequations}
Equation (\ref{lll-12}) represents  the pseudo-complex representation  (\ref{complexhopfsp}). The LLL eigenstates for $\pm g$ are simply related under the exchange: 
\be
u ~\longleftrightarrow ~-v^*, ~~~v ~\longleftrightarrow ~u^*, ~~~\eta ~\longleftrightarrow ~-\eta^*.
\ee
The probability densities  are  
\begin{subequations}
\begin{align}
&\rho_{|g|, m}^{g}(y_3)=|\Phi_{|g|, m}^g (y_3)|^2=\frac{(2|g|+1)!}{4\pi(|g|+m)!(|g|-m)!}\biggl(\frac{r+y_3}{2r}\biggr)^{|g|+\text{sgn}(g)m}\biggl(\frac{r-y_3}{2r}\biggr)^{|g|-\text{sgn}(g) m}, \\
&\varrho_{|g|, m}^g (y_3) 
=\frac{1}{2(2|g|+1)}\biggl((2|g|-2m+1 )~\rho_{|g|, m-\frac{1}{2}}^g(y_3)    + (2|g|+2m+1)~\rho_{|g|, m+\frac{1}{2}}^{g}(y_3)\biggr), \label{calrholll}
\end{align}
\end{subequations}
where (\ref{calrholll}) can also be  expressed in a concise form:  
\be 
\varrho_{|g|, m}^g (y_3) 
=\rho^{g-\frac{1}{2}\text{sgn}(g)}_{ |g|-\frac{1}{2}, m}(y_3).  \label{densprorelfer}
\ee 
The probability densities of the LLL fermionic harmonics thus coincide exactly with those of the original LLL monopole harmonics, provided the magnetic charge is shifted by $- \frac{1}{2}\text{sgn}(g)$. The extreme values (maxima) of $\rho_{ |g|, m}^g (y_3)$ and $\varrho_{ |g|, m}^g (y_3)$ occur at 
\be
\frac{y_3}{r}= \frac{m}{|g|} ,~~~\frac{m}{|g|-\frac{1}{2}}, \label{extmonolll}
\ee
which are distributed along the $y_3$-axis with constant intervals of $\Delta y_3 = \frac{r}{|g|}$ and $\frac{r}{|g| - 1/2}$, respectively.

\subsection{Superspherical harmonics ($g=\text{min}(|g|)=0$)}

The superspherical harmonics are the eigenstates of 
\be
H=\frac{1}{2Mr^2} (L_i^{0}L_i^{0}+C_{\alpha\beta}L_{\alpha}^{0}L_{\beta}^{0}),
\ee
with the eigenvalues (\ref{superspheene}). 
From (\ref{relorimobos}) and (\ref{fermoextheta}), the superspherical harmonics $(l=0,1,2,\cdots)$ are obtained as\footnote{
With  the Clebsch-Gordan coefficients, (\ref{fsupersphere}) can be written as 
\be
\mathcal{Y}^{0}_{l,m}(x_i, \theta_{\alpha}) =-\langle \frac{1}{2}, \frac{1}{2}; l-1, m-\frac{1}{2}|l-\frac{1}{2}, m \rangle ~\Phi^0_{l-1, m-\frac{1}{2}}(x_i)~\frac{\theta_1}{r} -\langle \frac{1}{2}, -\frac{1}{2}; l-1, m+\frac{1}{2}|l-\frac{1}{2}, m \rangle~\Phi^0_{l-1, m-\frac{1}{2}}(x_i)~\frac{\theta_2}{r}.
\ee
} 
\begin{subequations}
\begin{align}
&Y^{0}_{l, m}(x_i)
=\frac{1}{\sqrt{2l+1}} ~\Phi^0_{l,m}(x_i)=\frac{1}{\sqrt{2l+1}}(1-\frac{l}{2}\theta C\theta)~\Phi^0_{l,m}(y_i), 
\\
&\mathcal{Y}^{0}_{l,m}(x_i, \theta_{\alpha}) 
=-\sqrt{\frac{l+m-\frac{1}{2}}{2l-1}}~\Phi^0_{l-1, m-\frac{1}{2}}(x_i)~\frac{\theta_1}{r} -\sqrt{\frac{l-m-\frac{1}{2}}{2l-1}}~\Phi^0_{l-1, m+\frac{1}{2}}(x_i)~\frac{\theta_2}{r}, \label{fsupersphere}
\end{align}\label{supersph1}
\end{subequations}
where $\Phi^0_{l,m}$ denote the ordinary spherical harmonics. 
(Since the arguments $x_i$ are not c-numbers, $\Phi_{l,m}^0(x_i)$ should be understood as a formal expression.)  
The bosonic spherical harmonics depend only on the bosonic coordinates $x_i$. Low-dimensional examples are      
\begin{subequations}
\begin{align}
&l=0~~:~~Y^{0}_{0,0}=\frac{1}{\sqrt{4\pi}} ,  \\
&l=1~~:~~  Y^{0}_{1,1}=-\frac{1}{\sqrt{8\pi}}\frac{x_1+ix_2}{r},~~~~Y^{0}_{1,0}=\frac{1}{\sqrt{4\pi}}\frac{x_3}{r},~Y^{0}_{1,-1}=\frac{1}{\sqrt{8\pi}}\frac{x_1-ix_2}{r},  \nn\\
&~~~~~~~~~~~~~\mathcal{Y}_{1, 1/2}^0=-\frac{1}{\sqrt{4\pi}}\frac{\theta_1}{r}, ~~~~~\mathcal{Y}_{1, -1/2}^0=-\frac{1}{\sqrt{4\pi}}\frac{\theta_2}{r}, \label{l1spusp}\\
&l=2~~:~~ Y^{0}_{2,2}=\sqrt{\frac{3}{{32\pi}}}(\frac{x_1+ix_2}{r})^2,~~~~Y^{0}_{2,1}=-\sqrt{\frac{3}{{8\pi}}}\frac{(x_1+ix_2)x_3}{r^2},~~~~Y^{0}_{2,0}=\frac{1}{\sqrt{16\pi}}\frac{2{x_3}^2-{x_1}^2-{x_2}^2}{r^2},  \nn\\
&~~~~~~~~~~~~~Y^{0}_{2,-1}=\sqrt{\frac{3}{{8\pi}}}\frac{(x_1-ix_2)x_3}{r^2},~~~~Y^{0}_{2,-2}=\sqrt{\frac{3}{{32\pi}}}(\frac{x_1-ix_2}{r})^2, \nn\\
&~~~~~~~~~~~~~\mathcal{Y}^{0}_{2,3/2}=\sqrt{\frac{3}{{8\pi}}}\frac{(x_1+ix_2)\theta_1}{r^2},~~~~\mathcal{Y}^{0}_{2,1/2}=-\frac{1}{\sqrt{2\pi}}\frac{x_3 \theta_1}{r^2} +\frac{1}{\sqrt{8\pi}}\frac{(x_1+ix_2)\theta_2}{r^2},\nn\\
&~~~~~~~~~~~~~\mathcal{Y}^{0}_{2,-1/2}=-\frac{1}{\sqrt{8\pi}}\frac{(x_1-ix_2) \theta_1}{r^2} -\frac{1}{\sqrt{2\pi}}\frac{x_3\theta_2}{r^2},~~~~\mathcal{Y}^{0}_{2,-3/2}=-\sqrt{\frac{3}{{8\pi}}}\frac{(x_1-ix_2)\theta_2}{r^2}.\label{l2supersphe}
\end{align}
\end{subequations}
As expected, the components of the $l=1$ supermultiplet    
(\ref{l1spusp}) are given by  the  super-vector coordinates.  
 The probability densities for (\ref{supersph1}) are respectively given by 
\bse
\begin{align}
&\rho^0_{l,m}(y_3)=|\Phi^0_{l,m}(y_i)|^2, \\
&\varrho_{l, m}^0(y_3)
= \frac{1}{2(2l-1)} \biggl( (2l+2m-1) ~\rho_{l-1, m-\frac{1}{2}}^0(y_3)  + (2l-2m-1) ~\rho_{l-1, m+\frac{1}{2}}^0(y_3) \biggr). \label{probspherel}
\end{align}
\ese
Equation (\ref{probspherel}) indicates that the probability densities of the fermionic spherical harmonics are composed of the original ones with the angular momentum shifted by $-1$. (\ref{probspherel})  can also be expressed as  
\be
\varrho_{l, m}^0(y_i)=\frac{1}{2}(\rho_{ l-\frac{1}{2}, m}^{-\frac{1}{2}}+\rho_{ l-\frac{1}{2},m}^{\frac{1}{2}}).
\ee
The probability densities of  superspherical harmonics exhibit  reflection symmetry with respect to the $y_1y_2$ plane:  $\rho_{ l, m}^0(y_3)=\rho_{ l, m}^0 (-y_3)$ and $\varrho_{ l, m}^0(y_3)=\varrho_{ l, m}^0 (-y_3)$.

For half-integer $l$s ($l = 1/2, 3/2, \dots$), the superspherical harmonics are given by 
\begin{subequations}
\begin{align}
&Z^0_{l,m}(x_i, \theta_{\alpha}) =\sqrt{\frac{l-m+1}{2(l+1)}}~\Phi^0_{l+\frac{1}{2}, m-\frac{1}{2}}(x_i)~\frac{\theta_1}{r} - \sqrt{\frac{l+m+1}{2(l+1)}}~\Phi^0_{l+\frac{1}{2}, m+\frac{1}{2}}(x_i)~\frac{\theta_2}{r}, \\
&\mathcal{Z}^0_{l,m}(x_i, \theta_i)=\frac{1}{\sqrt{2l}}(1+l\theta C\theta)~\Phi^0_{l-\frac{1}{2}, m}(x_i)=\frac{1}{\sqrt{2l}}\biggl(1+\frac{1}{2}(l+\frac{1}{2})\theta C\theta\biggr) ~\Phi^{0}_{l-\frac{1}{2},m}(y_i). 
\end{align}\label{supersph2}
\end{subequations}
For instance, 
\begin{align}
l=1/2~~~:~~&Z^{0}_{1/2, 1/2} =\frac{1}{\sqrt{4\pi}r^2} (x_3\theta_1 +(x_1+ix_2)\theta_2), ~~~Z^{0}_{1/2, -1/2} =\frac{1}{\sqrt{4\pi}r^2} ((x_1-ix_2)\theta_1-x_3\theta_2 ), \nn\\
&\mathcal{Z}^{0}_{1/2, 0} =\frac{1}{\sqrt{4\pi}}(1+\frac{1}{2r^2}\theta C\theta ). \label{l1/2superspheh}
\end{align}
The probability densities for (\ref{supersph2}) are 
\begin{subequations}
\begin{align}
&\rho_{l,m}^0(y_3):=\int d\theta_1 d\theta_2 (1-\frac{1}{2}\theta C\theta) ~Z^{0 *}_{l,m} Z^{0}_{l,m} =\frac{l-m+1}{2(l+1)}|\Phi^0_{l+\frac{1}{2}, m-\frac{1}{2}}(y_3)|^2 +\frac{l+m+1}{2(l+1)}|\Phi^0_{l+\frac{1}{2}, m+\frac{1}{2}}(y_3)|^2, \\
&\varrho_{l,m}^0(y_3) :=-\int d\theta_1 d\theta_2 (1-\frac{1}{2}\theta C\theta) ~\mathcal{Z}^{0 *}_{l,m} \mathcal{Z}^{0}_{l,m} = |\Phi^0_{l-\frac{1}{2}, m}(y_3)|^2.  
\end{align}
\end{subequations}

To the best of the author's knowledge, the superspherical harmonics were first constructed in Ref.\cite{Daumens-1993}. 
While the specific forms may appear to differ, the present formulas (\ref{supersph1}) and (\ref{supersph2}) are  mathematically equivalent to those  in \cite{Daumens-1993}.

\subsection{Lowest half-integer Landau level eigenstates $(N=1/2)$}

From the general formulas (\ref{bosogoddpol}) and (\ref{ferhalinnnmo}), we readily obtain the eigenstates for the lowest half-integer LL $(N=1/2)$. In the case of a positive monopole charge $g > 0$, the eigenstates are given by 
\begin{subequations}
\begin{align}
Z^g_{g+\frac{1}{2}, m}& =\sqrt{\frac{2g+1}{4\pi}\frac{(2g+1)!}{(g+\frac{1}{2}+m)! (g+\frac{1}{2}-m)!}}~u^{g+\frac{1}{2}+m} v^{g+\frac{1}{2}-m}\eta^* \nn\\
&+\sqrt{\frac{1}{4\pi}\frac{(g+\frac{1}{2}+m)! (g+\frac{1}{2}-m)! } {(2g)!}}~\nn\\
&\cdot (\begin{pmatrix}
2g \\
g-m-\frac{1}{2}
\end{pmatrix}
u^{g+\frac{1}{2}+m} v^{g-\frac{1}{2}-m}u^*-\begin{pmatrix}
2g \\
g-m+\frac{1}{2}
\end{pmatrix}
u^{g-\frac{1}{2}+m} v^{g+\frac{1}{2}-m}v^* \biggr)\eta, \\ 
\mathcal{Z}^g_{g+\frac{1}{2}, m}&=\sqrt{\frac{1}{4\pi} \frac{(2g)!}{(g+m)!(g-m)!}}~(1-(2g+1)\eta^*\eta) ~u^{g+m}v^{g-m}.
\end{align}
\end{subequations}
Similarly, for a negative monopole charge $g = -|g| < 0$, we obtain  
\begin{subequations}
\begin{align}
Z^{-|g|}_{|g|+\frac{1}{2}, m}& =(-1)^{|g|+\frac{1}{2}+m}\sqrt{\frac{2|g|+1}{4\pi}\frac{(2|g|+1)!}{(|g|+\frac{1}{2}+m)! (|g|+\frac{1}{2}-m)!}}~{u^*}^{|g|+\frac{1}{2}-m} {v^*}^{|g|+\frac{1}{2}+m}\eta \nn\\
&+(-1)^{|g|+\frac{1}{2}+m}\sqrt{\frac{1}{4\pi}\frac{(|g|+\frac{1}{2}+m)! (|g|+\frac{1}{2}-m)! } {(2|g|)!}}~\nn\\
&\cdot (\begin{pmatrix}
2|g| \\
|g|+m+\frac{1}{2}
\end{pmatrix}
{u^*}^{|g|-\frac{1}{2}-m} {v^*}^{|g|+\frac{1}{2}+m}v-\begin{pmatrix}
2|g| \\
|g|+m-\frac{1}{2}
\end{pmatrix}
{u^*}^{|g|+\frac{1}{2}-m} {v^*}^{|g|-\frac{1}{2}+m}u \biggr)\eta^*, \\ 
\mathcal{Z}^{-|g|}_{|g|+\frac{1}{2}, m}&=(-1)^{|g|+\frac{1}{2}+m}\sqrt{\frac{1}{4\pi} \frac{(2|g|)!}{(|g|+m)!(|g|-m)!}}~(1-(2|g|+1)\eta^*\eta) ~{u^*}^{|g|-m}{v^*}^{|g|+m}.
\end{align}
\end{subequations}
Specifically, for $g = \pm 1/2$,  
\begin{subequations}
\begin{align}
&Z^{\frac{1}{2}}_{1,1} =\frac{1}{\sqrt{2\pi}}~({u}^2 \eta^* -uv^* \eta), ~~~Z^{\frac{1}{2}}_{1,0} =\frac{1}{\sqrt{4\pi}} (2 uv\eta^*+ (uu^*-vv^*)\eta), ~~~~Z^{\frac{1}{2}}_{1,-1} =\frac{1}{\sqrt{2\pi}} (u^*v\eta +{v}^2\eta^*), \nn\\
&\mathcal{Z}^{\frac{1}{2}}_{1,\frac{1}{2}} =\frac{1}{\sqrt{4\pi}}(1-2\eta^*\eta)u, ~~~~~\mathcal{Z}^{\frac{1}{2}}_{1,\frac{1}{2}} =\frac{1}{\sqrt{4\pi}}(1-2\eta^*\eta)v, \label{halfinlll+} \\
&Z^{-\frac{1}{2}}_{1,1} =\frac{1}{\sqrt{2\pi}}~({v^*}^2 \eta -uv^* \eta^*), ~~~Z^{-\frac{1}{2}}_{1,0} =-\frac{1}{\sqrt{4\pi}} (2 u^*v^*\eta- (uu^*-vv^*)\eta^*), ~~~~Z^{-\frac{1}{2}}_{1,-1} =\frac{1}{\sqrt{2\pi}} (u^*v\eta^* +{u^*}^2\eta), \nn\\
&\mathcal{Z}^{-\frac{1}{2}}_{1,\frac{1}{2}} =-\frac{1}{\sqrt{4\pi}}(1-2\eta^*\eta)v^*, ~~~~~\mathcal{Z}^{-\frac{1}{2}}_{1,\frac{1}{2}} =\frac{1}{\sqrt{4\pi}}(1-2\eta^*\eta)u^*. \label{halfinlll-}
\end{align}
\end{subequations}

\section{Useful formulas}

Here, we summarize the formulas for the monopole harmonics and  supermonopole harmonics used in the main text. 
 The products of the monopole harmonics are determined by the $SU(2)$ Clebsch-Gordan decomposition rule:  
\be
s \otimes s'=\oplus_{S=|s-s'|}^{s+s'} S~~~~~~~~(\Delta s=\Delta s'=\Delta S=1),
\ee
or 
\be
(2s+1)(2s'+1) =\sum_{S=|s-s'|}^{s+s'} (2S+1).
\ee
Similarly, the products of the supermonopole harmonics are determined by the $UOSp(1|2)$ Clebsch-Gordan decomposition  \cite{Pais-Rittenberg-1975, Scheunert-Nahm-Rittenberg-1977, Berezin-Tolstoy-1981, Minnaert-Mozrzymas-1990}: 
\be
l \otimes l'=\oplus_{L=|l-l'|}^{l+l'} L~~~~~~~(\Delta l=\Delta l'=\Delta L=1/2),
\ee
or 
\be
(4l+1)(4l'+1) =\sum_{L=|l-l'|}^{l+l'} (4L+1).
\ee

\subsection{For the monopole harmonics}\label{subsec:monoform}

The properties of monopole harmonics are  discussed in \cite{WuYang1977}.

\subsubsection{General formulas}

The product of two Wigner $D$-functions is expanded as  
\be
D_{l, m_1, m_2}\cdot  D_{l', m'_1, m'_2}  =\sum_{L, 
M_1, M_2}~\langle l, m_1 ; l', m_1' |L, M_1\rangle ~D_{L, M_1, M_2}~ \langle L, M_2|l, m_2; l', m_2'\rangle. \label{prodds}
\ee
In terms of the monopole harmonics (\ref{monohawig}), (\ref{prodds}) becomes 
\be
\Phi^g_{l,m} \cdot \Phi^{g'}_{l', m'} =\sum_{L, G, M}\sqrt{\frac{1}{4\pi}\frac{(2l+1)(2l'+1)}{2L+1}}~\langle l, g ; l', g' |L, G\rangle ~\Phi_{L, M}^G~ \langle L, M|l, m; l', m'\rangle.  \label{prodphiex}
\ee
The triple integral over $S^2$ is then given by 
\be
\int_{S^2}d\Omega_2 ~{\Phi_{L, M}^G}^*\cdot \Phi^g_{l,m} \cdot \Phi^{g'}_{l', m'}=\sqrt{\frac{1}{4\pi}\frac{(2l+1)(2l'+1)}{2L+1}}~\langle l, g ; l', g' |L, G\rangle ~\ \langle L, M|l, m; l', m'\rangle, \label{3phiint}
\ee
which can be expressed symmetrically using $3j$-symbols: 
\be
\int_{S^2}d\Omega_2 ~\Phi^{g_1}_{l_1, m_1}\cdot \Phi^{g_2}_{l_2, m_2}\cdot\Phi^{g_3}_{l_3, m_3} = \sqrt{\frac{(2l_1+1)(2l_2+1)(2l_3+1)}{4\pi}}\begin{pmatrix}
l_1 & l_2 & l_3 \\
g_1 & g_2 & g_3
\end{pmatrix}.\begin{pmatrix}
l_1 & l_2 & l_3 \\
m_1 & m_2 & m_3
\end{pmatrix}.
\ee

\subsubsection{Low-dimensional examples}

From (\ref{prodphiex}), we have 
\begin{subequations}
{\small
\begin{align}
&\Phi^{1/2}_{1/2, \pm 1/2}~ \Phi^g_{l,m} =\sqrt{\frac{(l+g+1)(l\pm m+1)}{{4\pi (l+1)(2l+1)}}}~\Phi^{g+\frac{1}{2}}_{l+\frac{1}{2}, m\pm \frac{1}{2}} \pm \sqrt{\frac{(l-g)(l\mp m)}{4\pi l(2l+1)}}~\Phi^{g+\frac{1}{2}}_{l-\frac{1}{2}, m\pm \frac{1}{2}} , \\
&\Phi^{-1/2}_{1/2,\pm 1/2}~ \Phi^g_{l,m} =\sqrt{\frac{(l-g+1)(l\pm m+1)}{4\pi (l+1)(2l+1)}}~\Phi^{g-\frac{1}{2}}_{l+\frac{1}{2}, m\pm \frac{1}{2}} \mp \sqrt{\frac{(l+g)(l\mp m)}{4\pi l(2l+1)}}~\Phi^{g-\frac{1}{2}}_{l-\frac{1}{2}, m\pm \frac{1}{2}} ,
\end{align}}
\label{prodmunumono}
\end{subequations}
and 
{\small
\begin{align}
&\Phi_{1, \pm1}^0~ \Phi_{lm}^g=  \mp \sqrt{\frac{3}{8\pi}} \frac{g}{l(l+1)}\sqrt{(l\pm m+1)(l \mp m)} ~\Phi^g_{l, m\pm 1}\nn\\
&~~~~~~~~~~~+\sqrt{\frac{3}{8\pi}} \frac{1}{l+1}\sqrt{\frac{((l+1)^2-g^2)(l\pm m +2)(l\pm m+1)}{(2l+3)(2l+1)}}~ \Phi^g_{l+1, m\pm 1} -\sqrt{\frac{3}{8\pi}} \frac{1}{l}\sqrt{\frac{(l^2-g^2)(l\mp m)(l\mp m-1)}{(2l+1)(2l-1)}}~\Phi^g_{l-1, m\pm 1}, \nn\\
&\Phi_{1,0}^0~\Phi_{lm}^g= \sqrt{\frac{3}{4\pi}} \frac{g}{l(l+1)} m ~\Phi^{g}_{l,m}  \nn\\
&~~~~~~~~~~~+\sqrt{\frac{3}{4\pi}}\frac{1}{l+1} \sqrt{\frac{((l+1)^2-g^2)((l+1)^2-m^2)}{(2l+3)(2l+1)}} ~\Phi^g_{l+1, m}+\sqrt{\frac{3}{4\pi}}\frac{1}{l}\sqrt{\frac{(l^2-g^2)(l^2-m^2)}{(2l+1)(2l-1)}}~\Phi^g_{l-1,m}.
\end{align}
}
Hence, 
\begin{subequations}
\begin{align}
&\int_{S^2} d\Omega_2 ~{\Phi_{l, m'}^g}^*\cdot  \Phi_{1,\pm 1 }^0 \cdot \Phi_{l , m}^g =\mp \sqrt{\frac{3}{8\pi}}\frac{g}{l(l+1)}\sqrt{(l\pm  m +1)(l \mp m)}~\delta_{m' , m \pm 1}, \\
&\int_{S^2} d\Omega_2 ~{\Phi_{l, m'}^g}^* \cdot \Phi_{1,0}^0 \cdot \Phi_{l , m}^g =\sqrt{\frac{3}{4\pi}}\frac{g}{l(l+1)} ~m~ \delta_{m' , m }. 
\end{align}\label{s2invec}
\end{subequations}
These formulas  are used in Sec.\ref{subsec:fuzzsphe}.

\subsection{For the supermonopole harmonics}\label{subsec:supermonoform}

We extend the formulas of the monopole harmonics  to the supermonopole harmonics. 

\subsubsection{General formulas}

The products of the supermonopole harmonics are expanded as 
{\small
\begin{subequations}
\begin{align}
&Y^g_{l,m} \cdot Y^{g'}_{l', m'} =\frac{1}{\sqrt{4\pi}} \biggl(1+(L-l-l')\frac{1}{2}\theta C\theta \biggr) \sum_{L, G, M} \langle l,g; l', g'|L, G\rangle~ Y_{L, M}^G~\langle L, M|l,m; l', m'\rangle, \\
&\mathcal{Y}^g_{l,m} \cdot \mathcal{Y}^{g'}_{l', m'} =-\frac{1}{2\sqrt{4\pi}}~ \theta C \theta ~\nn\\
&\cdot \sum_{L, G, M}
\biggl(
\sqrt{(l+g)(l'-g')}~\langle l-\frac{1}{2}, g-\frac{1}{2}; l'-\frac{1}{2}, g'+\frac{1}{2}|L, G \rangle-\sqrt{(l-g)(l'+g')}~\langle l-\frac{1}{2}, g+\frac{1}{2}; l'-\frac{1}{2}, g'-\frac{1}{2}|L, G \rangle
 \biggr)\nn\\
&~~~~~~~~~~\cdot Y_{L,M}^G ~\langle L, M|l-\frac{1}{2}, m ; l'-\frac{1}{2}, m'\rangle, \\
&Y_{l,m}^g \mathcal{Y}_{l', m'}^{g'}= \mathcal{Y}_{l', m'}^{g'}Y_{l,m}^g  \nn\\
&=\frac{1}{\sqrt{4\pi}}\sum_{L, G, M}\biggl( \sqrt{l'+g'} \langle l, g; l'-\frac{1}{2}, g'-\frac{1}{2}|L, G\rangle \eta -\sqrt{l'-g'} \langle l, g; l'-\frac{1}{2}. g'+\frac{1}{2}|L, G\rangle \eta^* \biggr)~ Y_{L,M}^G~\langle L,M|l,m; l'-\frac{1}{2}, m'\rangle.
\end{align}
\end{subequations}
}
Here, we used  (\ref{relorimobos}),   (\ref{genefermmon}) and (\ref{prodphiex}).
The triple integrals on $S^{2|2}$  are derived as 
{\small
\begin{subequations}
\begin{align}
&\int_{S^{2|2}}d\Omega_{2|2}~{Y^G_{L,M}}^* \cdot Y_{l,m}^g\cdot Y^{g'}_{l',m'} =\frac{1}{\sqrt{4\pi}}\frac{L+l+l'+1}{2L+1} \langle l,g; l', g'|L, G\rangle \langle L, M|l, m; l', m'\rangle, \\
&\int_{S^{2|2}}d\Omega_{2|2}~{Y^G_{L,M}}^* \cdot \mathcal{Y}_{l,m}^g\cdot \mathcal{Y}^{g'}_{l',m'} =\frac{1}{\sqrt{4\pi}}\frac{1}{2L+1}  \\
&~~\cdot \biggl(
\sqrt{(l+g)(l'-g')}~\langle l-\frac{1}{2}, g-\frac{1}{2}; l'-\frac{1}{2}, g'+\frac{1}{2}|L, G \rangle-\sqrt{(l-g)(l'+g')}~\langle l-\frac{1}{2}, g+\frac{1}{2}; l'-\frac{1}{2}, g'-\frac{1}{2}|L, G \rangle
 \biggr)\nn\\
&~~\cdot\langle L, M|l-\frac{1}{2}, m ; l'-\frac{1}{2}, m'\rangle, \\
&\int_{S^{2|2}}d\Omega_{2|2}~{\mathcal{Y}^G_{L,M}}^* \cdot Y_{l,m}^g\cdot \mathcal{Y}^{g'}_{l',m'}=
\int_{S^{2|2}}d\Omega_{2|2}~{\mathcal{Y}^G_{L,M}}^* \cdot \mathcal{Y}^{g'}_{l',m'}\cdot  Y_{l,m}^g
=\frac{1}{\sqrt{4\pi}}\frac{1}{2L}\nn\\
&~~\cdot \biggl( \sqrt{(L+G)(l'+g')} \langle l, g; l'-\frac{1}{2}, g'-\frac{1}{2}|L-\frac{1}{2}, G-\frac{1}{2}\rangle  +\sqrt{(L-G)(l'-g')} \langle l, g; l'-\frac{1}{2}. g'+\frac{1}{2}|L-\frac{1}{2}, G+\frac{1}{2}\rangle  \biggr)\nn\\
&~~\cdot \langle L-\frac{1}{2},M|l,m; l'-\frac{1}{2}, m'\rangle, \nn\\
&\int_{S^{2|2}}d\Omega_{2|2}~{\mathcal{Y}^G_{L,M}}^* \cdot Y_{l,m}^g\cdot {Y}^{g'}_{l',m'}=
\int_{S^{2|2}}d\Omega_{2|2}~{\mathcal{Y}^G_{L,M}}^* \cdot \mathcal{Y}_{l,m}^g\cdot \mathcal{Y}^{g'}_{l',m'}=0.
\end{align}\label{prodsupermonos}
\end{subequations}
} 
Other triple integrals are 
{\small
\begin{subequations}
\begin{align}
&\int_{S^{2|2}}d\Omega_{2|2}~{Y^G_{L,M}} \cdot {Y_{l,m}^g}^*\cdot  Y^{g'*}_{l',m'}
=\int_{S^{2|2}}d\Omega_{2|2}~{Y^G_{L,M}}^* \cdot Y_{l,m}^g\cdot Y^{g'}_{l',m'}, \\
&\int_{S^{2|2}}d\Omega_{2|2}~{\mathcal{Y}^G_{L,M}} \cdot {Y_{l,m}^g}^*\cdot \mathcal{Y}^{g' *}_{l',m'}=\int_{S^{2|2}}d\Omega_{2|2}~{\mathcal{Y}^G_{L,M}}^* \cdot Y_{l,m}^g\cdot \mathcal{Y}^{g'}_{l',m'}, \\
&\int_{S^{2|2}}d\Omega_{2|2}~{\mathcal{Y}^G_{L,M}} \cdot {Y_{l,m}^g}^*\cdot {Y}^{g' *}_{l',m'}=-\int_{S^{2|2}}d\Omega_{2|2}~{\mathcal{Y}^G_{L,M}}^* \cdot Y_{l,m}^g\cdot {Y}^{g'}_{l',m'},
\end{align}\label{otherintegth}
\end{subequations}
}where  the right-hand sides  are given by (\ref{prodsupermonos}). In terms of the $3j$-symbols, the  triple integrals are represented as 
\begin{subequations}
\begin{align}
&\int_{S^{2|2|}}d\Omega_{2|2}~Y^{g_1}_{l_1, m_1}\cdot Y^{g_2}_{l_2, m_2}\cdot Y^{g_3}_{l_3, m_3}=\frac{1}{\sqrt{4\pi}}(l_1+l_2+l_3+1) \begin{pmatrix}
l_1 & l_2 & l_3 \\
g_1 & g_2 & g_3 
\end{pmatrix} \begin{pmatrix}
l_1 & l_2 & l_3 \\
m_1 & m_2 & m_3 
\end{pmatrix}, \\ 
&\int_{S^{2|2|}}d\Omega_{2|2}~Y^{g_1}_{l_1, m_1}\cdot \mathcal{Y}^{g_2}_{l_2, m_2}\cdot \mathcal{Y}^{g_3}_{l_3, m_3}=\frac{1}{\sqrt{4\pi}}\begin{pmatrix}
l_1 & l_2-\frac{1}{2} & l_3 -\frac{1}{2} \\
m_1 & m_2 & m_3 
\end{pmatrix}\nn\\
&~~~\cdot \biggl(  \sqrt{(l_2+g_2)(l_3-g_3)}\begin{pmatrix}
l_1 & l_2-\frac{1}{2} & l_3-\frac{1}{2} \\
g_1 & g_2-\frac{1}{2} & g_3 +\frac{1}{2}
\end{pmatrix}-\sqrt{(l_2-g_2)(l_3+g_3)} \begin{pmatrix}
l_1 & l_2-\frac{1}{2} & l_3 -\frac{1}{2}\\
g_1 & g_2 +\frac{1}{2} & g_3-\frac{1}{2} 
\end{pmatrix}\biggr) , \\ 
&\int_{S^{2|2|}}d\Omega_{2|2}~{Y}^{g_1}_{l_1, m_1}\cdot {Y}^{g_2}_{l_2, m_2}\cdot \mathcal{Y}^{g_3}_{l_3, m_3}=\int_{S^{2|2|}}d\Omega_{2|2}~\mathcal{Y}^{g_1}_{l_1, m_1}\cdot \mathcal{Y}^{g_2}_{l_2, m_2}\cdot \mathcal{Y}^{g_3}_{l_3, m_3}=0.
\end{align}
\end{subequations}
For other supermonopole harmonics,  we have   
\begin{subequations}
\begin{align}
&\int_{S^{2|2|}}d\Omega_{2|2}~\mathcal{Z}^{g_1}_{l_1, m_1}\cdot \mathcal{Z}^{g_2}_{l_2, m_2}\cdot \mathcal{Z}^{g_3}_{l_3, m_3}=-\frac{1}{\sqrt{4\pi}}(l_1+l_2+l_3+\frac{1}{2}) \begin{pmatrix}
l_1-\frac{1}{2} & l_2-\frac{1}{2} & l_3-\frac{1}{2} \\
g_1 & g_2 & g_3 
\end{pmatrix} \begin{pmatrix}
l_1-\frac{1}{2} & l_2-\frac{1}{2} & l_3-\frac{1}{2} \\
m_1 & m_2 & m_3 
\end{pmatrix}, \\  
&\int_{S^{2|2|}}d\Omega_{2|2}~\mathcal{Z}^{g_1}_{l_1, m_1}\cdot {Z}^{g_2}_{l_2, m_2}\cdot {Z}^{g_3}_{l_3, m_3}=\frac{1}{\sqrt{4\pi}}\begin{pmatrix}
l_1-\frac{1}{2} & l_2 & l_3  \\
m_1 & m_2 & m_3 
\end{pmatrix}\nn\\
&~~~\cdot \biggl(  \sqrt{(l_2+g_2+\frac{1}{2})(l_3-g_3+\frac{1}{2})}\begin{pmatrix}
l_1-\frac{1}{2} & l_2 & l_3 \\
g_1 & g_2+\frac{1}{2} & g_3 -\frac{1}{2}
\end{pmatrix}-\sqrt{(l_2-g_2+\frac{1}{2})(l_3+g_3+\frac{1}{2})} \begin{pmatrix}
l_1-\frac{1}{2} & l_2 & l_3 \\
g_1 & g_2 -\frac{1}{2} & g_3+\frac{1}{2} 
\end{pmatrix}\biggr) , \\ 
&\int_{S^{2|2|}}d\Omega_{2|2}~{Z}^{g_1}_{l_1, m_1}\cdot \mathcal{Z}^{g_2}_{l_2, m_2}\cdot \mathcal{Z}^{g_3}_{l_3, m_3}=\int_{S^{2|2|}}d\Omega_{2|2}~{Z}^{g_1}_{l_1, m_1}\cdot {Z}^{g_2}_{l_2, m_2}\cdot {Z}^{g_3}_{l_3, m_3}=0.
\end{align}
\end{subequations}

\subsubsection{Low-dimensional examples}

Equations (\ref{prodsupermonos}) yield  
\begin{subequations}
\begin{align}
&\int_{S^{2|2}}d\Omega_{2|2}~{Y^g_{l,m'}}^* \cdot Y_{1,\pm 1}^0\cdot Y^{g}_{l,m} =\mp  \frac{1}{\sqrt{2\pi}}\frac{g}{l(2l+1)}\sqrt{l(l+1) \mp m(m\pm 1)}  ~\delta_{m', m \pm 1}, \label{sf1} \\
&\int_{S^{2|2}}d\Omega_{2|2}~{Y^g_{l,m'}}^* \cdot Y_{1,0}^0\cdot Y^{g}_{l,m} = \frac{1}{\sqrt{\pi}} \frac{g}{l(2l+1)}m ~\delta_{m', m }, \label{sf2}\\
&\int_{S^{2|2}}d\Omega_{2|2}~{\mathcal{Y}^g_{l,m'}}^* \cdot Y_{1,\pm 1}^0\cdot \mathcal{Y}^{g}_{l,m} =\mp \frac{1}{\sqrt{2\pi}} \frac{g}{l(2l+1)}\sqrt{(l-\frac{1}{2})(l+\frac{1}{2}) \mp m(m\pm 1)}  ~\delta_{m', m \pm 1}, \label{sf3} \\
&\int_{S^{2|2}}d\Omega_{2|2}~{\mathcal{Y}^g_{l,m'}}^* \cdot Y_{1,0}^0\cdot \mathcal{Y}^{g}_{l,m} =\frac{1}{\sqrt{\pi}}\frac{g}{l(2l+1)} m~\delta_{m', m}, \label{sf4} \\
&\int_{S^{2|2}} d\Omega_{2|2}~ {Y_{l , m'}^g}^* \cdot \mathcal{Y}_{1, \pm \frac{1}{2}}^0 \cdot \mathcal{Y}_{l,m}^g =-\frac{1}{\sqrt{\pi}}\frac{g}{2l(2l+1)} \sqrt{l\pm m +\frac{1}{2}}~\delta_{m', m\pm \frac{1}{2}}, \label{sf5} \\
&\int_{S^{2|2}} d\Omega_{2|2}~ {\mathcal{Y}_{l , m'}^g}^* \cdot \mathcal{Y}_{1,\pm \frac{1}{2}}^0  \cdot {Y}_{l, m}^g=\mp \frac{1}{\sqrt{\pi}} \frac{g}{2l(2l+1)} \sqrt{l\mp m}~\delta_{m', m\pm \frac{1}{2}}. \label{sf6}
\end{align}\label{threenewforsusy}
\end{subequations}
These formulas are used in Sec.\ref{subsec:matrcoorintegerll}.

\section{The lowest Landau level supergeometry}\label{append:lllgeo}

Here, we provide  concise derivations of the fuzzy supersphere geometry for the LLL, employing  the Lagrangian and Hamiltonian formalisms, respectively.

\subsection{From Lagrange formalism}

The SUSY  Lagrangian for the present system may be given by 
\be
L=\frac{M}{2}(\dot{x}_i \dot{x_i}+C_{\alpha\beta}\dot{\theta}_{\alpha}\dot{\theta}_{\beta})-\dot{x}_iA_i-\dot{\theta}_{\alpha}A_{\alpha}. 
\ee
In the LLL, the kinetic terms are quenched, and the Lagrangian reduces to a first-order form: 
\be
L_{\text{LLL}}=-\dot{x}_iA_i-\dot{\theta}_{\alpha}A_{\alpha}=-2ig \psi^{\ddagger}\dot\psi. 
\ee
The canonical momenta for  $\psi_{a}=(u,v,\eta)$ are derived as  
\be
\pi_{a} =\frac{{\partial}}{\partial \dot{\psi}_{a}} L_{\text{LLL}} =-2ig \psi^{*}_{a}. 
\ee
Imposing the canonical super-commutation relations $[\pi_{a}, \psi_{b}\} = -i \delta_{ab}$, we obtain 
\begin{equation}
\pi_{a} =-i\frac{\partial}{\partial \psi_{a}} 
\end{equation}
or 
\be
\psi^*_{a} =\frac{1}{2g}\frac{\partial}{\partial \psi_{a}}.  \label{psicomp} 
\ee
By substituting (\ref{psicomp}) into the graded Hopf map (\ref{Hopfmapsusy}), we promote the coordinates of the supersphere to operators:
\be
X_i =\frac{r}{g} \psi^tl_i \frac{\partial} {\partial \psi}, ~~~~\Theta_\alpha =\frac{r}{g} \psi^tl_\alpha \frac{\partial} {\partial \psi}.
\ee
They obviously satisfy the super-algebra of the fuzzy supersphere;  
\be
[X_i, X_j] =i\frac{r}{g} \epsilon_{ijk}X_k, ~~~[X_i, \Theta_{\alpha}] = \frac{r}{2g}(\sigma_i)_{\beta\alpha}\Theta_{\beta}, ~~~\{\Theta_{\alpha}, \Theta_{\beta}\} = \frac{r}{2g}(C\sigma_i )_{\alpha\beta} X_i, \label{llleffeal}
\ee 
and 
\be
X_iX_i+C_{\alpha\beta}\Theta_{\alpha}\Theta_{\beta} =(\frac{r}{g})^2~ L(L+\frac{1}{2}) =r^2 ~(1+\frac{1}{2|g|}), 
\ee
where $L$ is given by (\ref{opll}) with eigenvalue  $|g|$.

\subsection{From Hamilton formalism}

In the LLL, the kinetic part of the Hamiltonian is quenched, which effectively sets the covariant derivatives to zero: $D_i \rightarrow 0$ and $D_\alpha \rightarrow 0$. With this replacement, the $L_A$ (\ref{llds}) reduce to the following forms \cite{Hasebe-Kimura-2005}:
\be
L_i ~~\rightarrow ~~r^2 B_i=\frac{g}{r}x_i, ~~~~~~~ L_\alpha ~~\rightarrow ~~r^2 B_\alpha=\frac{g}{r}\theta_\alpha,
\ee
implying that the superspin index $l$  reduces to $|g|$. 
Then,   the coordinates on the supersphere can be identified with the $UOSp(1|2)$ generators:  
\be
X_i =\frac{r}{g}L_i, ~~~~\Theta_{\alpha} =\frac{r}{g}L_{\alpha},
\ee
which obey the super-commutation relations (\ref{llleffeal}) and satisfy the fuzzy supersphere constraint:
\be
X_iX_i+C_{\alpha\beta}\Theta_{\alpha}\Theta_{\beta} =(\frac{r}{g})^2~ (L_iL_i+C_{\alpha\beta}L_{\alpha}L_{\beta}) =r^2 ~(1+\frac{1}{2|g|}).
\ee

\bibliographystyle{ytamsalpha}

\bibliography{refs} 


\end{document}